\documentclass[a4paper,structabstract]{aa}

\usepackage{aas_macros}
\usepackage{natbib}
\usepackage{txfonts}
\usepackage{array}
\usepackage{wasysym} 
 \usepackage[squaren,Gray,thinspace,thickqspace]{SIunits}
 \usepackage{mathtools} 
\usepackage{nicefrac}

\usepackage{graphicx,subfigure}

\usepackage[lofdepth,lotdepth]{subfig} 

\def\NRe{N_{\rm R_{\rm e}}}
\def\sa{s_{\rm a}}

\def\chim{\langle\chi\rangle}
\def\chit{\bar{\chi}}

\def\ie{{\it i.e.}}
\def\eg{{\it e.g.},}

\def\aap{{\it Astron. \& Astrophys}}

\begin{document}

\title{3D mixing in hot Jupiters atmospheres. \\ I. Application to the day/night cold trap in HD 209458b}
\titlerunning{3D mixing in hot Jupiters atmospheres I}
\authorrunning{Parmentier et al.}

\author{Vivien Parmentier\inst{1}
\and
Adam P. Showman\inst{2}
\and
Yuan Lian\inst{3} }

\offprints{V. Parmentier}

\institute{Universit\'e de Nice-Sophia Antipolis, Observatoire de la C\^ote d'Azur, CNRS UMR 6202, B.P. 4229, 06304 Nice Cedex 4, France\\
           \email{vivien.parmentier@oca.eu}
           \and 
Department of Planetary Sciences, Lunar and Planetary Laboratory,University of Arizona, Tucson AZ, USA
	 \and 
Ashima Research, Suite 104, 600 South Lake Ave., Pasadena, CA 91106, USA	 
}

\date{Received January 18, 2013 / Accepted August 12, 2013}

\abstract
{Hot Jupiters exhibit atmospheric temperatures ranging from hundreds
  to thousands of Kelvin.  Because of their large day-night
  temperature differences, condensable species that are stable in the
  gas phase on the dayside---such as TiO and silicates---may condense
  and gravitationally settle on the nightside.  Atmospheric
  circulation may counterbalance this tendency to gravitationally
  settle. This three-dimensional (3D) mixing of {condensable} species has
  not previously been studied for hot Jupiters, yet it is crucial to
  assess the existence and distribution of TiO and silicates in the
  atmospheres of these planets.}
{We investigate the strength of the nightside cold trap in hot Jupiters
  atmospheres by investigating the mechanisms and strength of
  the vertical mixing in these stably stratified atmospheres. We apply
  our model to the particular case of TiO to address the question of
  whether TiO can exist at low pressure in sufficient abundances to
  produce stratospheric thermal inversions despite the nightside cold trap.}
{We modeled the 3D circulation of HD 209458b including passive (\ie radiatively inactive) tracers that advect with the
  3D flow, with a source and sink term on the nightside to
  represent their condensation into haze particles and their gravitational settling.}
{We show that global advection patterns produce strong vertical
  mixing that can keep condensable species aloft as long as they are
  trapped in particles of sizes of a few microns or less on the night
  side. We show that vertical mixing results not from small-scale
  convection but from the large-scale circulation driven by the
  day-night heating contrast. Although this vertical mixing is not
  diffusive in any rigorous sense, a comparison of our results with
  idealized diffusion models allows a rough estimate of the effective
  vertical eddy diffusivities in these atmospheres. The parametrization
  $K_{zz}=\unit{\frac{5\times
      10^{4}}{\sqrt{P_{bar}}}}\meter\squared\reciprocal\second$, valid from $\sim$1 bar to a few
$\mu$bar, can be used
  in 1D models of HD 209458b. Moreover, our models exhibit strong
  spatial and temporal variability in the tracer concentration that
  could result in observable variations during either transit or
  secondary eclipse measurements. Finally, we apply our model to the case
  of TiO in HD 209458b and show that the day-night cold trap would
  deplete TiO if it condenses into particles bigger than a few microns
  on the planet's nightside, keeping it from creating the observed
  stratosphere of the planet.}
{}


\maketitle
%
%
\section{Introduction}

The year 1988 was marked by the discovery of the first substellar object
outside our solar system \citep{Becklin1988}. This discovery was
followed by numerous other discoveries of cool brown
dwarfs, whose spectra differ significantly from stars.  This led to the
definition of three new spectral classes beyond the stellar classes O
to M: the L, T, and Y dwarfs, ranging from temperatures of
$~\unit{2300}\kelvin$ to $\unit{350}\kelvin$ \citep{Cushing2011}. To
understand of these brown dwarfs and the mechanisms that
cause the transition from one spectral class to another, it is
fundamental to take condensation processes into account: in cold
atmospheres molecules can form, condense, and rain out, affecting
the observed spectral features of these objects. In particular, the M
to L transition is marked by the disappearance of the titanium oxide
(TiO) bands, which is understood by its transformation into titanium dioxide (TiO$_{2}$)
and condensation into perovskite (CaTiO$_{3}$) \citep{Lodders2002}. The
L to T transition is understood as the switch from a cloudy atmosphere
to a cloud-free atmosphere as the cloud layer migrates into the deep,
unobservable regions of the star (see~\citet{Kirkpatrick2005} for a
review). Because brown dwarfs produce their own light, high-quality spectra
of their atmosphere can be measured. Numerous atmospheric models
have been built to fit the data, from simple models with a reduced set of
free parameters \citep{Ackerman2001} to sophisticated models
including the detailed physics of condensation, growth, and settling of
particles \citep{Woitke2004}.

The discovery and characterization of exoplanets followed close
behind. The first detection of a planet around a main sequence star
by \citet{Mayor1995} opened the trail for discovering hundreds of
exoplanets. Some years later, atmospheric characterization of these
objects became available, both from transit spectroscopy
\citep{Charbonneau2002} and from direct detection of the planet's
thermal emission
\citep{Charbonneau2005}. Although the global-mean effective
temperatures of these planets are similar to those on brown dwarfs, a
major difference is that hot Jupiters are strongly irradiated.
Depending on the incident stellar flux, planetary rotation rate, and
other factors, atmospheric circulation models show that this day-night
heating gradient can lead, at low pressures, to nightside temperatures
that are at least $\sim$1000 K colder than dayside temperatures
\citep[\eg][]{Showman2008, Showman2009, Dobbs-Dixon2010,
  Rauscher2010, Rauscher2012b,  Heng2011a, Heng2011, Perna2012}.  As a result, a
wide variety of condensable species that are stable in the gas phase
on the dayside may condense on the nightside, leading to the formation
of particles there. In the absence of atmospheric vertical mixing,
such particles would gravitationally settle, depleting the atmosphere
of these species on both the dayside and the nightside.
Sufficiently strong vertical mixing, anywhere on the planet, however, may keep these particles
suspended in the atmosphere, allowing them to sublimate into the gas
phase in any air transported from nightside to dayside.  Thus,
the existence of this nightside ``cold trap'' is crucial for
understanding not only the existence of hazes on hot Jupiters
\citep[\eg][]{Pont2013} but also the gas-phase composition of
the atmosphere on both the dayside and the nightside.

These arguments are relevant to a wide range of titanium and vanadium
oxides, silicate oxides, and other species.  In particular,
chemical-equilibrium calculations show that, at temperatures of
$\sim$1000--2000 K, there exist a wealth of condensates including
Ti$_2$O$_3$, Ti$_3$O$_5$, and Ti$_4$O$_7$ (among other titanium
oxides), MgAl$_2$O$_4$, Mg$_2$SiO$_4$, MgSiO$_3$, NaAlSi$_3$O$_8$,
KAlSi$_3$O$_8$, and several phosphorus oxides \citep{Burrows1999, Lodders2002}.  Understanding the possible existence of
gas-phase TiO, Na, K, and other species on the dayside therefore requires an understanding of the nightside cold trap.

A particularly interesting problem in this regard is the existence of
exoplanet stratospheres.  For some transiting hot Jupiters, Spitzer IRAC
secondary-eclipse observations indicate the presence of thermal
inversions (stratospheres) on these planets' daysides \citep{Knutson2008}.  These stratospheres are generally thought to result
from absorption of starlight by strong visible/ultraviolet absorbers,
but debate exists about the specific chemical species that are
responsible. \citet{Hubeny2003} and \citet{Fortney2008} showed that,
because of their enormous opacities at visible wavelengths, the
presence of gaseous titanium and vanadium oxides can lead to
stratospheres analogous to those inferred on hot Jupiters. This hypothesis is supported by the possible detection of TiO by~\citet{Desert2008}
in the atmospheric limb of HD 209458b.
Then, a crucial question is whether the nightside
cold trap would deplete the atmosphere of TiO, preventing this
species from serving as the necessary absorber \citep{Showman2009,
Spiegel2009}.

As pointed out by several authors, there exists another possible
cold trap. 1D radiative-transfer models suggest
that, on some hot Jupiters, the global-mean temperature-pressure
profile becomes sufficiently cold for condensation of gaseous TiO to
occur at pressures of tens to hundreds of bars
\citep[\eg][]{Fortney2008}.  Even though a stratosphere on such a
planet would be sufficiently hot for TiO---if present---to exist in
the gas phase, the condensation of TiO and downward settling of the
resulting grains at $\sim$10--100 bars might prevent the existence
of TiO in the atmosphere \citep{Showman2009, Spiegel2009}.
The strength of these two cold traps is given by a competition between
gravitational settling and upward mixing. The vertical mixing results
from complex 3D flows and can be inhomogeneous over the planet \citep{Cooper2006,Heng2011a}. When using a 1D-model, the vertical mixing is
usually considered to be diffusive only, and parametrized by a
diffusion coefficient $K_{zz}$. Several studies give an estimate for
this vertical diffusion coefficient in hot Jupiters
atmospheres. \citet{Heng2011a} uses the magnitude of the Eulerian mean
streamfunction as a proxy for the strength of the vertical motions and
derived a vertical mixing coefficient of the order of $K_{zz}\approx
\unit{10^{6}}\meter\squared\reciprocal\second$. Other authors used an estimate based on the root
mean squared vertical velocity either the local value
\citep{Cooper2006} or the planet-averaged value
\citep{Moses2011}. However these estimates are crude and 
there is a need for theoretical work to more rigorously characterize the
vertical mixing rates in hot Jupiters atmospheres.

In this study we focus in the dynamical mixing of condensable species. We therefore neglect all the potential feedback of the condensable species on the atmospheric flow such as the radiative effects of the condensates \citep{Heng2012,Dobbs-Dixon2012}, non-equilibrium chemistry due to the depletion of a particular species, latent heat release during condensation among others.  We present three-dimensional general circulation model (GCM) experiments
of HD 209458b
to model a chemical species that condenses and settles on the nightside
of the planet. We show that mixing in hot Jupiters atmospheres is
dominated by large-scale circulation flows resolved by the
GCM. Finally, from the 3D model, we derive the values of an effective
vertical mixing coefficient, representing the averaged vertical mixing in the
planet atmosphere. We also compare these 3D models to an idealized 1D model
parameterizing the mixing using an eddy diffusivity, with the goal of
estimating an effective eddy diffusivity for the mixing rates in 
the 3D models. These are the first circulation models of hot Jupiters to
include the influence of the dynamics on condensable species.

\section{3D model}

Here, we use a state-of-the-art 3D circulation model, coupled to a
passive tracer representing a condensable species, to determine how
the interplay between dynamical mixing and vertical settling controls
the spatial distribution of condensable species on hot Jupiters.
Although the day/night cold trap should be present in most hot
Jupiters, for concreteness, we must select a particular system to
investigate. HD 209458b is among the best studied hot Jupiters. It is believed to harbor a stratosphere \citep{Knutson2008} and a strong day-night
temperature contrast \citep{Showman2009}. We decided to use it as our
reference model in this study, keeping in mind that most of the
mechanisms discussed here should apply to all hot Jupiters. To
model the atmosphere of HD 209458b we use the 3D Substellar and
Planetary Atmospheric Radiation and Circulation (SPARC/MITgcm) model
of \citet{Showman2009}, which couples the plane-parallel, multi-stream
radiative transfer model of \citet{Marley1999} to the MITgcm
\citep{Adcroft2004}.

\subsection{Dynamics}
\label{sec::Dynamics}
To model the dynamics of the planet we solve the global,
three-dimensional primitive equations in spherical geometry using the
MITgcm, a general circulation model for atmosphere and oceans developed
and maintained at the Massachusetts Institute of Technology. The
primitive equations are the standard equations used in stably
stratified flows where the horizontal dimensions greatly exceed the
vertical ones. In hot Jupiters, the horizontal scales are
$10^{7}-10^{8}\meter$ whereas the vertical scale height fall between
$200$ and $500\kilo\meter$ leading to an aspect ration of $20$ to
$500$. In order to minimize the constraints on the timestep by the CFL
criterion, we solve the equations on the cubed-sphere
grid as described in \citet{Adcroft2004}. The simulations do not contain any explicit
viscosity nor diffusivity. However, in order to smooth the grid noise
and ensure the stability of the code we use a horizontal fourth-order
Shapiro filter \citep{Shapiro1970}. In the vertical direction, no
filtering process is applied. \citet{kalnay2003} provide a more detailed description of the
equations (see \citet{Showman2009} for the numerical method used to solve them).

We use a gravity of $\unit{9.81}\meter\usk\rpsquare\second$, a
planetary radius of $\unit{9.44\times 10^{7}}\meter$, and a rotation
rate of $\unit{2.06\times10^{-5}}.\reciprocal\second$ (implying a
rotation period of 3.5 days).  The average pressure ranges from
200 bars at the bottom of the atmosphere to $p_{\rm top}$ at the top of
the second-highest level, with the uppermost level extending from a
pressure of $p_{\rm top}$ to zero.  In most models, $p_{\rm top}$ is
$\unit{2}\micro\bbar$ with 53 vertical levels. In some
models---particularly those with the largest cloud particle size---we
adopt $P_{\rm top}$ of $20\micro$bar with 47 vertical levels.  In
either case, this leads to a resolution of almost three levels per
scale height. We use a horizontal resolution of C32, equivalent to an
approximate resolution of 128 cells in longitude and 64 in latitude and a timestep of $\unit{15}\second$.
We reran some models at C64 resolution (equivalent to an approximate
resolution of $256\times 128$ in longitude and latitude) to check
convergence.

It is worth mentioning that, although the primitive equations are hydrostatic
in nature, it does not imply an absence of vertical motion ~\citep[see section 3.5 of][]{Holton1992}.  In the primitive equations, the horizontal divergence is
generally non-zero, which requires the presence of vertical motions.  Indeed,
the dayside radiative heating and nightside radiative cooling are balanced by a
combination of horizontal and vertical thermal advection. In many cases, including Earth's tropics, the vertical advection dominates; in such cases, vertical motions play a crucial, zeroth-order role in the thermodynamic energy balance. For conditions relevant to hot Jupiters, GCMs and order-of-magnitude calculations show that vertical motions of $\unit{10-100}\meter\reciprocal\second$ or more are expected, despite the fact that the atmosphere is stably stratified.  Further discussion can be found in~\citet{Showman2002} and in Sect. 6.2 of \citet{Showman2008}.

\subsection{Radiative transfer}
\label{sec::RadTrans}
The radiative transport of energy is calculated with the
plane-parallel radiative transfer code of \citet{Marley1999}. The code
was first developed for Titan's atmosphere \citep{McKay1989} and since
then has been extensively used for the study of giant planets \citep{Marley1996}, brown
dwarfs \citep{Marley2002,Burrows1997}, and hot Jupiters \citep[\eg][]{Fortney2005, Fortney2008, Showman2009}. We use the
opacities developed by \citet{Freedman2008}, including more recent
updates, and the molecular abundances described by \citet{Lodders2002a}
and \citet{Visscher2006}.

As in the models of HD 209458b presented by
\citet{Showman2009}, our opacity
tables include gas-phase TiO and VO whenever temperatures are locally
high enough for TiO to reside in gas phase.  Note that, because
of the assumption of local chemical equilibrium, 
we do not consider the effect of cold traps on the
atmospheric composition, and thus on the opacities. TiO opacities are
always taken into account where temperatures are high enough for
TiO to exist in the gas phase. This
causes a warm stratosphere on the dayside of our modeled planet,
regardless of the 3D distribution of our tracers (to be described below).

Metallicities are solar, as given by~\citet{Freedman2008}. A metal enhanced atmosphere should modestly change the day-night temperature difference (at a given pressure) and the pressure of the photosphere itself, as shown in~\citet{Showman2009}. Cloud opacities are ignored in this study and might alter the flow more significantly~\citep{Dobbs-Dixon2012a}.

Opacities are described using the correlated-$k$ method \citep[\eg][]
{Goody1961}.  We consider 11 frequency bins for the opacities ranging
from $0.26$ to $\unit{300}\micro\meter$; within each bin, opacity
information from typically 10,000 to 100,000 frequency intervals is
represented statistically over 8 $k$-coefficients. This is radically different from other methods in the literature. In particular, ~\citet{Dobbs-Dixon2012a} uses a multi-bin approach for the opacity, but inside each of their 30 frequency bins they use a single average opacity. However, inside each bin of frequency, the line by line opacity varies by several orders of magnitudes and no mean, neither the Planck mean nor the Rosseland mean opacity (see~\citet{Parmentier2013b} and \citet{Parmentier2013c} for more details) can take into account this large variability. Thus, as stated in~\citet{Dobbs-Dixon2012a}, our correlated-$k$ radiative transfer is today's most sophisticated radiative transfer approach implemented in a hot Jupiters GCM. \citet{Showman2009} provide a detailed
description of the radiative-transfer model and its implementation in
the GCM.

\subsection{Tracer fields} 
Our target species is represented by a passive tracer field. Tracer
fields are often used in GCMs to follow the concentration of a
chemical species such as water vapor or cloud amount in the atmosphere
or salinity in the oceans. \citet{Cooper2006} were the first to
include a passive tracer in a circulation model of hot Jupiters, in
their case to investigate the quenching of CO and CH$_4$ due to
atmospheric mixing.

The tracer field is advected by the flow
calculated in the GCM. Thus the tracer abundance $\chi$ is given by
the continuity equation:
 \begin{equation}
\frac{D\chi}{Dt}=S
\end{equation}
Here, $\chi$ represents the mole fraction of the tracer, i.e., the
number of molecules of the species in a given volume, either in
gaseous phase or trapped in condensed particles, with respect to the
total number of atmospheric molecules in that volume. For simplicity the value of $\chi$ is normalized to its initial value in the deep layers of the planet. In the equation,
$S$ is a source term and the total derivative is defined by
$D/Dt = {\partial/\partial t} + {\bf v}\cdot\nabla + \omega{\partial\over
\partial p}$, where $t$ is time, ${\bf v}$ is the horizontal velocity,
$\nabla$ is the horizontal gradient operator on the sphere, $\omega 
= DP/Dt$ is the vertical velocity in pressure coordinates, and $P$ is 
pressure. {We model the simplest possible horizontal cold trap in hot Jupiters atmospheres : a situation where the day/night temperature contrast is so strong that our target species is gaseous in the dayside of the planet but condenses and is incorporated in particles of size $a$ in the nightside}. Thus the source term $S$ represents the
gravitational settling of these particles and is given by
\begin{equation}
S= 
\begin{cases}
  0  &\text{on the dayside} \\
  \frac{1}{\rho}\frac{\partial (\rho\chi V_{\rm f})}{\partial z}&  \text{on the nightside}
  \end{cases} 
  \label{eq::Source}
\end{equation}
where $z$ is the height, increasing upward, $\rho$ is the density of the air and $V_{\rm f}$ is the settling velocity of the particles defined by Eq.~\eqref{eq::Vf}. This
velocity depends on the size of the particles, which is determined by
the complex microphysics of condensation, out of the scope of this
study~\citep[see][]{Woitke2003}. Thus we treat $a$ as a free parameter in our model. We model
spherical particles of radii $0.1$, $0.5$, $1$, $2.5$, $5$, and
$\unit{10}\micro\meter$.  Equation \eqref{eq::Source} describes a
simple, bimodal mechanism for the condensation of chemical species on
the nightside. Although this scheme is highly simplified (ignoring the
detailed temperature and pressure dependence of the condensation
curves of possible condensates in hot Jupiters atmospheres), it
reflects the fact that a wide range of species will reside in gaseous
form on the dayside yet condensed form at low pressures on the
nightside.

We consider several independent passive tracers, representing species that condenses in different particle size. Thus, they do not influence
either the dynamics or the radiative transfer of the
simulation and does not interact with each others. This ignores a role for possible radiative
feedback mechanisms -- discussed in Sect.
\ref{sec::Applications} -- or particle growth but represents a necessary first step
toward understanding how dynamics controls the 3D distribution
of a condensable species.

\subsection{Settling velocity}
\label{sec::Velocity}
We assume that the target chemical species condenses, in the nightside only, into spherical particles of radius $a$ that reach immediately their terminal fall speed, which is given by \citep{Pruppacher}:
\begin{equation}
V_{\rm f}=\frac{2\beta a^2g(\rho_{\rm p}-\rho)}{9\eta}
\label{eq::Vf}
\end{equation}
where $\eta$ is the viscosity of the gas, $g$ is the gravitational acceleration of the planet, $\rho_{\rm p}$ is the density of the particle, and $\rho$ the density of the atmosphere. $V_{f}$ is positive when the particles goes downward. The Cunningham slip factor, $\beta$, accounts for gas kinetic effects that become relevant when the mean free path of the atmospheric molecules is bigger than the size of the falling particle. This factor has been measured experimentally by numerous experiments. We adopt the expression from \citet{Li2003} as done by \citet{Spiegel2009}.

\begin{equation}
 \beta={1+K_{\rm N}(1.256+0.4 e^{-1.1/K_{\rm N}})}
 \label{eq::beta}
 \end{equation} 
where the Knudsen number $K_{\rm N}$ is the ratio of the mean free path to the size of the particle :
\begin{equation}
K_{\rm N}=\frac{\lambda}{a}.
\label{Kn}
\end{equation}

For a perfect gas, the mean free path can be expressed as \citep{Chapman1970}:
\begin{equation}
\lambda=\frac{k_{\rm B}T}{\sqrt{2}\pi  d^2}{\frac{1}{P}}
\label{lambdaP}
\end{equation}
with $d$ the diameter of the gas molecules, $P$ and $T$ the pressure and temperature of the gas, and $k_{B}$ the Boltzmann constant.
In the limit of a high-density atmosphere, $\beta\to1$ and the terminal
speed $V_{f}$ becomes the Stokes velocity $V_{\rm Stokes}$.

For low density gases, the dynamical viscosity is independent of pressure and can be expressed as a power law of the local temperature with an exponent varying between 1/2 in the hard-sphere model to near unity, depending on the strength of the interactions between the molecules. Following \citet{Ackerman2001}, we use the analytical formula given by \citet{Rosner} for the viscosity of hydrogen :
\begin{equation}
\eta=\frac{5}{16}\frac{\sqrt{\pi m k_BT}}{\pi d^2}\frac{(k_BT/\epsilon)^{0.16}}{1.22}
\label{viscosity}
\end{equation}
with $d$ the molecular diameter, $m$ the molecular mass, and $\epsilon$ the depth of the Lennard-Jones potential well (for H$_{2}$ we use \unit{2.827 \times 10^{-10}}{\meter} and \unit{59.7k_{B}}{\kelvin} respectively). The power law behavior of the viscosity remains valid for temperatures ranging from $\unit{300}K$ to $\unit{3000}K$ and for pressures less than $\unit{100}\bbar$ \citep{Stiel1963}. At higher temperature, ionization of hydrogen becomes relevant and the viscosity reaches a plateau. However the temperatures of the model are everywhere less than $\unit{3000}K$ and so Eq.~\eqref{viscosity} remains valid.

\begin{figure}[t]
\includegraphics[width=\linewidth]{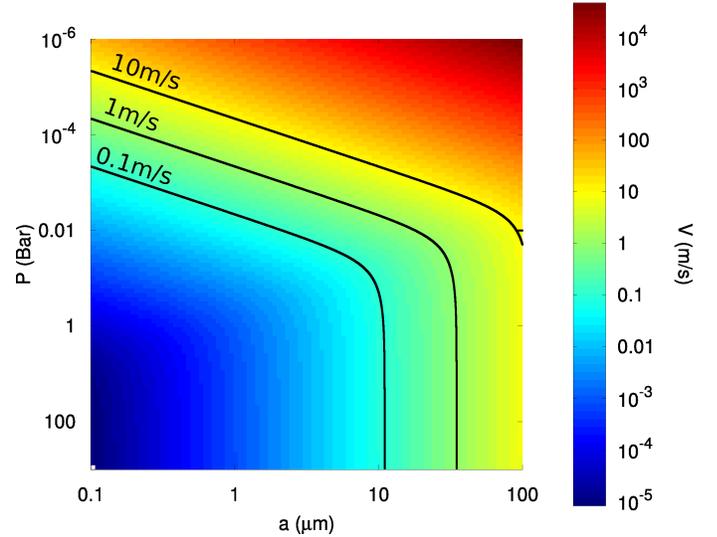}
\caption{Terminal velocity of a falling particle as a function of
  pressure and particle size for an H$_{2}$-atmosphere at
  \unit{1000}\kelvin\usk and a value of the gravitational acceleration
  of \unit{15}\meter\usk\second\rpsquared.
\label{fig::Vf}
}
\end{figure}

Figure \ref{fig::Vf} displays the resulting terminal velocity as
a function of pressure and particle size. Two different regimes are
observed. For Knudsen numbers smaller than unity, the terminal
velocity is independent of pressure, whereas for Knudsen numbers
exceeding unity, the terminal velocity is inversely proportional to
the pressure. At low pressure and for particles bigger than a few tens
of micrometers, the Reynolds number becomes higher than unity, and 
Eq.~\eqref{eq::Vf} is no longer valid. However, we show in Appendix \ref{sec::Appendix} that these differences remain
smaller than one order of magnitude and confined to a small parameter
space thus we decided to neglect them for this work.

\subsection{Integration time---limitation of the study}
\label{sec::Timemachin}
\begin{figure}[t]
\includegraphics[width=\linewidth]{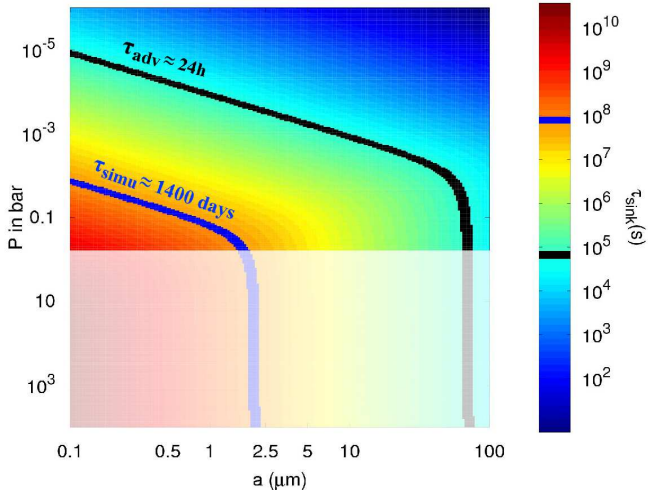}
\caption{Settling timescale as a function of pressure and particle size for the same conditions as in Fig.~\ref{fig::Vf}. Also shown are the advective timescale (black curve) and the simulation timescale (blue curve). In the  shaded region (below 1bar), the tracers are considered gaseous and the settling timescale is not relevant for our study.}
\label{fig::Timescale}
\end{figure}

A challenge for any 3D numerical integrations of hot Jupiters dynamics
is the wide range of timescales exhibited by these atmospheres.
This is true for the radiative time constant, which varies significantly
from low to high pressure \citep{Iro2005, Showman2008}.  
Nevertheless, \citet{Showman2009} showed that the computed light-curve of
HD189733b changes little for integration time longer than hundreds
of days, indicating that the dynamics in the millibar
regime has stabilized. As we integrate $\sim$1400 days, we
consider the dynamics of the planet to be spun up at pressures less
than $\sim$100 mbar. For the settling of particles, another
timescale must be taken into account. The particle settling timescale 
can be defined as the time for the particles to fall one
atmospheric scale height. To obtain a correct picture of the
problem at a given location of the planet, we need to integrate the
simulation for at least several times longer than the settling
timescale at this location. As we can see in
Fig.~\ref{fig::Timescale} the settling timescale ranges from tens of
seconds for big particles at low pressure to tens of years for small
particles at high pressure. Due to computational limitation it is not
possible to run the simulation long enough to ensure that every
considered tracer field has reached a statistical steady state for
the full range of particle sizes we consider. The
integration during 1400 days allows us to calculate the steady state for
every particle size at pressure lower than $\unit{10}\bbar$ and at every
pressure for nightside condensates bigger than $\unit{2}\micro\meter$.

We define the advective timescale as the time for a parcel in the main
jet stream (see Sec.~\ref{sec::GlobalCirculation}) to cross one hemisphere of the planet. If the advective
timescale exceeds the settling timescale at a given level, the
particles at these levels will fall several scale heights while on the
nightside. We thus expect that these levels will be depleted.
 Conversely when the advective timescale is
shorter than the settling timescale, the coupling between the flow and
the particle is essential and particles' behavior cannot be predicted
easily.  Typical advective times in our models are $\sim$24 hours;
this is marked by the upper thick black curve in Fig.~\ref{fig::Timescale}. 
We expect depletion to occur above this line.

Our initial conditions for the tracer field correspond to a tracer that is spatially constant everywhere at a normalized mole fraction of 1. The final tracer abundance in our models at low pressures (less than $\approx\unit{0.1-1}\bbar$) is qualitatively insensitive to the initial abundance at those low pressures. This results from the short vertical mixing and settling timescales at low pressures (\eg~Fig.~\ref{fig::Timescale}).

\begin{figure*}
   \begin{minipage}[c]{.48\linewidth}
\includegraphics[width=\linewidth]{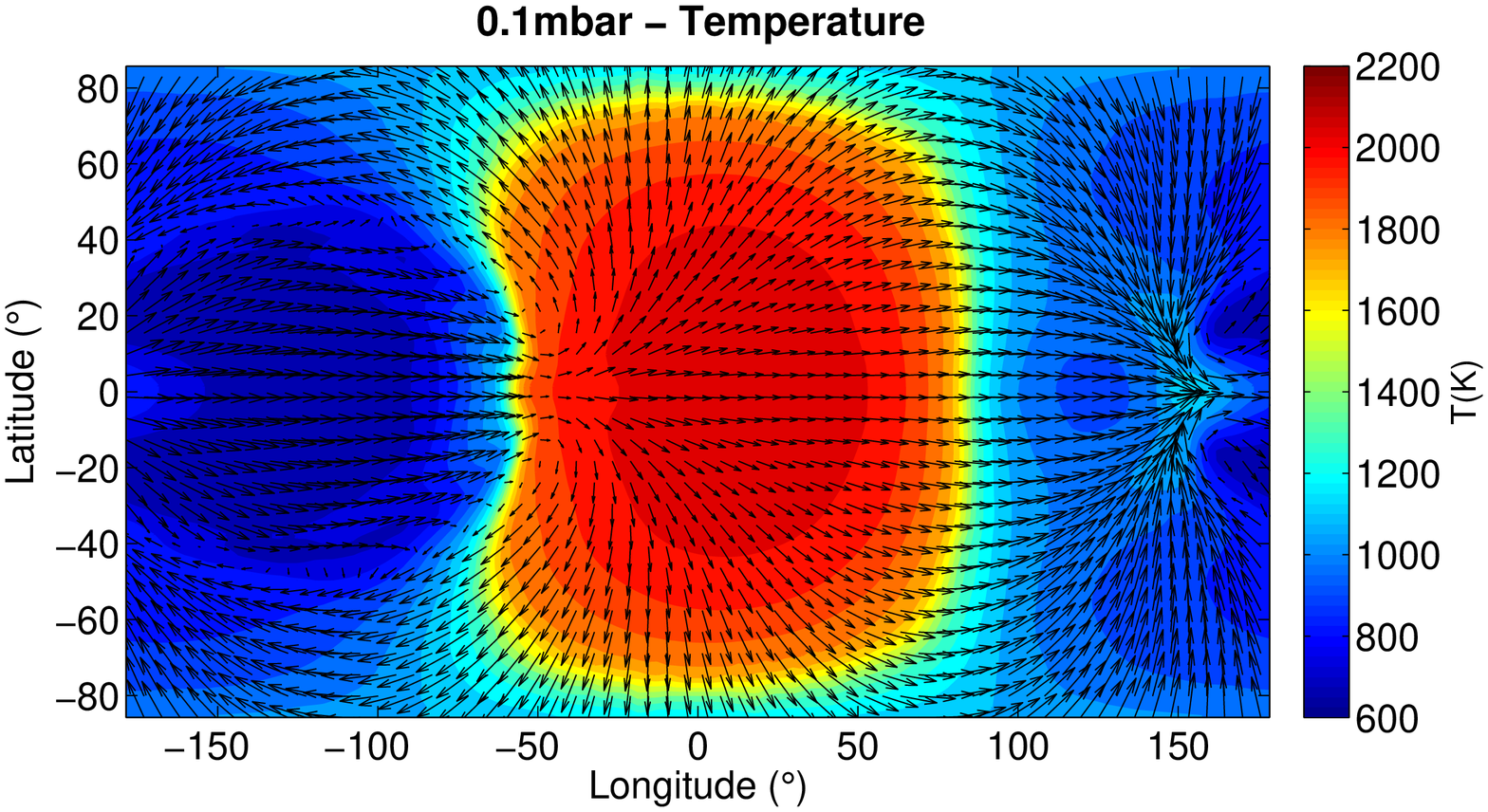}
\includegraphics[width=\linewidth]{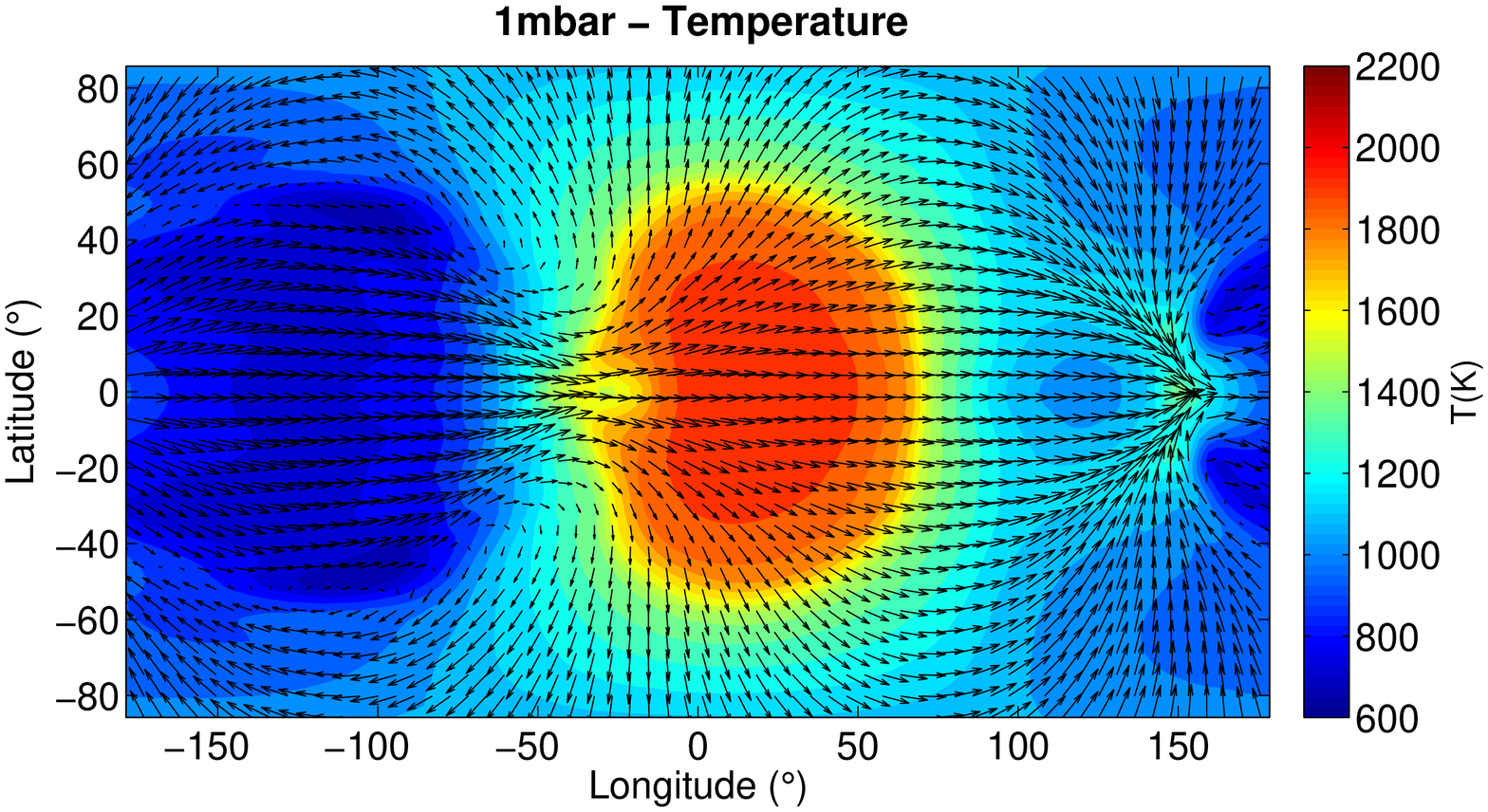}
\includegraphics[width=\linewidth]{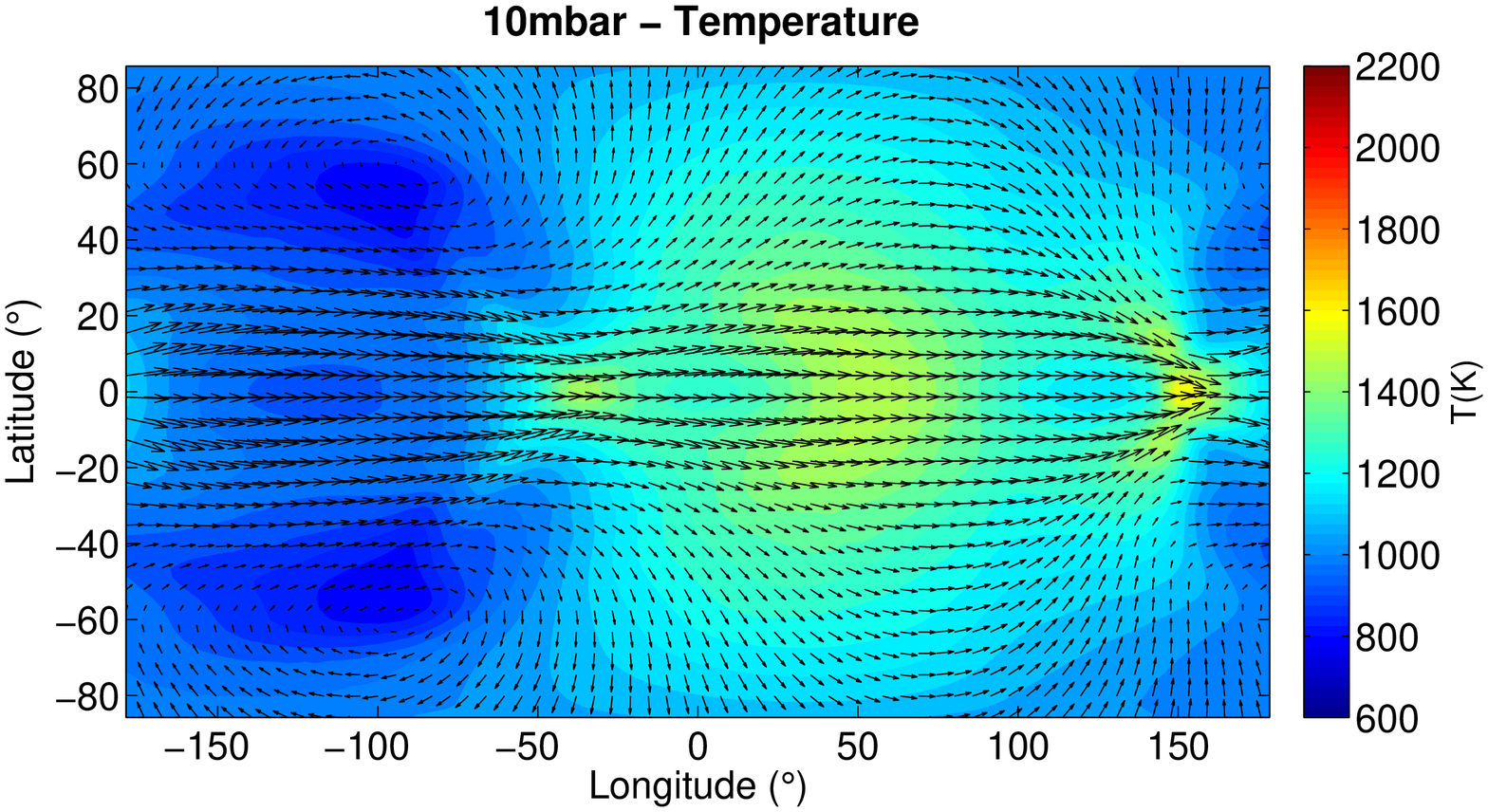}
\includegraphics[width=\linewidth]{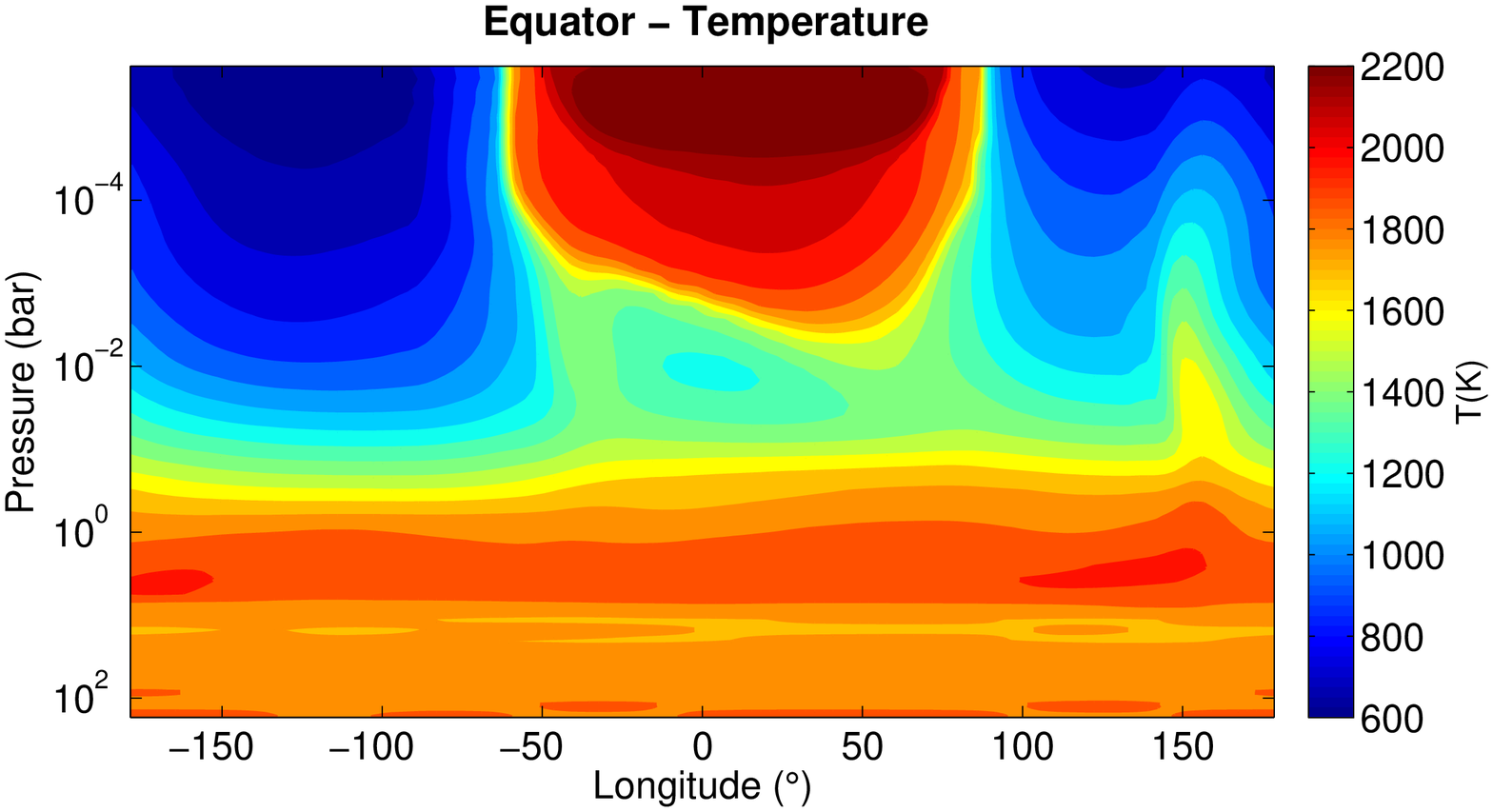}
   \end{minipage} \hfill
   \begin{minipage}[c]{.48\linewidth}
\includegraphics[width=\linewidth]{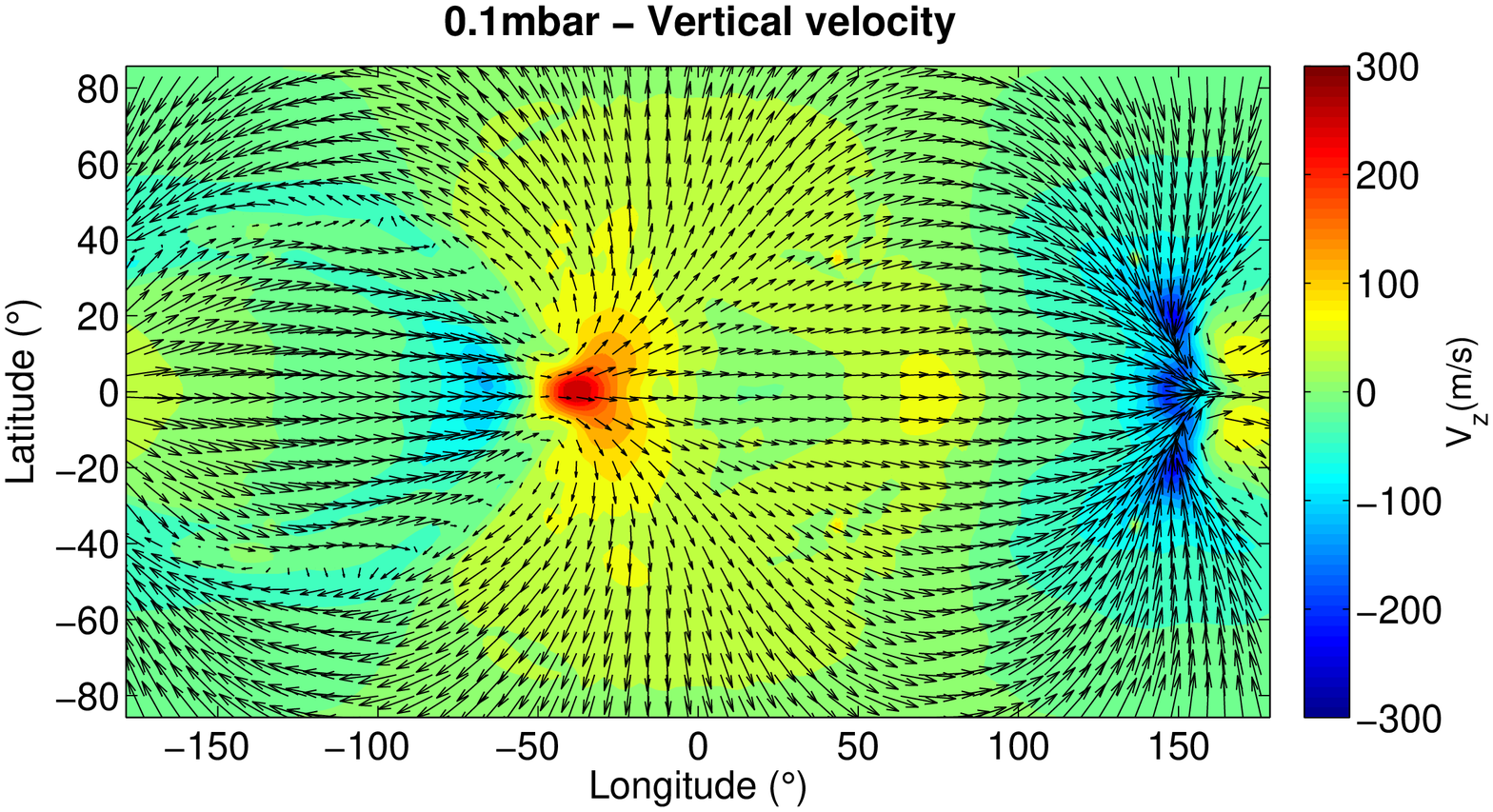}
\includegraphics[width=\linewidth]{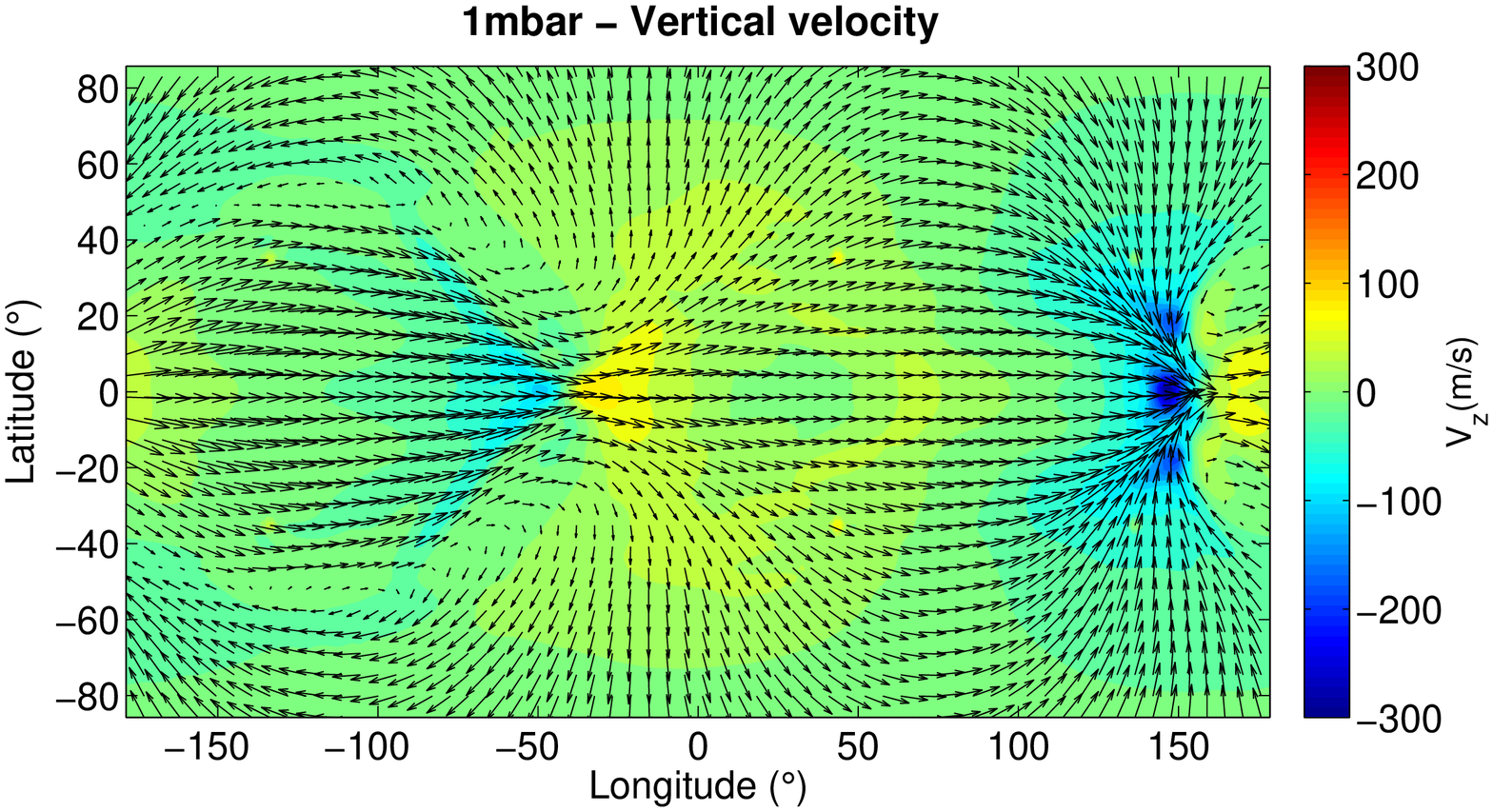}
\includegraphics[width=\linewidth]{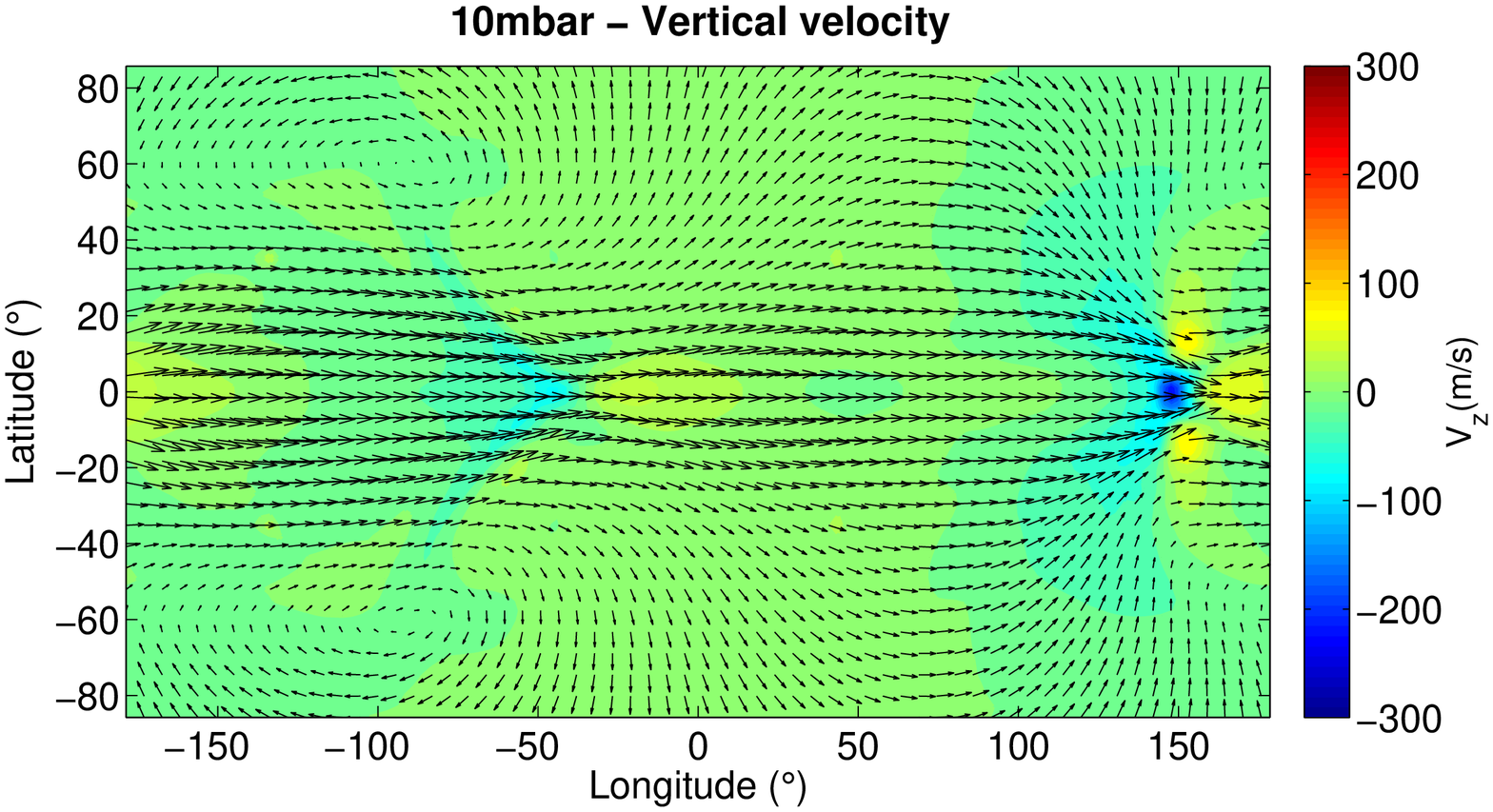}
\includegraphics[width=\linewidth]{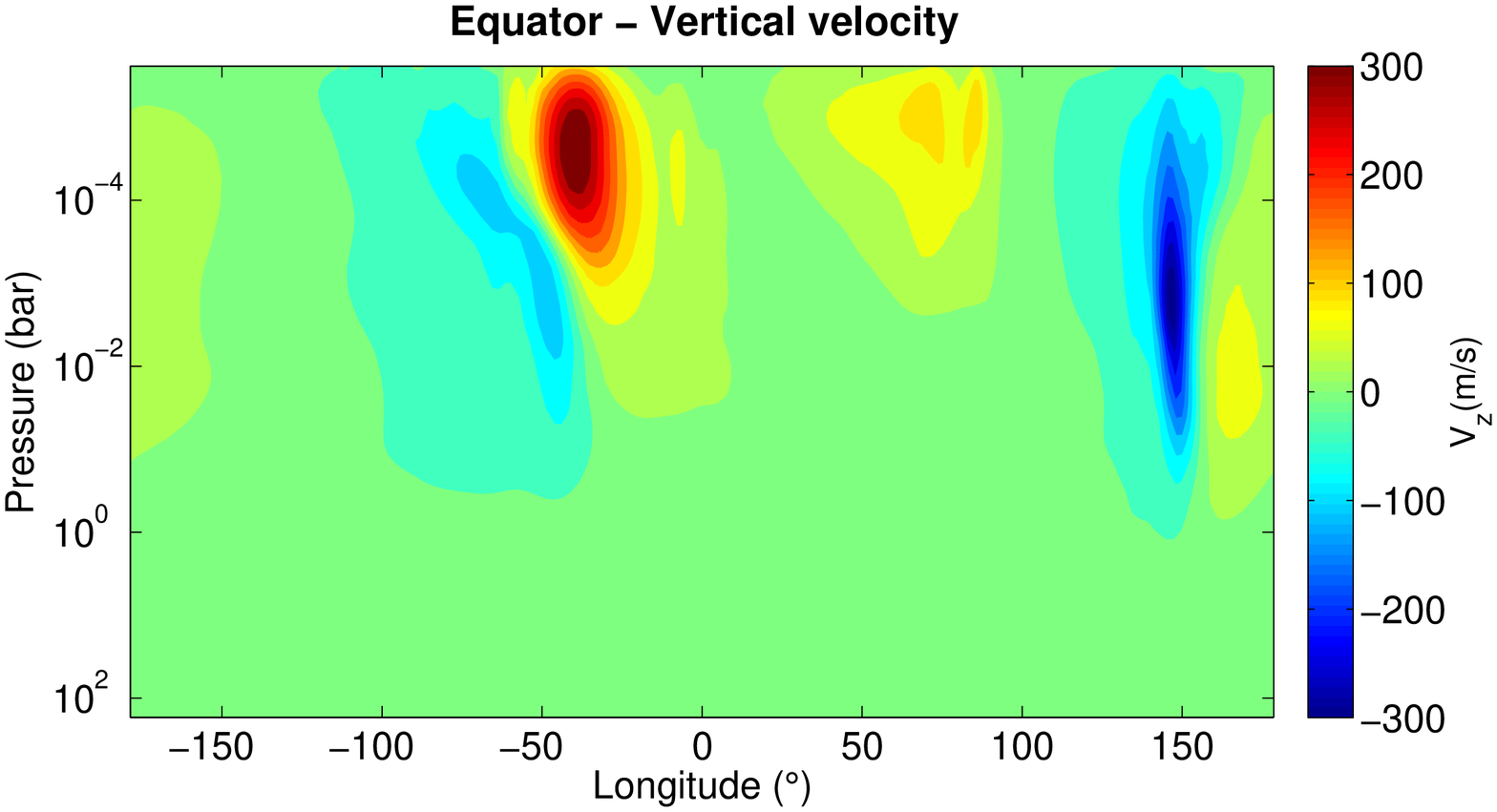}
\end{minipage}
 \caption{Temperature (left panel, colorscale), vertical velocities (right panel, colorscale), and horizontal winds (arrows) in our model of HD 209458b. Positive velocities are
  upward. The top three panels show the flow at three different
  pressure (0.1mbar, 1mbar, and 10mbar). The bottom panel show the
  vertical velocities versus longitude and pressure along the
  equator. The substellar point is at longitude, latitude ($0\degree$,
  $0\degree$), the dayside is between $-90\degree$ and
  $+90\degree$. All the quantities are time averaged.}
   \label{fig::TempVelocityField}
\end{figure*}
\section{Results}
\subsection{Dynamical regime}
We run the simulations for 1400 days and calculate the time average
for all the variables over the last 400 days only, once the simulation
 reaches a statistical steady state at upper levels.
\label{sec::GlobalCirculation}
Understanding the flow structure is essential to understanding the
Lagrangian advection of particles, thus we first present a brief
description of the dynamics. A more complete description of the
circulation in our simulations is presented in \citet{Showman2009}. The
temperature structure is shown in Fig.~\ref{fig::TempVelocityField}. A hot stratosphere is visible at altitudes above
the 10-mbar level, due to the strong visible-wavelength absorption by titanium
oxide present in solar abundances in the simulation. Titanium oxide abundances should be affected by the horizontal cold trap. However, as explained in Sect. \ref{sec::RadTrans}, we consider its radiative effects as if it was in local chemical equilibrium, regardless of the behavior of our passive tracers. Temperatures reach $\sim\unit{2200}K$ at low pressures near the substellar point.  By contrast, the temperatures deeper than 10 mbar in the dayside are relatively temperate. This could lead to the presence of a vertical cold trap, not considered in this study. The day-night temperature contrast becomes significant at pressures less than $\sim$100 mbar, reaching 1600 K near the top of the model. This large temperature difference results from the short radiative timescale at low pressures in comparison to dynamical timescales~\citep{Iro2005, Cooper2005, Showman2008}.

The horizontal flow on isobars comprises an eastward (superrotating)
jet close to the equator and a day-to-night flow pattern at higher
latitude. As the pressure increases, the radiative time constant
increases and the jet extends to higher latitudes. As seen in Fig.~\ref{fig::TempVelocityField}, the mean vertical
velocities exhibit planet-wide variations. The highest velocities coincide
with strong horizontal convergence of the flow and occur mostly at the
equator. West of the anti-stellar point, the convergence of the
day-to-night circulation forces strong downwelling motions. As
described in \citet{Rauscher2010}, this convergence point appears in a
range of hot Jupiters GCMs of varying complexity. It is usually
associated with a shock-like feature \citep{Heng2012b}. Our simulations assume local hydrostatic equilibrium and thus can not treat properly the physics of
shocks. To date, no global model of hot Jupiters atmospheric dynamics can
handle shocks properly. Yet, a similar wind convergence pattern
appears when considering the non-hydrostatic nature of the flow as can
be seen in \citet{Dobbs-Dixon2010}. Between the substellar point and
the west terminator, there are additional convergence/divergence
points associated with the jet, leading to a region of strong
ascending motion $\sim$$40^{\circ}$ of longitude west of the
substellar point, and a broad region of descending motion west of
that. These vertical flows remain coherent
over several orders of magnitude in pressure, giving them the
potential to transport vertically large quantities of
material. Outside of these points of strong vertical motions, the
vertical velocities are more than one order of magnitude smaller. They
are mostly upward on the dayside and downward on the nightside.

\subsection{Spatial distribution of condensable species}
\label{sec::SpatialVar}
\begin{figure}[h!]
\includegraphics[width=\linewidth]{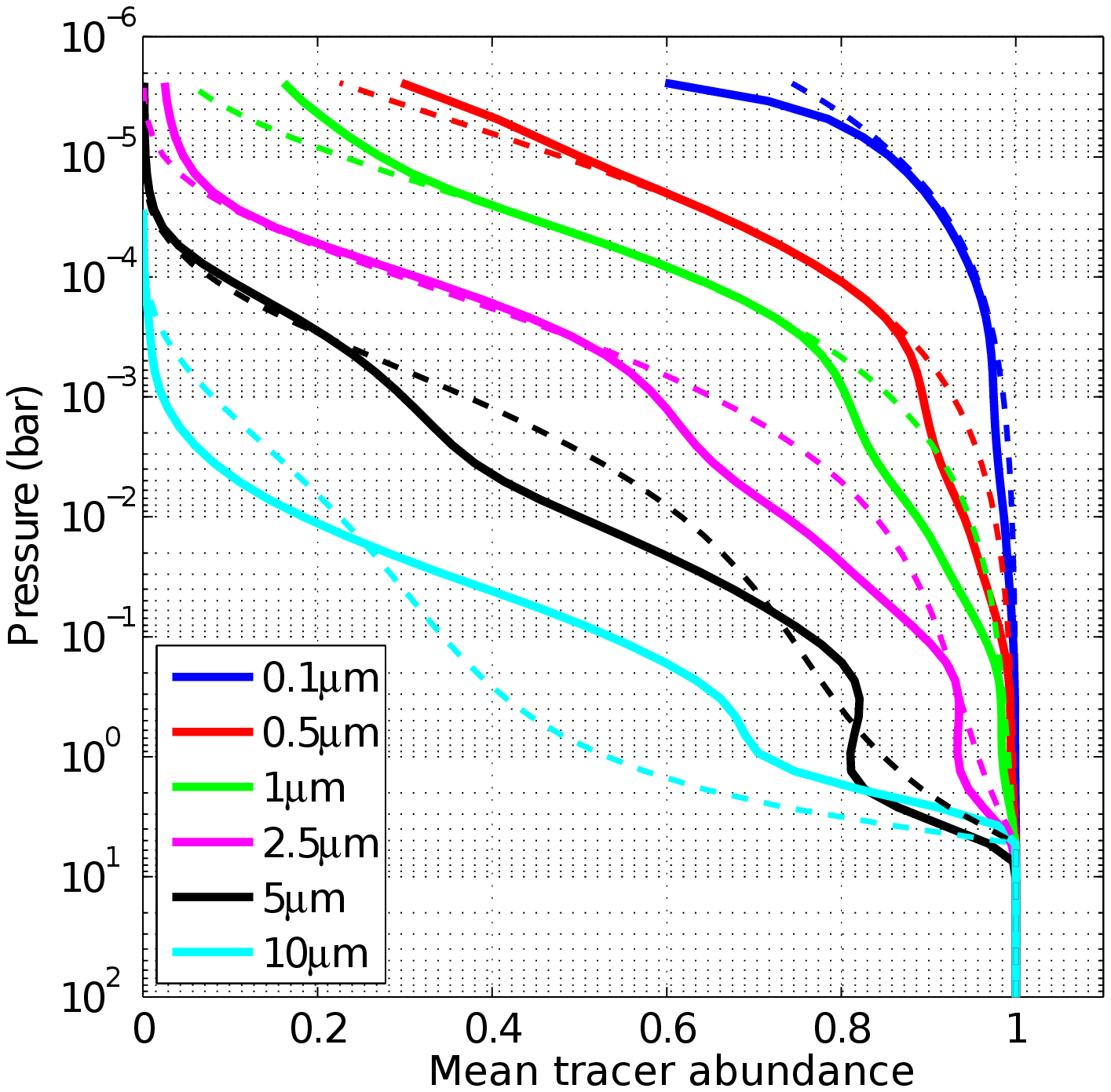}
\caption{Planet-wide, time averaged tracer abundances on isobars. The abundances are normalized to the abundance in the deepest layer. We compare the value of the 3D model (solid lines) and the fit using the 1D model (dashed lines).}
\label{fig::TracerMeanProfile}
\end{figure}

Our simulations show that, as expected, condensation and particle
settling on the nightside deplete the tracer from upper levels
relative to the abundances at depth---an effect that is stronger for
larger particles. This is illustrated in
Fig.~\ref{fig::TracerMeanProfile} (solid curves), which shows the
global-mean tracer abundance (averaged horizontally on isobars) versus
pressure for simulations with nightside condensates sizes ranging from $\unit{0.1}\micro\meter$
to $\unit{10}\micro\meter$.  In all cases, the horizontally averaged tracer
abundance decreases with altitude.  The depletion is modest for the
smallest particle size ($\unit{0.1}\micro\meter$), but for the largest particle
sizes, tracer abundances at the top are almost two orders of magnitude
smaller than abundances at the bottom.  A useful metric is the
``$50\%$ depletion pressure,'' that is, the pressure above which the
global-mean tracer abundance is less than $50\%$ of the deep abundance.
This pressure is only $\unit{10}\micro\bbar$ for a particle size of $\unit{0.5}\micro\meter$ but is $\unit{0.1}\bbar$ for a particle size of $\unit{10}\micro\meter$. We also note that,
for all the models shown in Fig.~\ref{fig::TracerMeanProfile},
the particles settling times near the top of the model are much less
than our integration times; at low pressures, the tracer abundances have reached
a statistical equilibrium where downward transport of tracer due
to particle settling is balanced by upward mixing of tracer by the large-scale
dynamics.

The tracer abundance on isobars exhibits a strong spatial variation as
can be seen in Fig.~\ref{fig::Time-average}. Although a day/night
pattern is imposed in the tracer source/sink (with particle settling
on the nightside but not the dayside), the three-dimensional advection
of the tracer field by the atmospheric winds leads to a complex tracer
distribution that does not exhibit an obvious day-night geometry.
The main pattern appears to be an equator-to-pole gradient, with large
zonal-mean abundances at the poles, and smaller zonal-mean abundances at
the equator. This is particularly true around $\unit{0.1}\milli\bbar$. 
Significant longitudinal tracer variations also occur; at $\sim$1 mbar
(Fig.~\ref{fig::Time-average}), these variations are particularly
prominent at high latitudes. Interestingly, these variations are
phase shifted in longitude relative to the day-night pattern
with maximum (minimum) peak tracer abundances occurring $\sim$60--80$^{\circ}$
of longitude east of the substellar (antistellar) point.
 
On top of these main patterns, we clearly see two points depleted in
tracers at the equator. These two points correspond to points of
horizontal convergence of the flow and high downwelling motions as
discussed in the previous section. This correlation between strong
downwelling motions and low tracer abundances arises naturally in the
presence of a background vertical gradient of tracer abundances. Due to their
settling on the nightside, the local tracer abundance generally
decreases with height and thus any downwelling motion would carry
parcels of gas depleted in tracers whereas any upwelling motion should
carry parcels of gas enhanced in tracers.

\begin{figure}[h!]
\includegraphics[width=\linewidth]{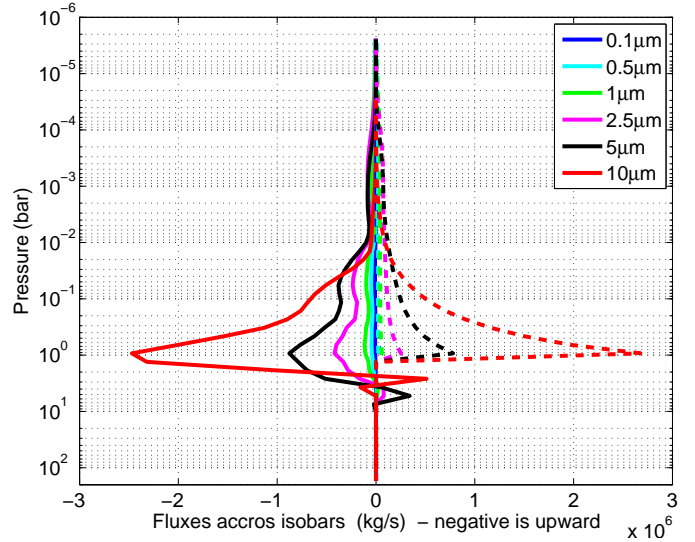}
\caption{Time averaged fluxes of tracers across isobars. Upwelling
  fluxes due to the dynamics (solid lines) balance the downwelling
  fluxes due to the gravitational settling of the tracers (dashed
  lines). Negative values are upward fluxes.}
\label{fig::TracerMeanFluxes}
\end{figure}

\begin{figure*}
   \begin{minipage}[c]{.46\linewidth}
\includegraphics[width=\linewidth]{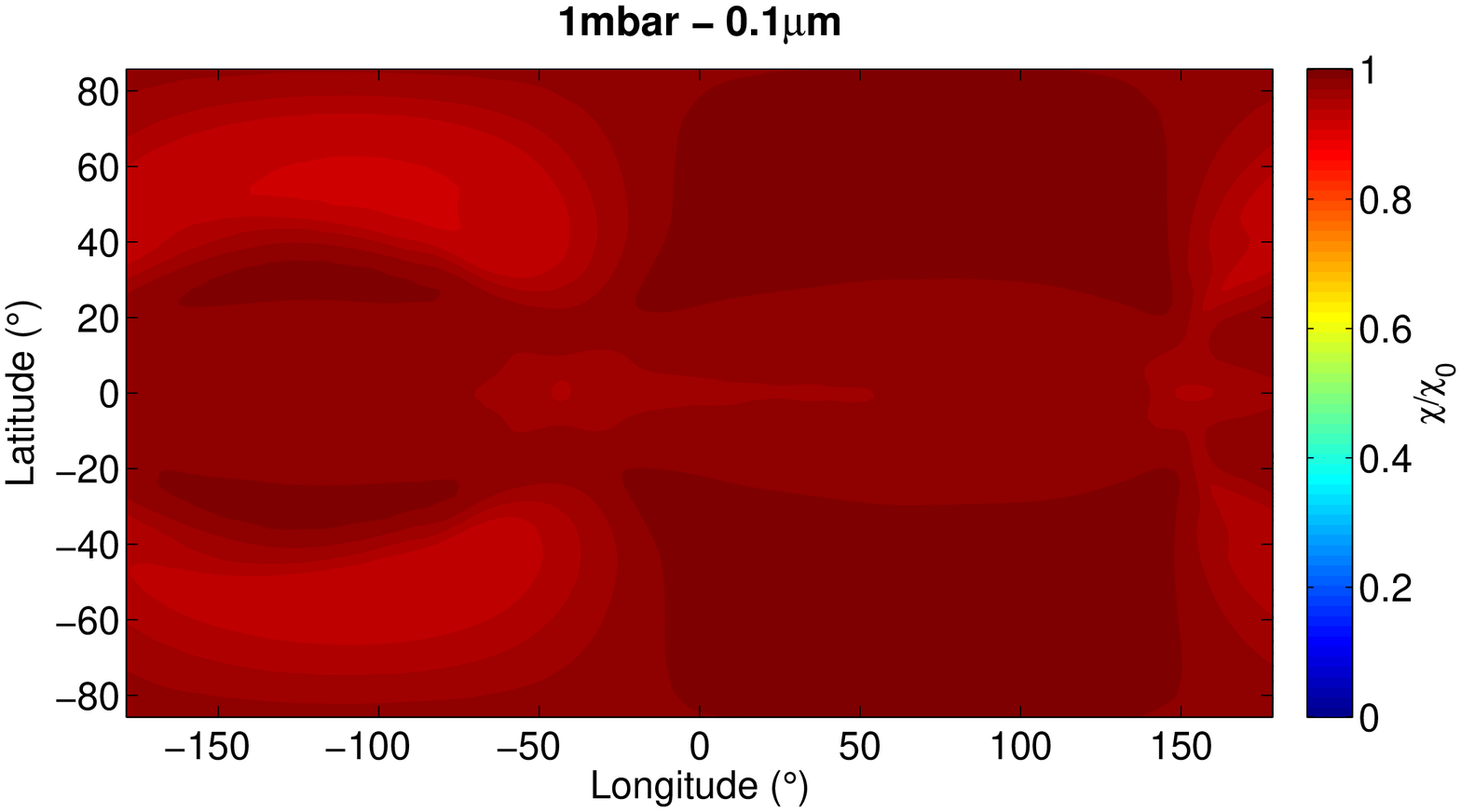}
\includegraphics[width=\linewidth]{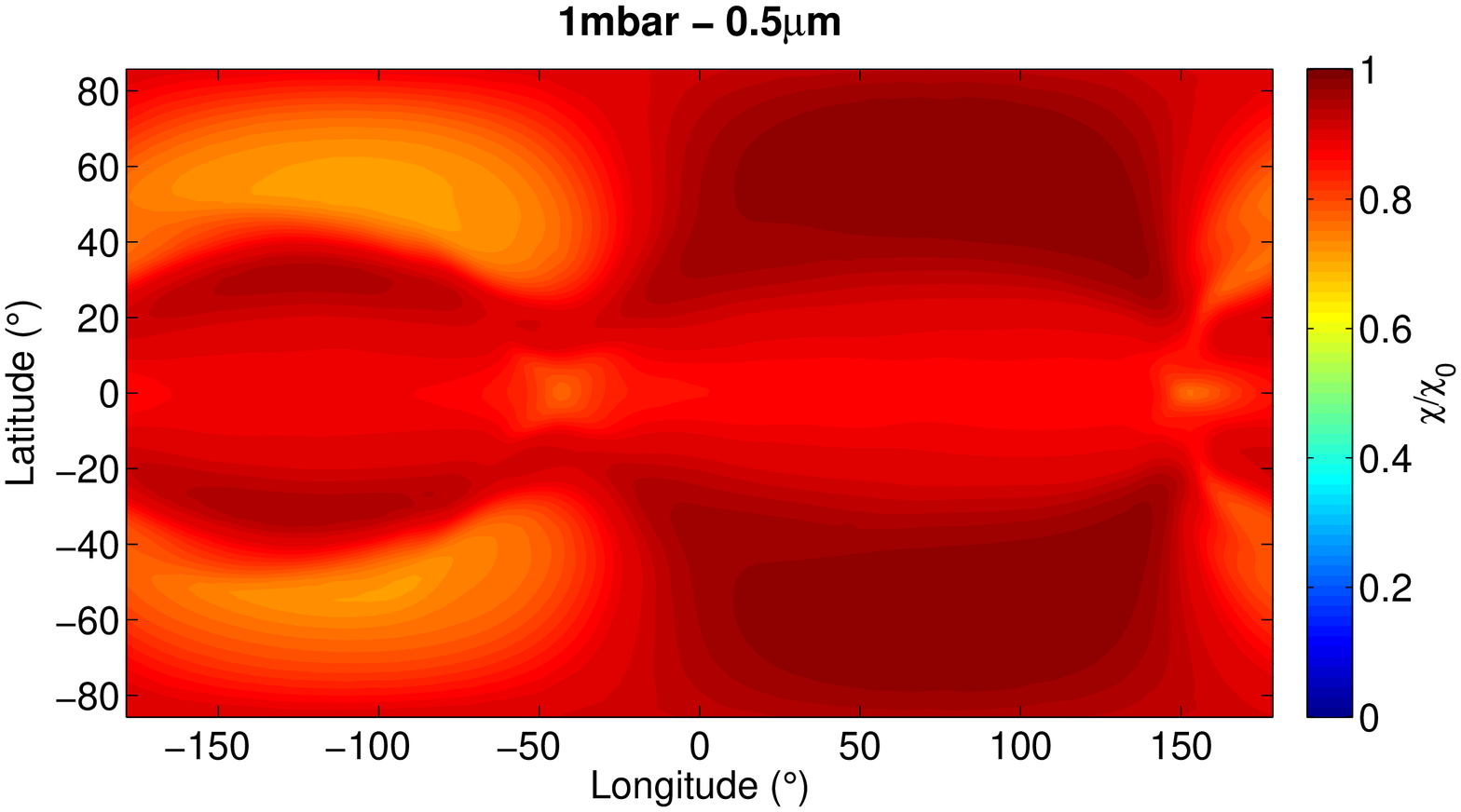}
\includegraphics[width=\linewidth]{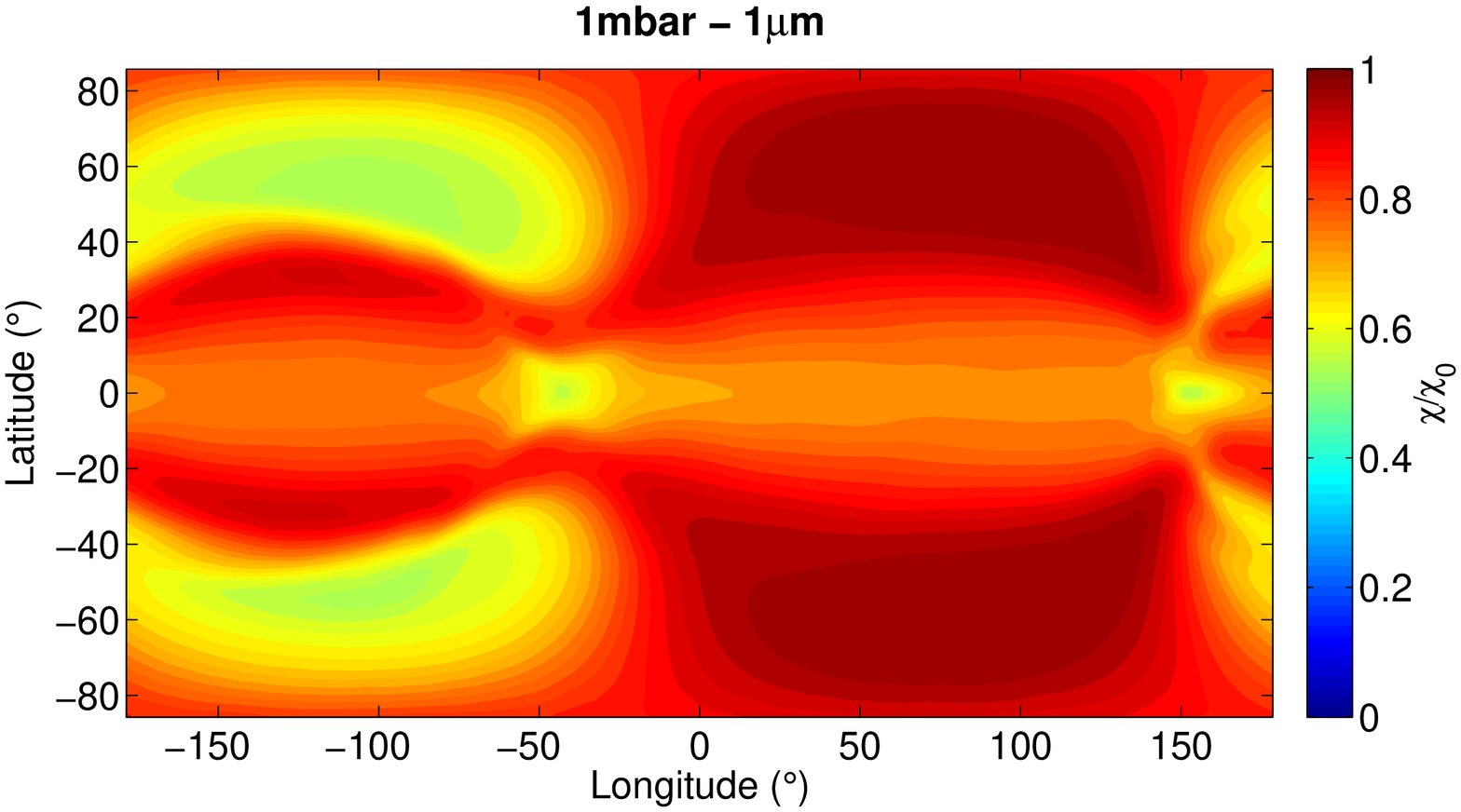}
\includegraphics[width=\linewidth]{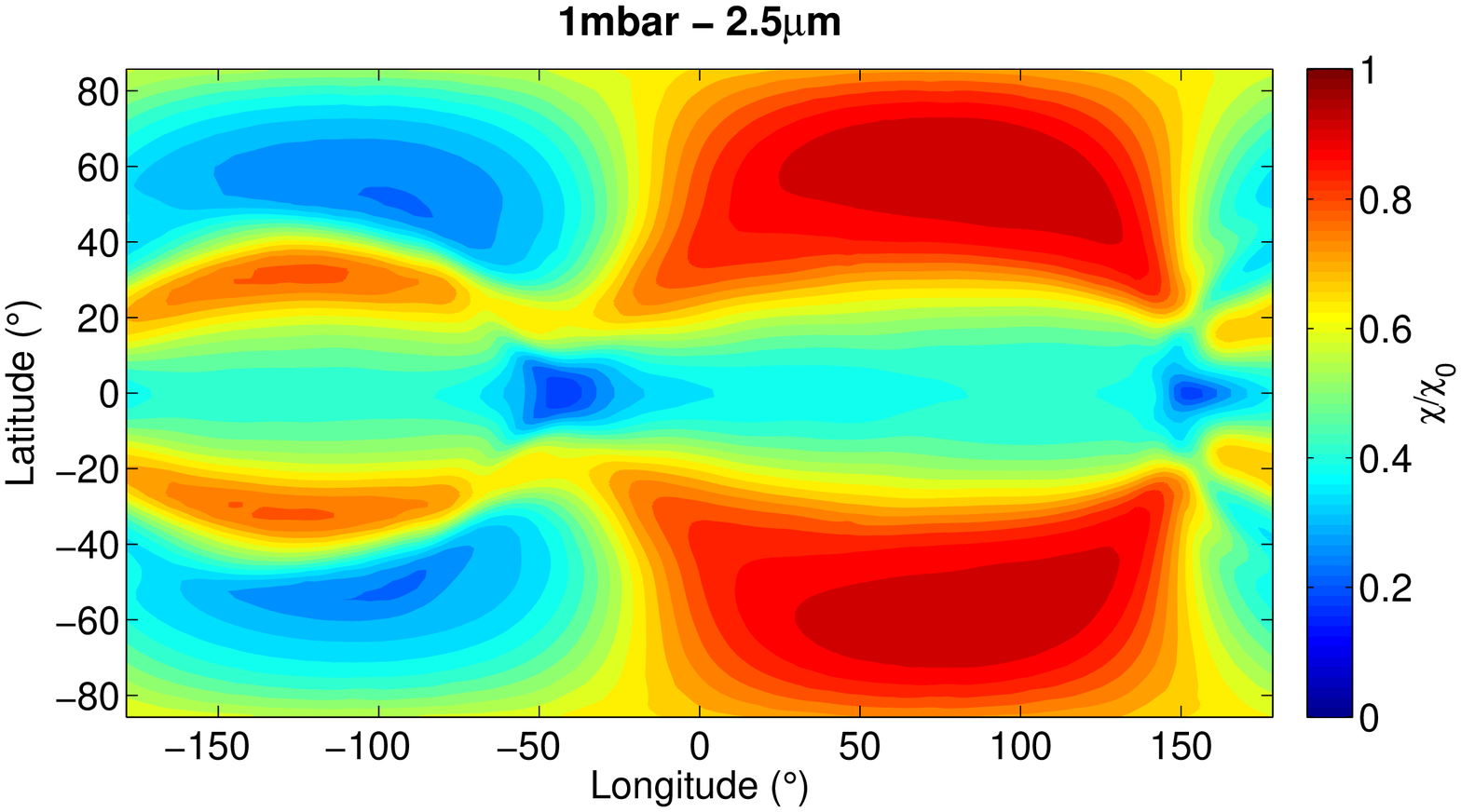}
\includegraphics[width=\linewidth]{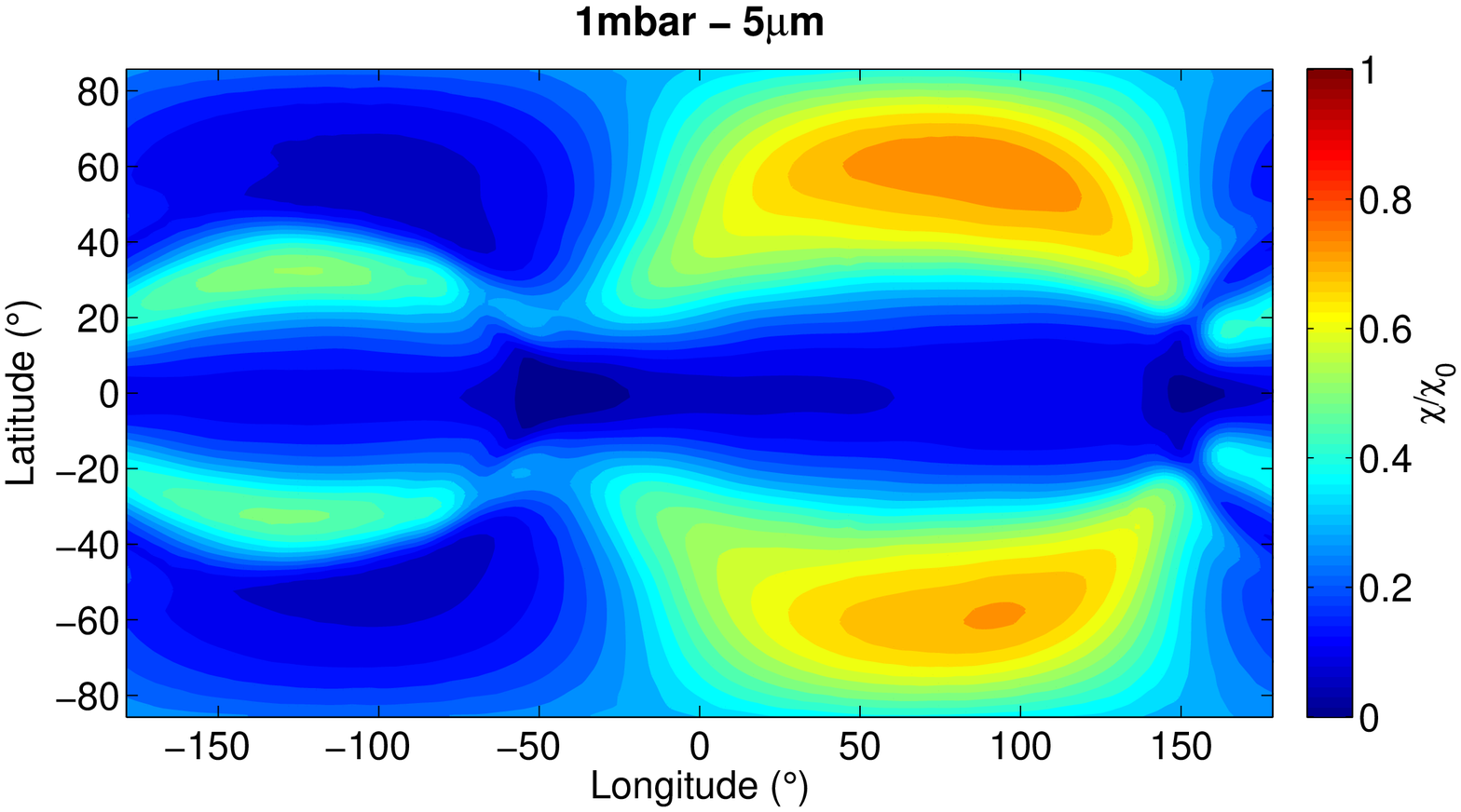}
   \end{minipage} \hfill
   \begin{minipage}[c]{.46\linewidth}
\includegraphics[width=\linewidth]{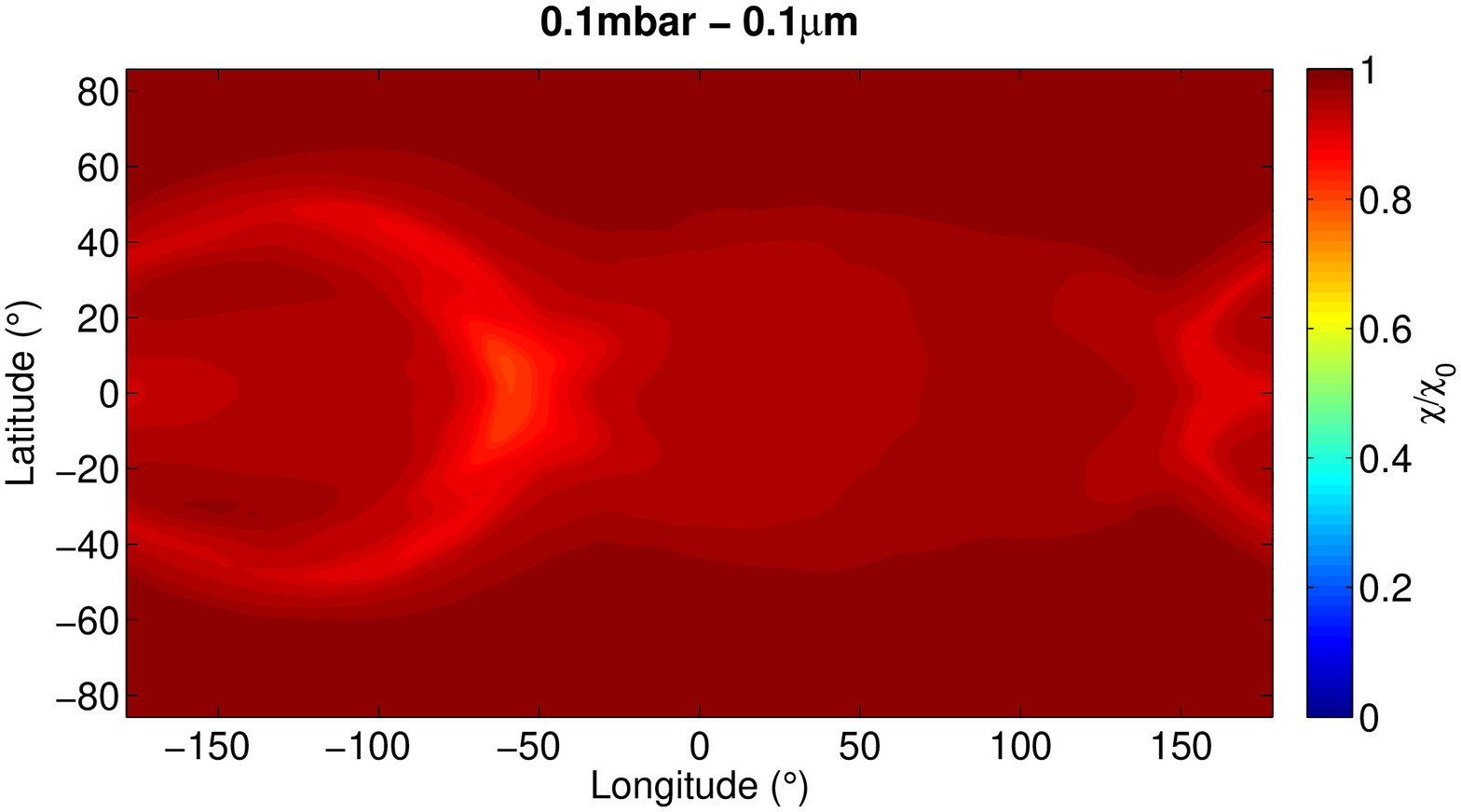}
\includegraphics[width=\linewidth]{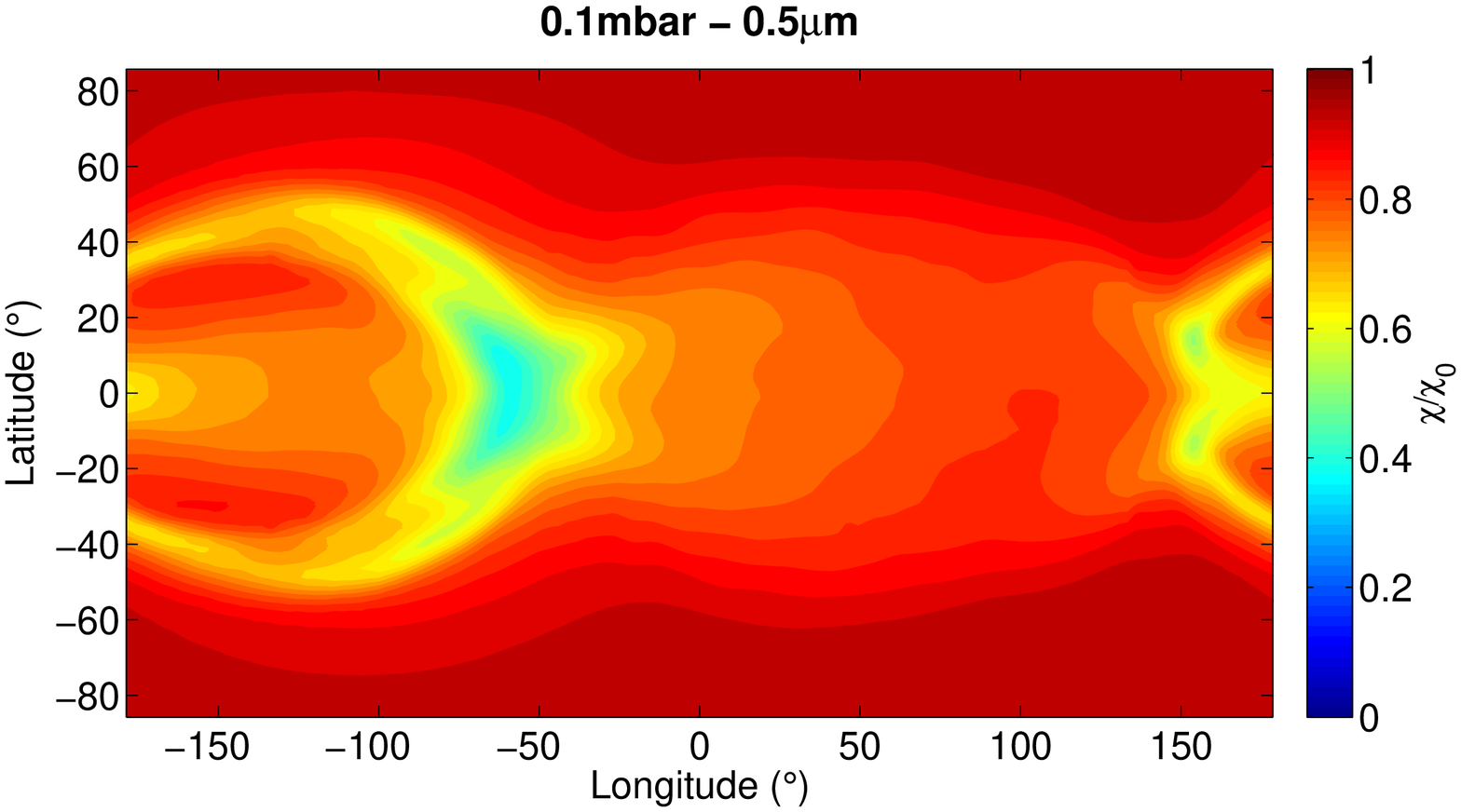}
\includegraphics[width=\linewidth]{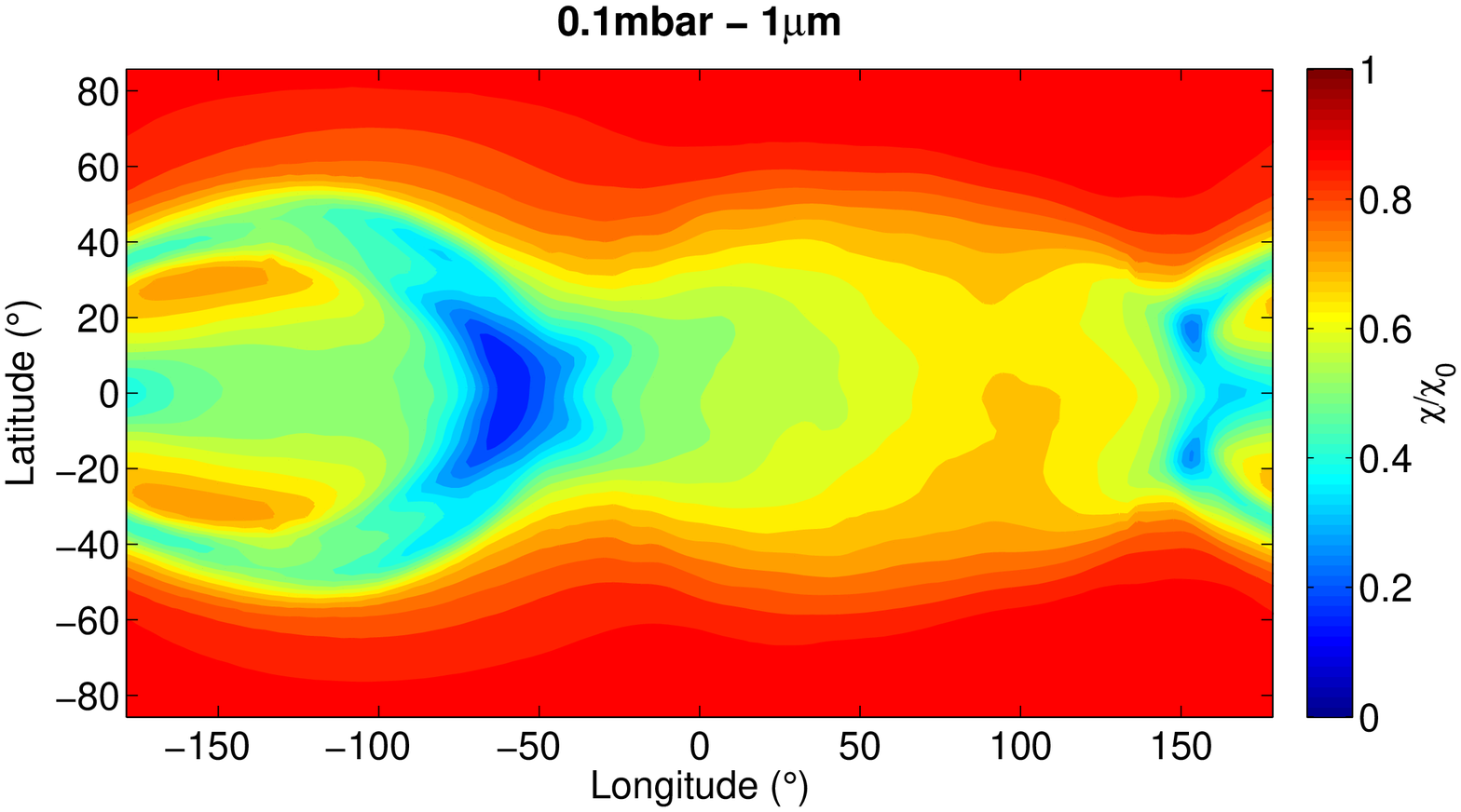}
\includegraphics[width=\linewidth]{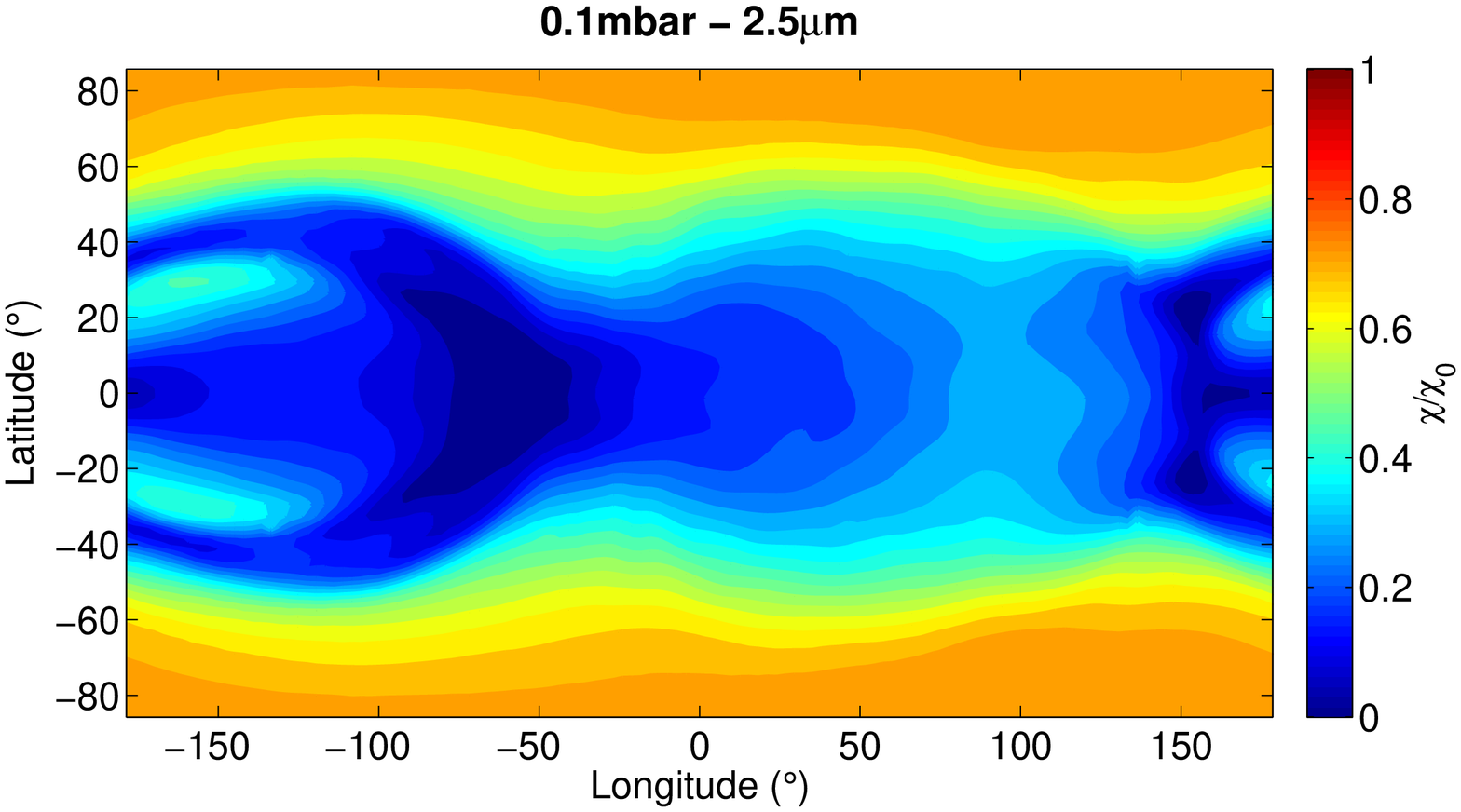}
\includegraphics[width=\linewidth]{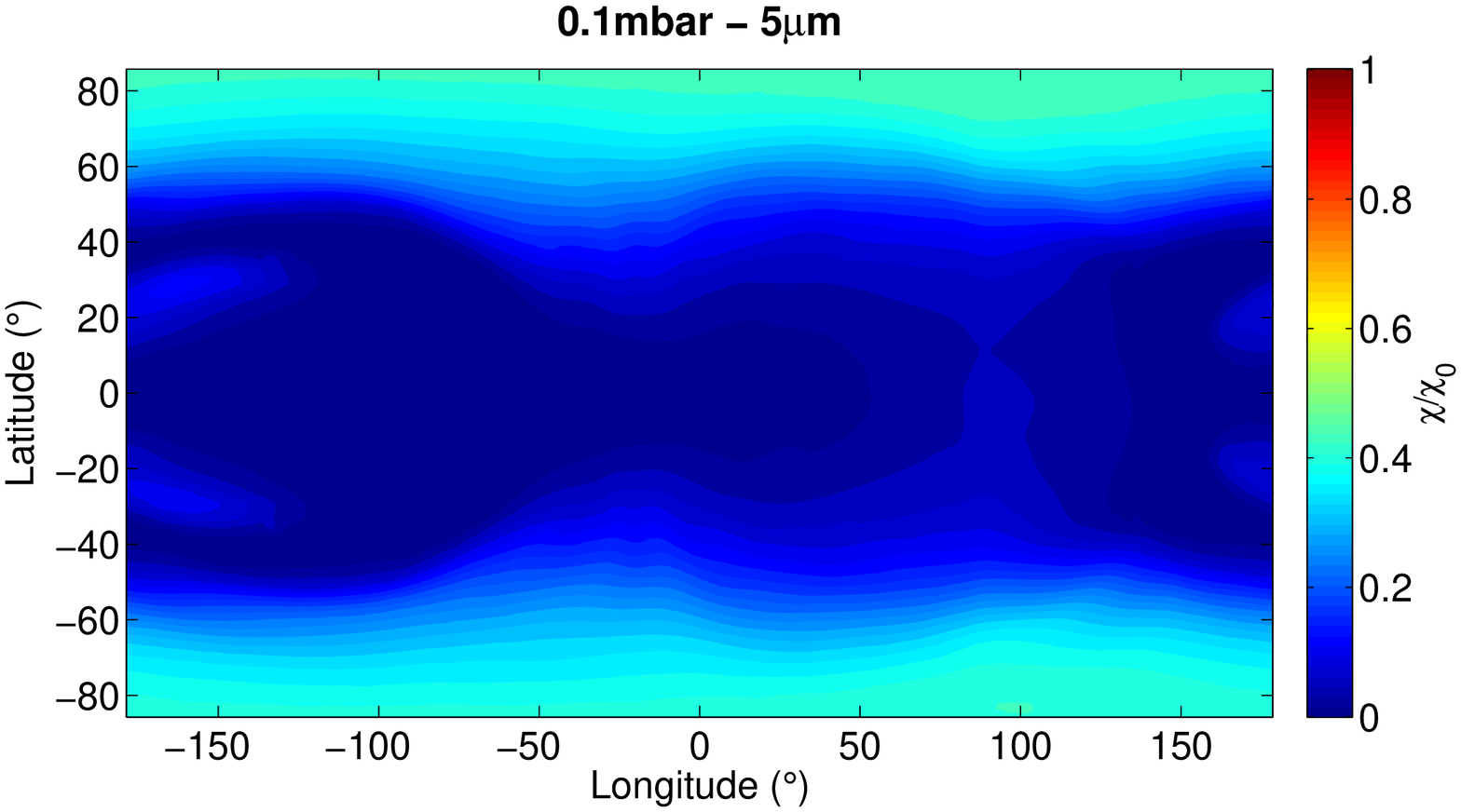}   \end{minipage}
 \caption{Time averaged tracer abundance relative to the deep abundance ($\chi/\chi_{0}$) at two different pressures and for different particle sizes.}
   \label{fig::Time-average}
\end{figure*}

 \subsection{Geometry of the mixing}

As discussed in the previous section, the tracer abundance is not null
everywhere in the planet, which implies that tracers did not rain out
during the simulation time. Yet the integration time greatly
exceeds the fall times at low pressure for all particle sizes considered,
and everywhere throughout the domain for particle sizes exceeding
a few $\mu m$.  Therefore vertical mixing must happen in order
to keep these particles aloft. This vertical mixing is characterized
by an upward dynamical flux of tracers  that balances the
downward flux due to the gravitational settling in the nightside. The
upward dynamical flux of tracers across isobars can be calculated as
$\left\langle\frac{-\omega\chi}{g}\right\rangle$ where $\omega$ is the
vertical velocity in pressure coordinates,~$\chi$ the tracer
abundance,~$g$ the gravity of the planet and the brackets denote the average on isobars. As seen in Fig.~\ref{fig::TracerMeanFluxes}, the upward flux of tracers due to the
dynamics (solid lines) balances nicely the downward flux due to
settling (dashed lines), showing that the simulation did reached a quasi
steady-state.

Hot Jupiters atmospheres are heated from above and thus are believed
to be stably stratified. Then, vertical mixing cannot be driven by
small scale convection as it is the case in the deep atmosphere of
brown dwarfs \citep{Freytag2010} and the gas giants of the solar
system. As explained in Sect. \ref{sec::Dynamics}, our model does not use any
parametrization of sub-grid scale mixing. Thus we do not account for
mixing induced by small-scale turbulence and gravity wave breaking,
two mechanisms that are believed to dominate the mixing in the
radiative part of brown dwarfs atmospheres \citep{Freytag2010}.
Rather, the upward flux of tracers in our model is due to the
large-scale, resolved flow of the simulation.

\begin{figure}[h!]
\includegraphics[width=\linewidth]{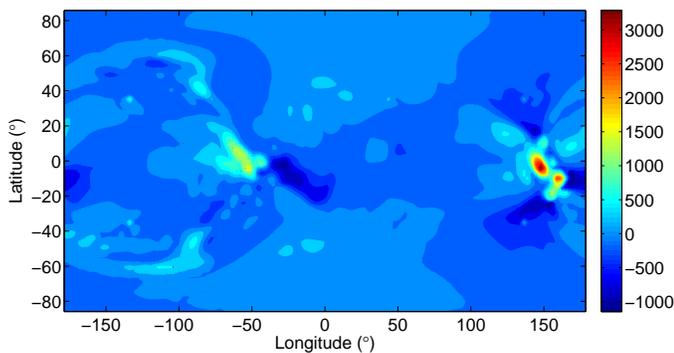}
\caption{ 2D distribution of the mixing efficiency on the 1mbar
  isobar. We plot
  $F=\frac{\omega\chi-\omega\langle\chi\rangle}{\langle\omega\chi\rangle}$. Positive
  is either an upward flux of gas enhanced in tracers compared to the
  horizontal mean or a downward flux of gas depleted in tracer
  compared to the horizontal mean value. Negative is either a downward
  flux of gas enhanced in tracer or an upward flux of gas depleted in
  tracer. Thus a positive value enhances the (horizontally averaged)
  upward flux, whereas a negative value diminishes it. These fluxes
  are normalized to the mean upward flux of tracer and the global-mean
  on an isobar of the plotted quantity is 1.}
\label{fig::UpFlux}
\end{figure}

Given that mass is conserved, any upward flux of gas is
compensated by a downward flux of gas. Thus, if the tracers
concentrations were horizontally homogeneous on isobars, there would be no net
upward flux of tracer through that isobar. For a net upward flux of
tracers across isobars to occur, there must be a correlation between
the horizontal distribution of the tracers and the vertical
velocities. Dynamics will produce an upward flux if---on an
isobar---ascending regions exhibit greater tracer abundance than
descending regions.  In other words, an upward tracer flux due to
dynamics will occur only if $\left(\chi-\chim\right) \omega<0$ where
$\omega$ is the upward velocity in pressure coordinates, $\chi$ is the
tracer abundance and the brackets are the mean over one isobar (note
that negative $\omega$ implies upward motion). Given
a vertical gradient of $\langle\chi\rangle$ such that the abundance
of $\langle\chi\rangle$ decreases upward, an upward flux of gas will
naturally bring enhanced material whereas a downward flux will
naturally advect parcels of gas depleted in tracers, thereby creating the
correlation between $\omega$ and $\chi$ favorable for upward tracer
transport. Figure \ref{fig::UpFlux} shows the relative contribution
to the upward mixing versus longitude and latitude at a given isobar, 
which can be estimated by the quantity :
\begin{equation}
F\equiv\frac{\omega(\chi-\chim)}{\langle\omega\chi\rangle}
\label{eq::F}
\end{equation}
Mass conservation in the primitive equations implies that
$\left\langle\omega\chim\right\rangle=0$: the advection of $\chim$
does not contribute to the net (horizontally averaged) upward flux of
material and so we remove the contribution of this term when
defining $F$ in Eq.~\eqref{eq::F}. The quantity
$-\frac{1}{g}\langle\omega\chi\rangle$ is the mean upward flux of
material across isobars, thus the quantity $F$ represents the local
contribution to the total upward flux on isobars.  It is normalized
such that $\langle
F\rangle=1$.

The strength of the mixing varies significantly with longitude and
latitude. Both upward and downward fluxes are one order of
magnitude greater than typical values in a handful of specific small
areas across the planet---particularly at the two points of horizontal
convergence and strong vertical velocities described in Sect.
\ref{sec::GlobalCirculation}. This vertical flow remains coherent over
several order of magnitude in pressure (see Fig.
\ref{fig::TempVelocityField}), acting like a vertical ``chimney'' where
efficient transport of material can be achieved.

In summary, the mechanism by which the large-scale, resolved
atmospheric circulation transports tracers upward is extremely simple
and straightforward. The settling of particles leads to a mean
vertical gradient of tracer abundance, with, on average, small tracer
mixing ratios aloft and large tracer mixing ratios at depth. Given this
background gradient, advection by vertical atmospheric
motions---whatever their geometry---{\it autmomatically} produces a
correlation between $\omega$ and ($\chi - \langle \chi\rangle$) on
isobars, with ascending regions exhibiting larger values of $(\chi -
\langle\chi\rangle)$ than descending regions. In turn, this
correlation automatically causes an upward dynamical net flux of tracers
when averaged globally over isobars. In statistical steady state,
this upward dynamical flux balances the downward transport due to
particle settling and allows the atmospheric tracer abundance to equilibrate at
finite (non-zero) values despite the effect of particle settling.  
The mechanism does not require convection, and indeed, the vertical
motions that cause the upward transport in our models are resolved,
large-scale motions in the stably stratified atmosphere. These
vertical motions are a key aspect of the global-scale
atmospheric circulation driven by the day-night heating contrast.

\subsection{Time variability}
\label{sec::TimeVariability}
Besides the spatial variability at a given time, the model exhibits
significant temporal variability, both in the 3D flow and especially
in the tracer field. The equatorial jet exhibits an oscillation
pattern at planetary scale as seen in Fig.~\ref{fig::jet}. At the convergence
point west of the substellar point, the jet orientation can be toward the north, the
south or well centered on the equator. Whereas \citet{Dobbs-Dixon2010}
described a variation in longitude with time of the convergence point
of the flow in the nightside, we see a variation in latitude of this
convergence point and interpret it as a result of the larger
oscillation of the jet itself.

The tracers abundances at specific locations on the planet exhibit
strong temporal variability. This is illustrated in
Fig.~\ref{fig::jet}, which shows the tracer abundance at $0.1$ and $\unit{1}\milli\bbar$ over the globe at several snapshots in time for a model where the
radius of particles on the nightside is $2.5\,\mu$m.  Significant
variations in tracer abundance are advected by the equatorial jet and,
at high latitudes, by the day-to-night flow, leading to large local
variations in time. In many cases, the strongest tracer variability seems
to involve regional-scale structures with typical sizes of
$\sim$1--3$\times10^4\,$km but also includes hemispheric-scale
fluctuations (\eg~in the abundance averaged over the day or night)
and between the northern and southern hemispheres. Around the
substellar point, the tracer abundance can vary by up to $50\%$,
whereas along the terminator, this temporal variation can reach
$75\%$ relative to the mean value. Such variability---if
it occurs in radiatively active species like TiO---has important 
implications for secondary-eclipse and transit observations, which
probe the dayside and terminator, respectively.

\begin{figure}[h!]
\includegraphics[width=\linewidth]{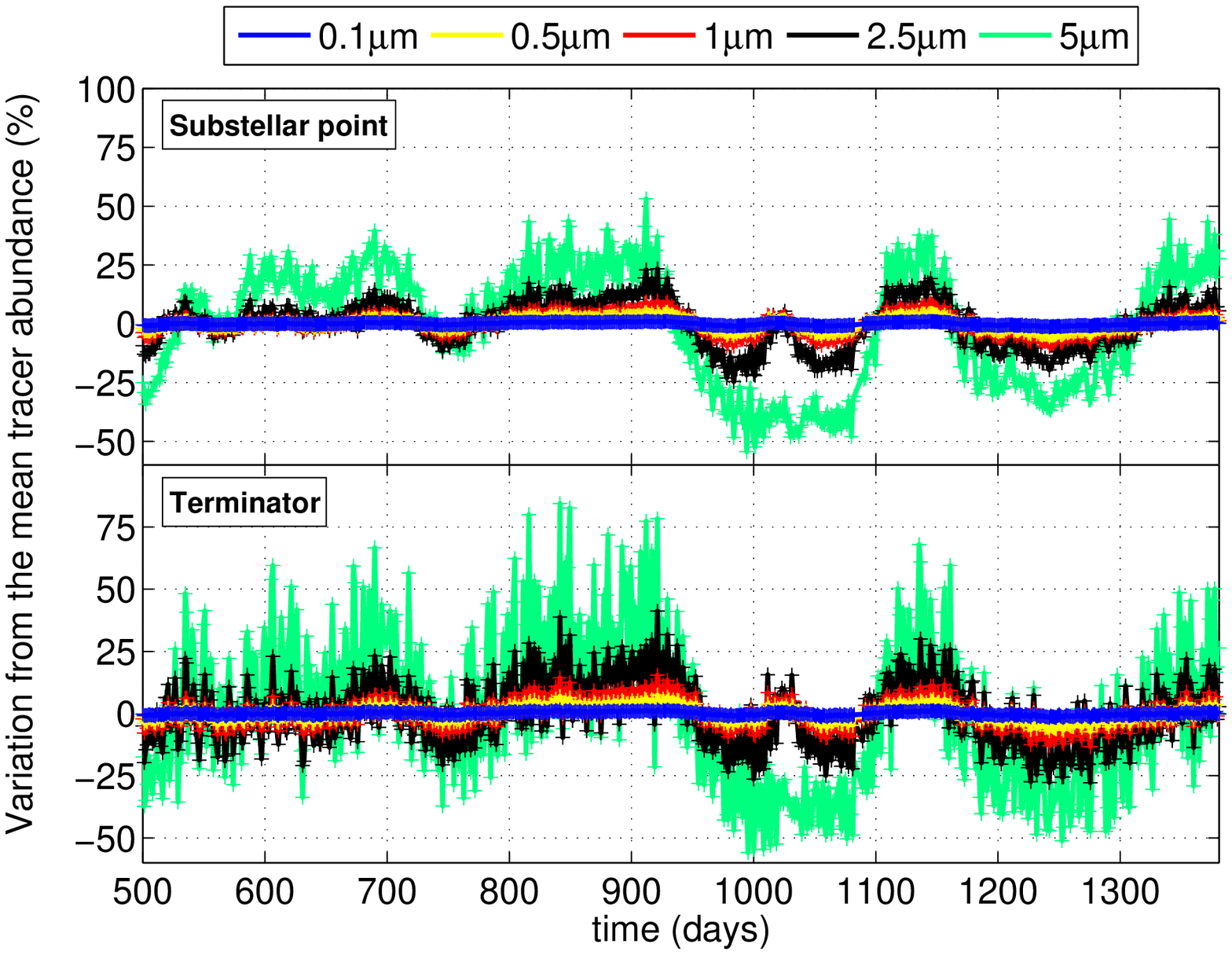}
\caption{Time dependence of the mean tracer abundance averaged between
  $1\milli\bbar$ and $0.1\milli\bbar$. The abundances are averaged horizontally over a
  circular region of radius $45\degree$ centered on the substellar
  point (top) and averaged horizontally around the terminator,
  including all regions within $\pm5^{\circ}$ of the terminator 
  (bottom). The top panel is relevant for inferring the presence of a
  stratosphere (see Sect. \ref{sec::Stratosphere}), albedo
  variations (see Sect. \ref{sec::Clouds}) or secondary eclipse
  measurements (see Sect. \ref{sec::Parameters}). The bottom panel
  is relevant for transit spectroscopy measurements.}
\label{fig::TracerTimeDependance}
\end{figure}

\begin{figure*}
   \begin{minipage}[c]{.46\linewidth}
\includegraphics[angle=-90,width=\linewidth]{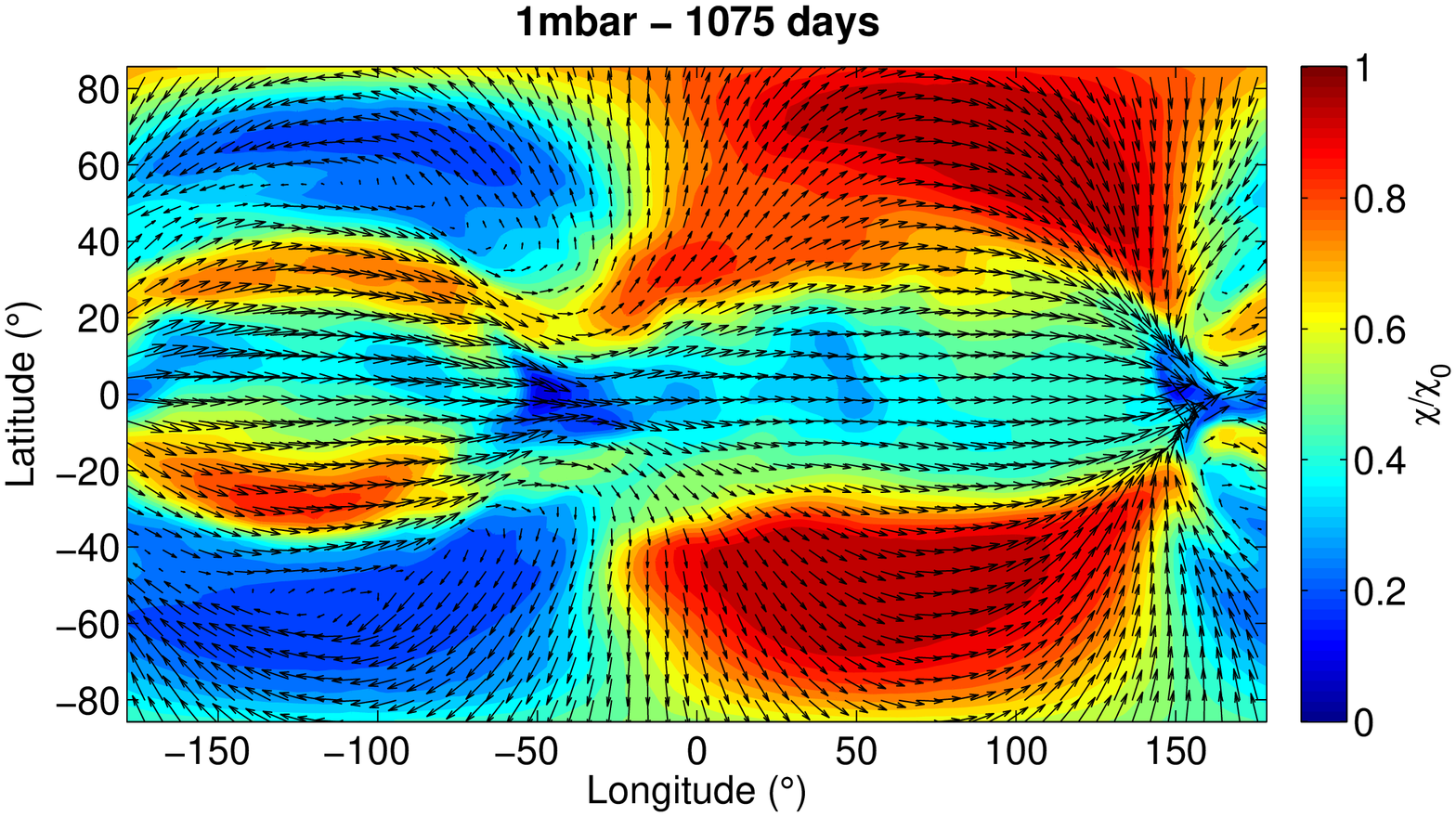}
\includegraphics[angle=-90,width=\linewidth]{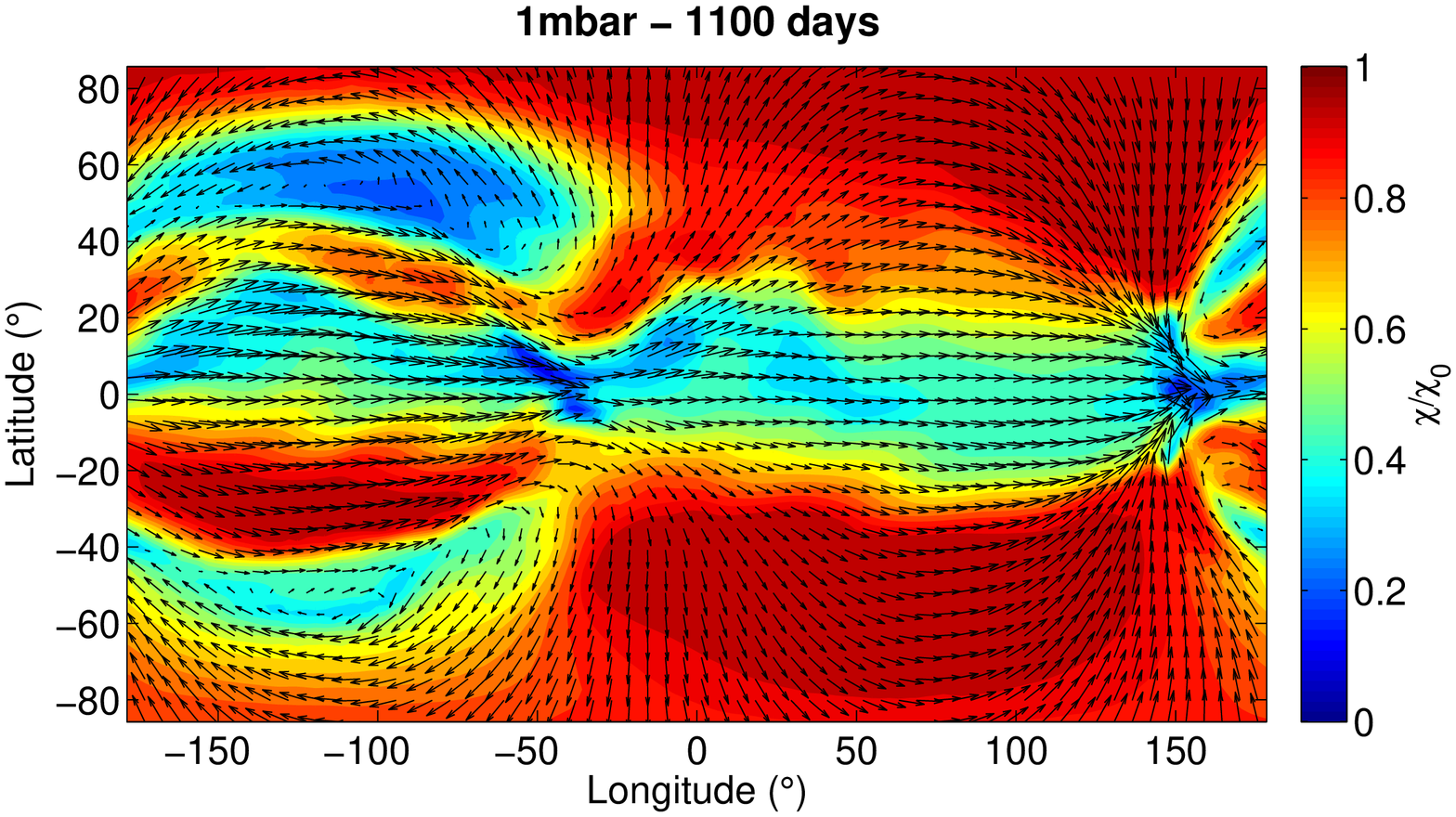}
\includegraphics[angle=-90,width=\linewidth]{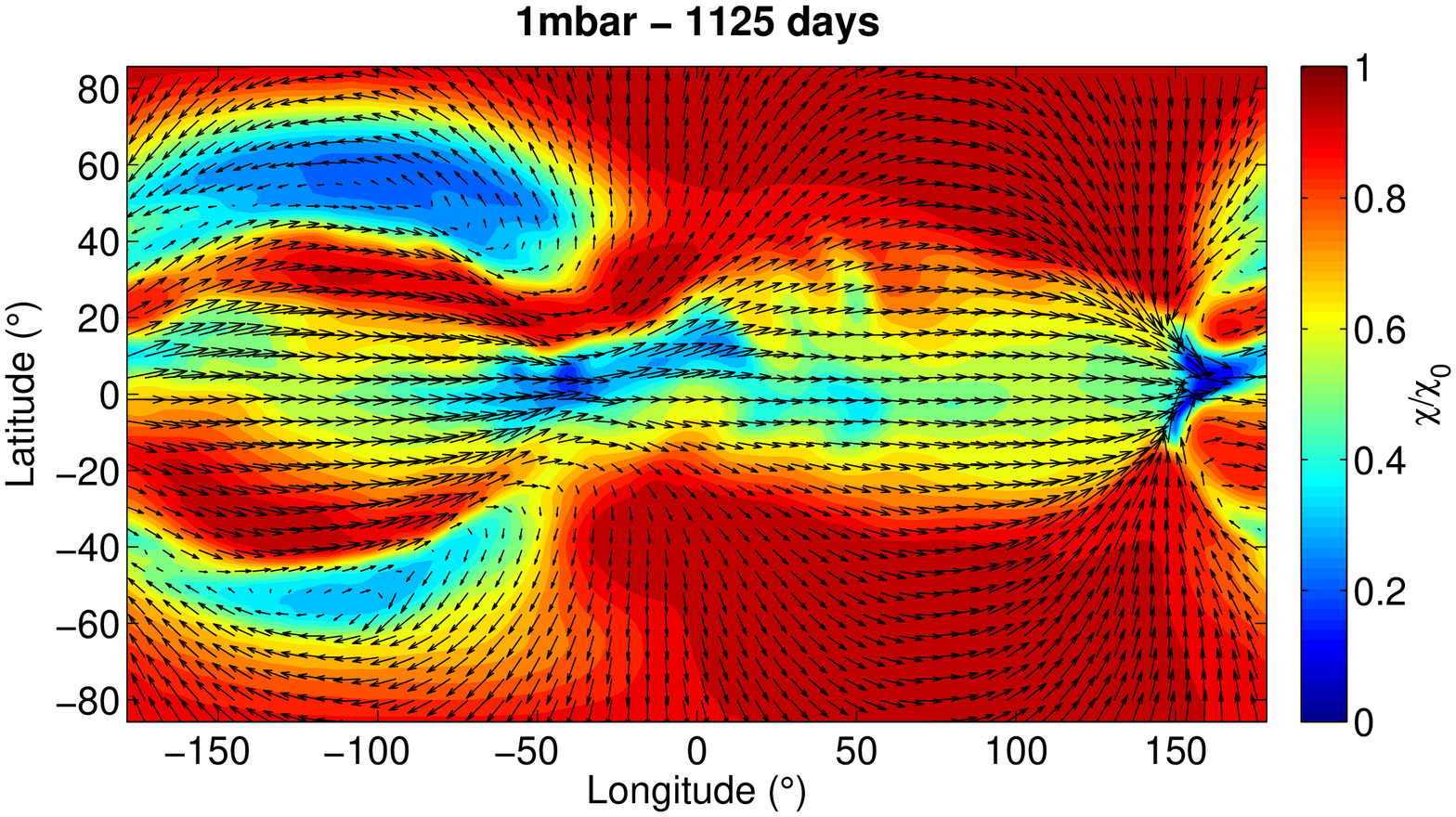}
   \end{minipage} \hfill
   \begin{minipage}[c]{.46\linewidth}
\includegraphics[angle=-90,width=\linewidth]{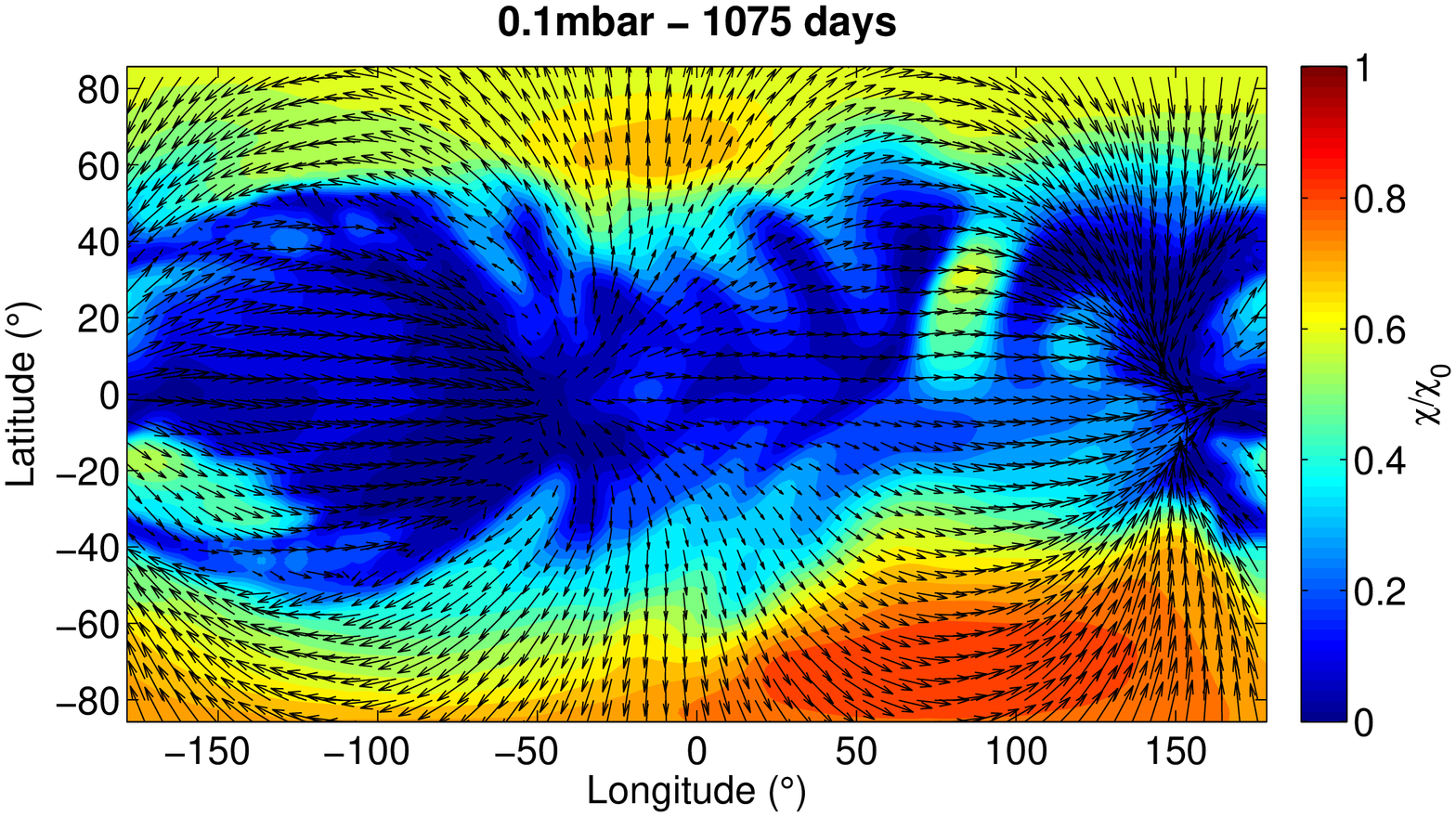}
\includegraphics[angle=-90,width=\linewidth]{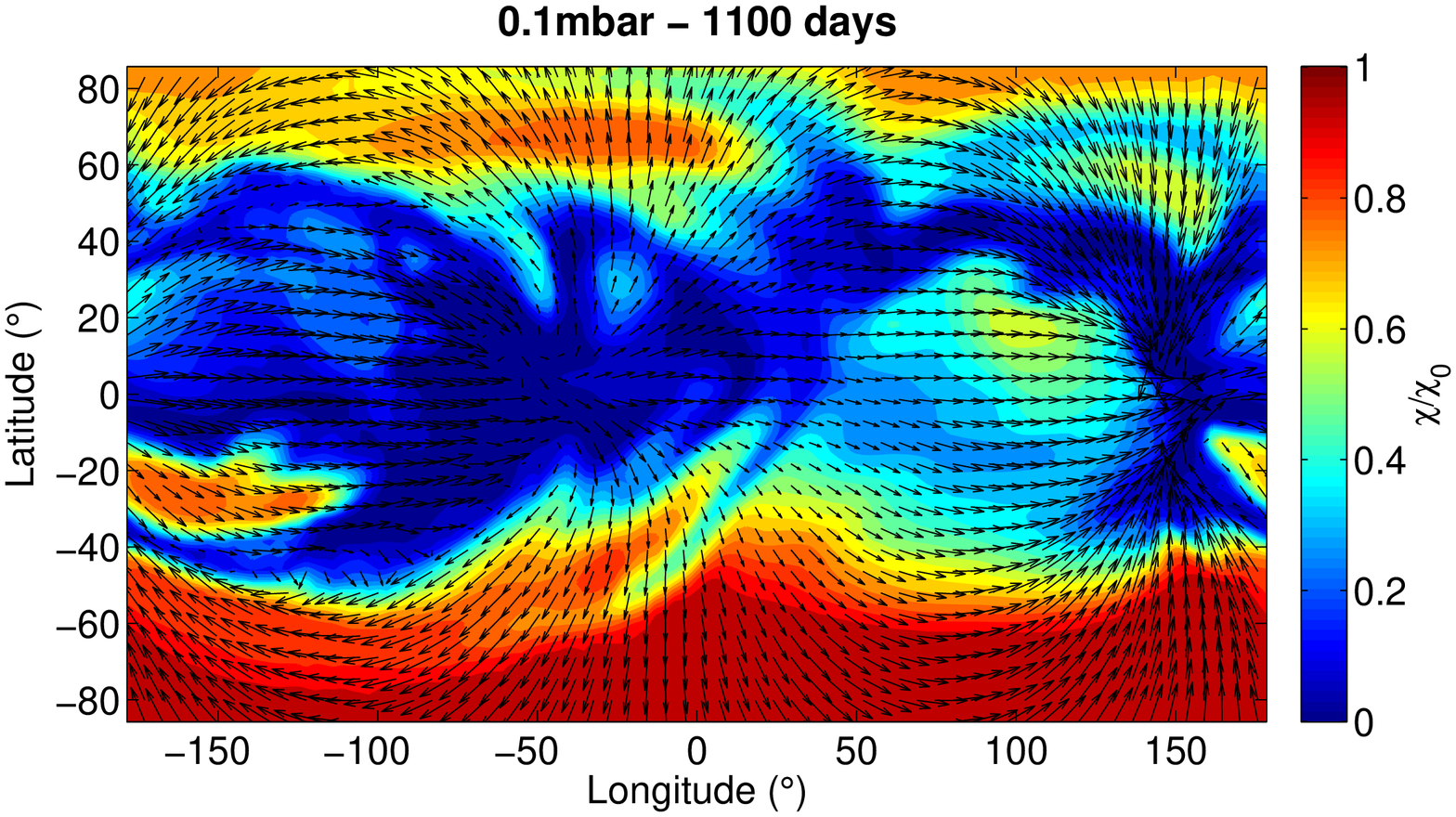}
\includegraphics[angle=-90,width=\linewidth]{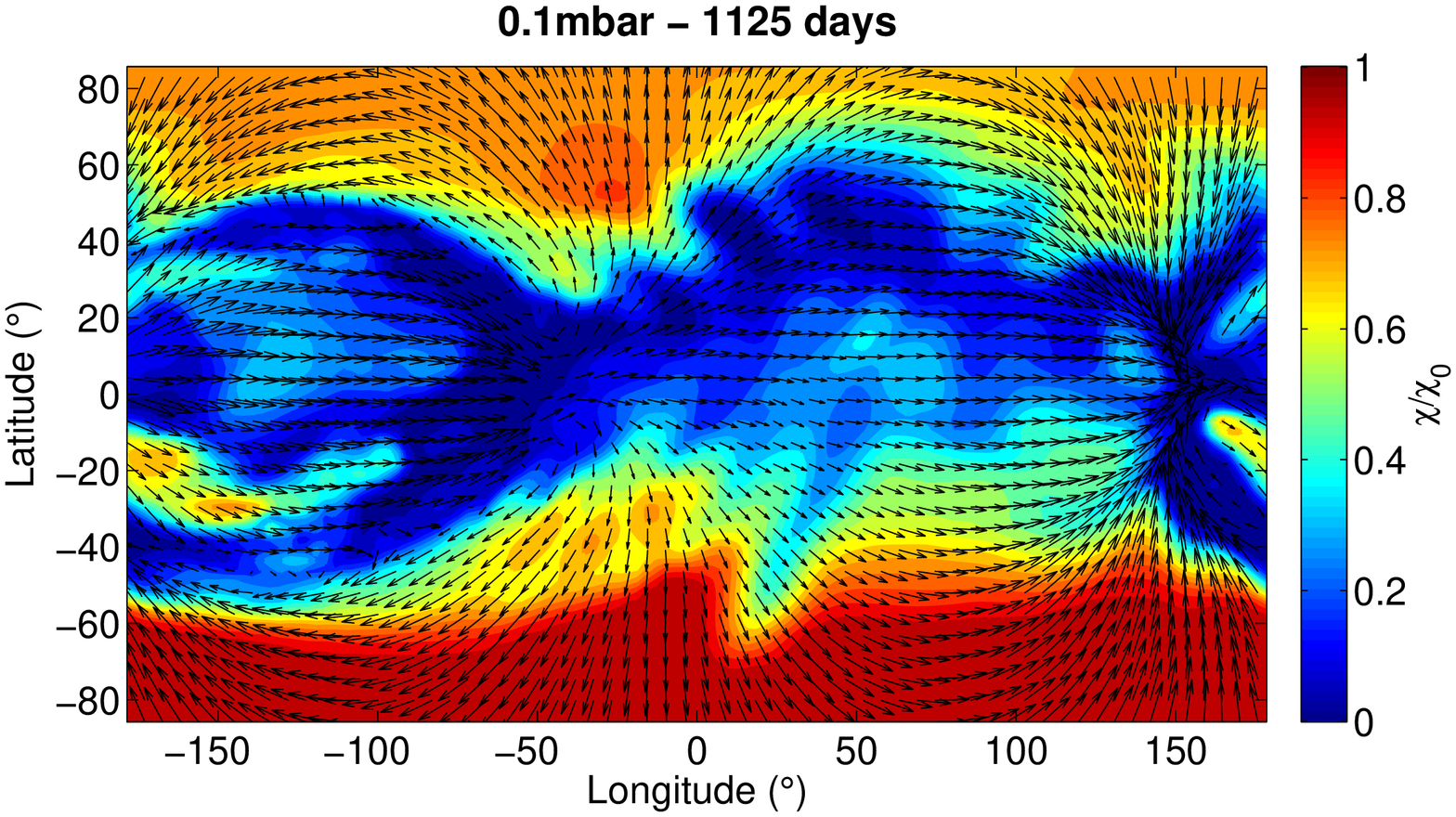}
   \end{minipage}
  \caption{Tracer abundance (colorscale) and winds (arrows) at two different pressure levels and three different times of the simulation for the case $a=\unit{2.5}\micro\meter$. }
   \label{fig::jet}
\end{figure*}

Figure \ref{fig::TracerTimeDependance} sheds light on the different
timescales at which this variability occurs. The top panel shows the
tracer abundance averaged vertically between $0.1$ and $\unit{1}\milli\bbar$ and
horizontally over a circular patch of $45^{\circ}$
centered on the substellar point; this gives a sense of how the tracer
abundance would vary in secondary-eclipse measurements probing the
dayside. The bottom panel shows the tracer abundance averaged
vertically between 0.1--1 mbar and horizontally around the terminator,
including all regions within $5^{\circ}$ of the terminator itself;
this gives a sense of how the tracer might vary in transit
measurements.  The variability exhibits two characteristic
timescales: a short (fast) timescale of order of
days and a long (slower) timescale of $\sim$50 to 100 days. The bigger the
particles on the nightside, the bigger the amplitude of the
variations. This comes from the smaller settling timescale of bigger particles. In these models, the amplitude of the long-period
variations exceeds those of the short-period variations by a factor of
$\sim$2--3. The long-timescale variations exhibit similar amplitudes
in the dayside and terminator time series. The short-period
oscillations exhibit stronger amplitude at the terminator and seem
related to variations of the flow itself, such as the oscillation of
the jet described previously. Figure \ref{fig::TracerTimeDependance}
suggests that radiatively active tracer species that can condense on
the nightside, such as TiO or silicates, could lead to detectable time
variations in transit or secondary eclipse spectra.  The amplitude of
this variability will depend on the type of tracer being considered
(see Sect. \ref{sec::Applications}) and may vary from planet to
planet depending on the availability of the considered species.
However we can predict the expected period of these variations: some
days for the small amplitude ones and fifty to one hundred days for
the biggest ones.

\subsection{Limb profile}

Transit observations are sensitive to atmospheric composition near the
terminator, thus we want to characterize the
distribution of our tracer species at the terminator. The possibility
of variations in chemical composition between the leading and trailing
limbs (as seen during transit) has been discussed in a variety of
studies \citep[\eg][]{Iro2005, Fortney2010}.  However, these studies
did not investigate the particular depletion of species due to the
interaction between their condensation and the atmospheric dynamics.
Our model leads to the first quantitative estimate of how dynamics affects the spatial distribution of condensable species at the day-night terminators, relevant to the interpretation of transit observations.
Figure \ref{fig::Abundance-limb} shows the tracer abundance at the
terminator at a snapshot in time for our models with particle sizes of
0.1, 0.5, 1, 2.5, 5, and $10\,\mu$m.  Angle represents angle around
the terminator and the radial coordinate represents the logarithm of the pressure. In
agreement with Fig.~\ref{fig::TracerMeanProfile}, the tracers tend
to be depleted from upper levels, particularly in models where the
particles on the nightside are larger.  Moreover, Fig.~\ref{fig::Abundance-limb}
demonstrates that significant spatial variations occur along the
terminator.  Depletion occurs first at the equator along the leading
limb, corresponding to the terminator $90^{\circ}$ west of the
substellar point.  The superrotating jet carries air depleted in
tracer from the nightside directly to this region of the terminator,
explaining why abundances are particularly depleted there.  In
contrast, air along most of the remainder of the terminator has
arrived from the dayside, where no particle settling occurs, so
depletion is less strong---particularly for particle sizes $\lesssim
1\,\mu$m.  Once the particle size becomes sufficiently large, however,
depletion occurs everywhere along the terminator at upper levels
regardless of whether the air arrived there from the dayside or
nightside.

\begin{figure*}
   \begin{minipage}[c]{.46\linewidth}
\includegraphics[width=\linewidth]{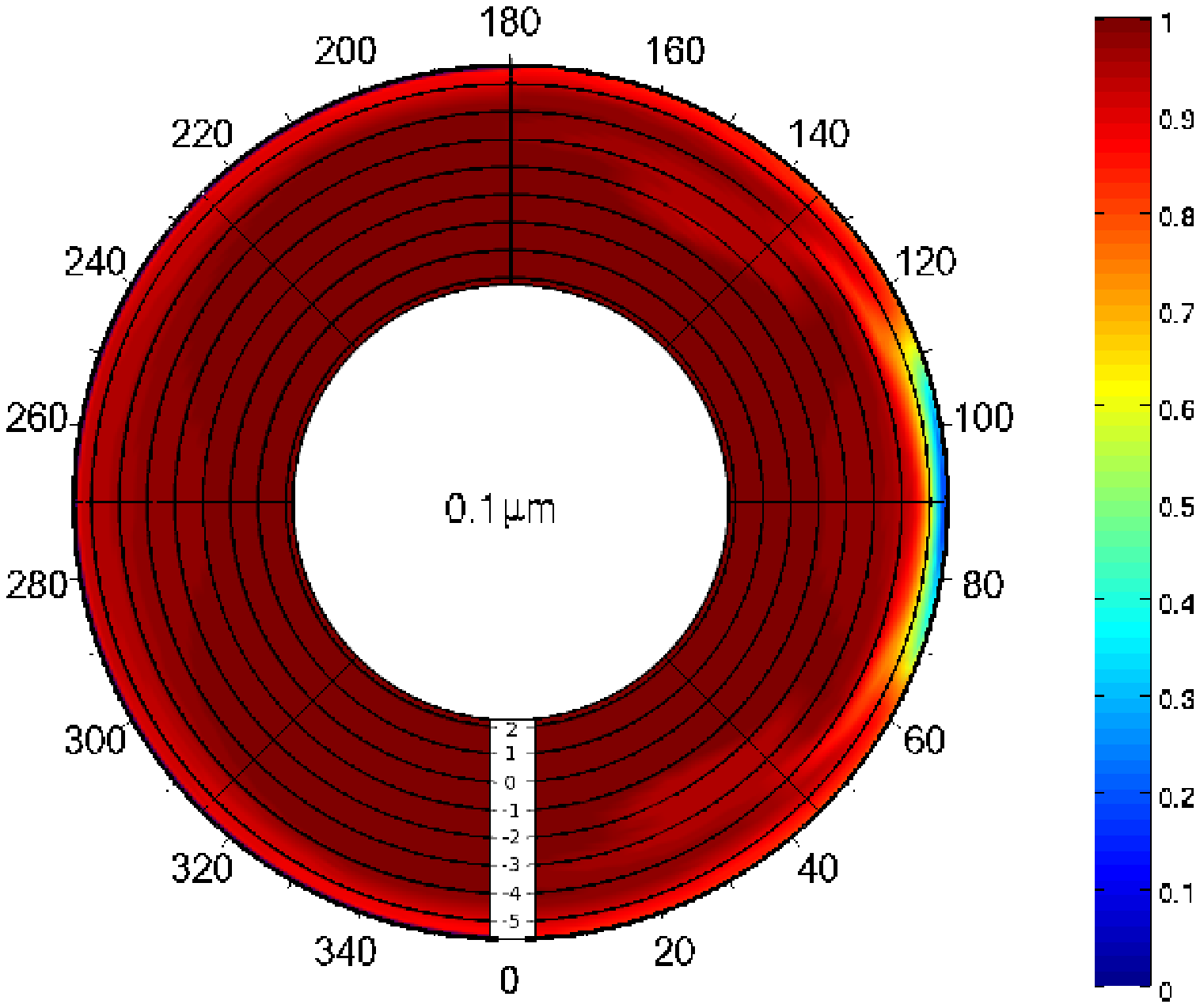}
\includegraphics[width=\linewidth]{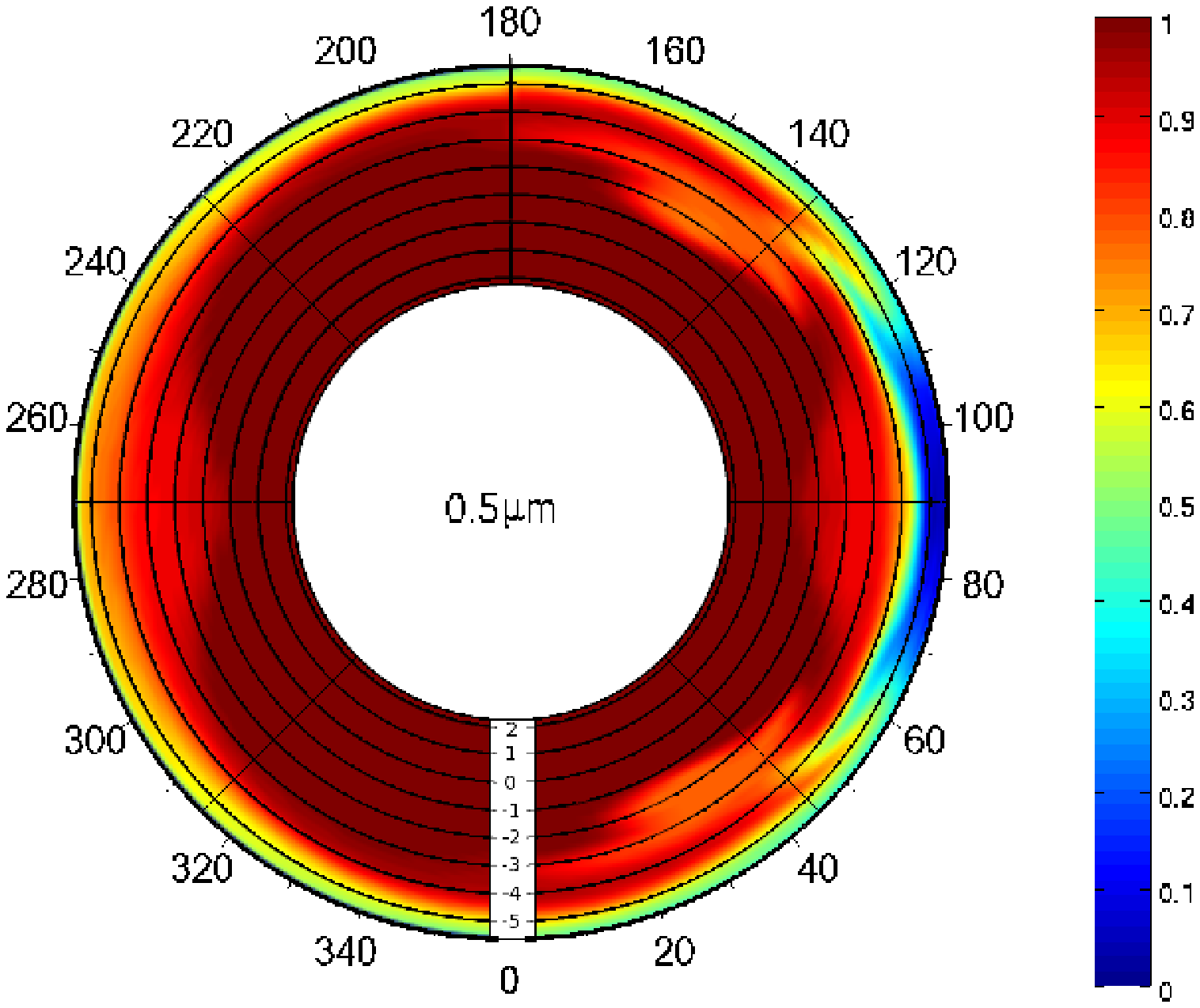}
\includegraphics[width=\linewidth]{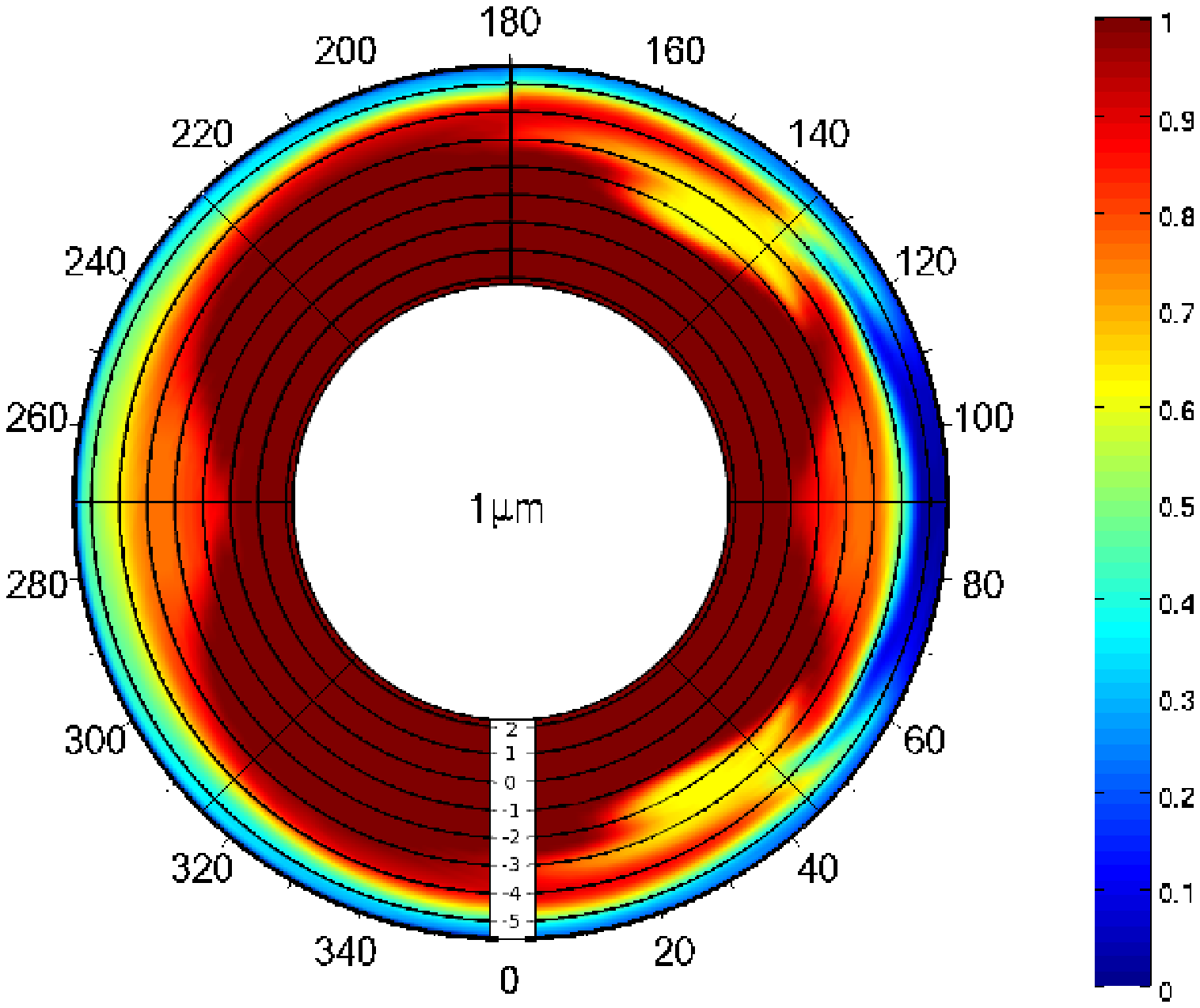}
   \end{minipage} \hfill
   \begin{minipage}[c]{.46\linewidth}
\includegraphics[width=\linewidth]{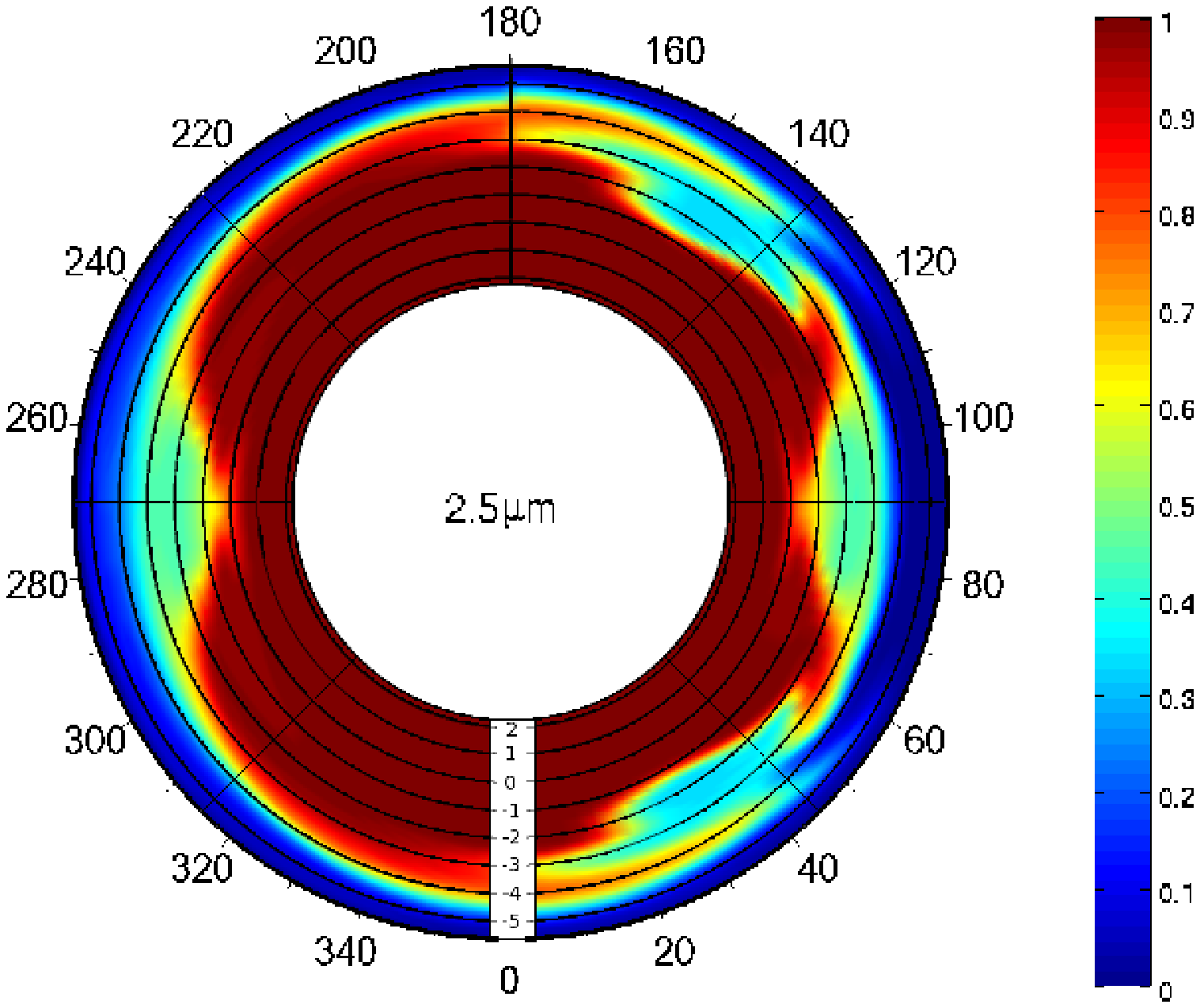}
\includegraphics[width=\linewidth]{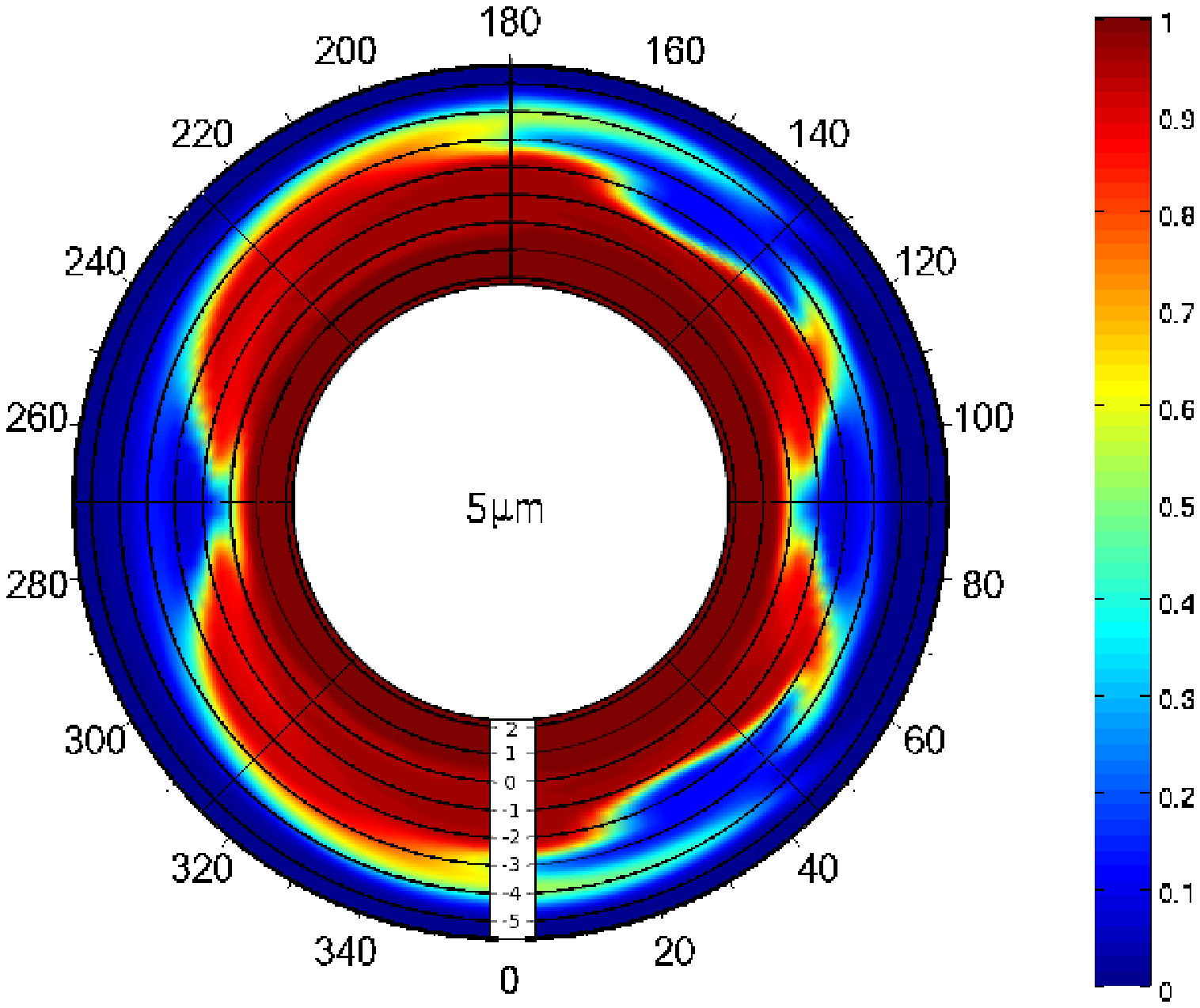}
\includegraphics[width=\linewidth]{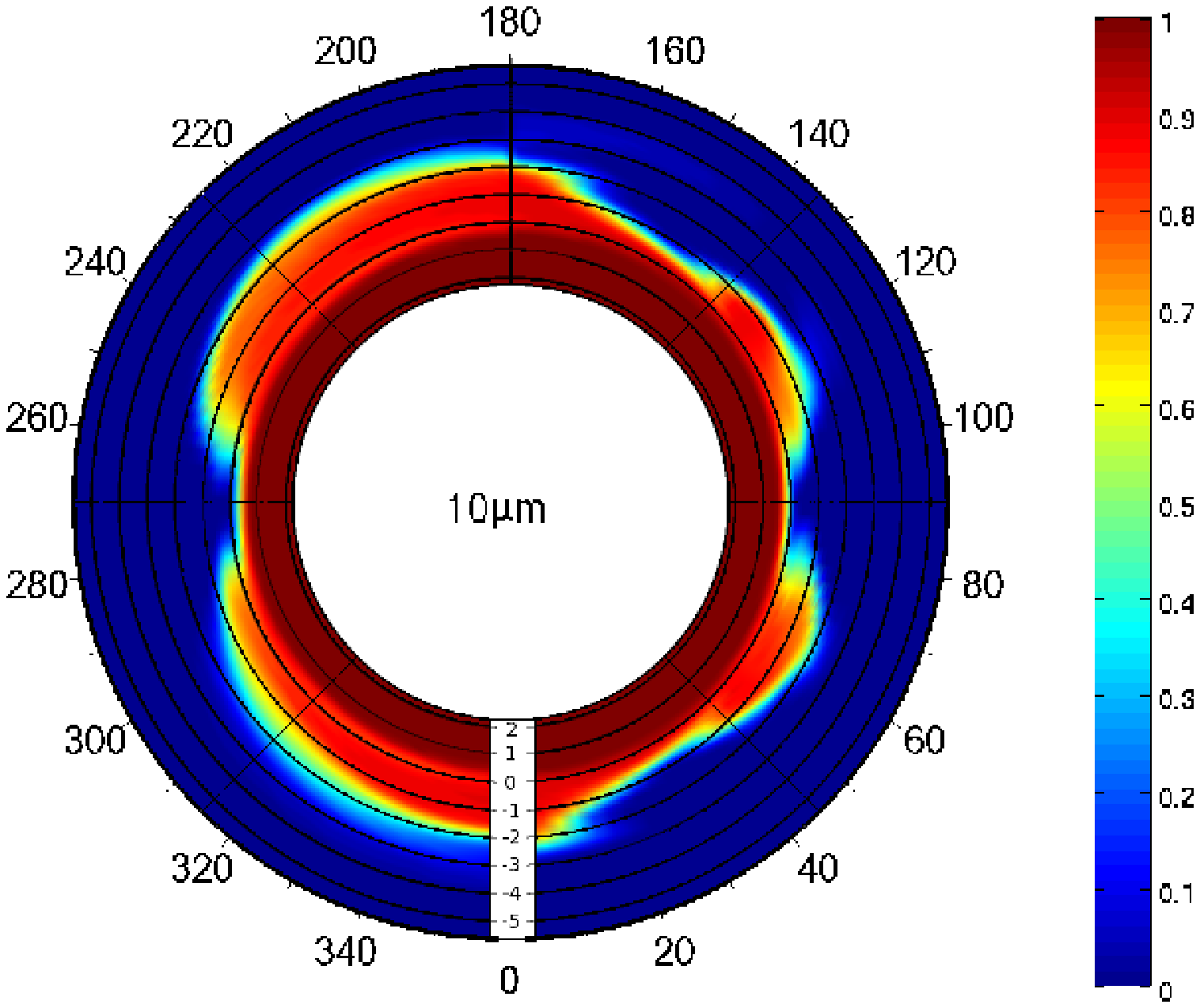}
   \end{minipage}
   \caption{Limb profiles of our tracer field from $200\bbar$ to $1\micro\bbar$ as seen during transit. Black circles are situated at $100\bbar$, $10\bbar$, $1\bbar$, $0.1\bbar$, $0.01\milli\bbar$, $1\milli\bbar$, $0.1\milli\bbar$, $0.01\milli\bbar$, and $1\micro\bbar$. The north pole is on top and the leading limb on the right. The abundances are normalized to the abundance of the deepest level.}
   \label{fig::limb}
\end{figure*}

\begin{figure}[h!]
 \center
\includegraphics[width=\linewidth]{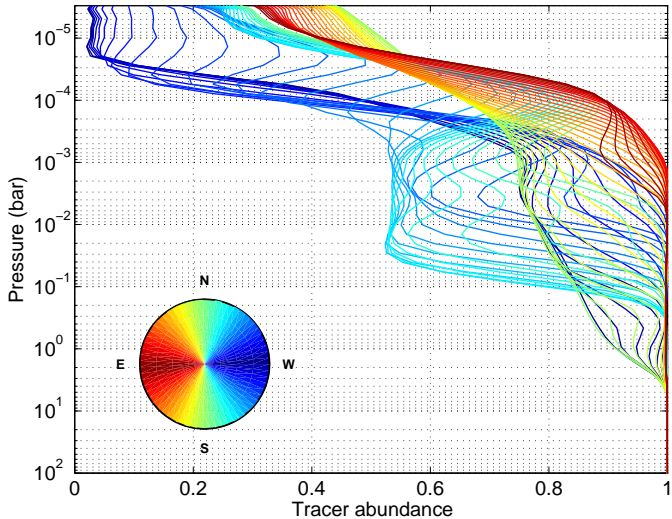}
\caption{Mean tracer abundance along the terminator for the case with nightside condensates of $\unit{1}\micro\meter$. The different colors correspond to different latitudes along the terminator as shown in the legend. The colorwheel represents the terminator of the planet as seen during a transit, with the eastern hemisphere to the left. The mean profiles are almost symmetrical with respect to the equator, so only the profiles of the northern hemisphere are shown here.}
\label{fig::Abundance-limb}
\end{figure}

Considering now the depth dependence of the terminator abundances, our results
suggest two different zones (see Figs.
\ref{fig::limb} and \ref{fig::Abundance-limb}) :
\begin{itemize}
\item At altitudes above the 1-mbar level, the tracer abundance is homogeneous over most of the limb except for the east (trailing) equatorial limb that is strongly depleted.
\item At altitudes below the 1-mbar level, the east and west equatorial limb are rather homogeneous; however, the east/west dichotomy shifts to higher latitudes and the west limb above $\unit{45}\degree$ is more depleted than the equivalent region of the
east limb.
\end{itemize}

Moreover, it should be noted that due to the shift of the hot spot,
the temperatures at the east limb are higher than at the west limb,
thus a given species is more likely to be gaseous and detectable on
the trailing limb than on the leading limb.

\section{1D model of the day-night cold trap}
\label{sec::1DModel}

Although hot Jupiters atmospheres are inherently three-dimensional, 1D
models continue to play a useful role for understanding the vertical
thermal and chemical structure of these atmospheres.  In particular,
many groups have explored the chemistry of hot Jupiters using 1D
models in which the vertical mixing caused by the large-scale dynamics
is parameterized by a specified eddy diffusivity \citep[\eg][]{Spiegel2009, Zahnle2009a, Zahnle2009, Youdin2010, Line2010, Line2011, Madhusudhan2011a}.  In these studies, the chosen eddy diffusivity
is ad hoc, with no convincing theoretical support.  Although the
vertical mixing in our 3D models is not diffusive in any rigorous
sense, there is merit in comparing the results of our 3D models with
1D models parameterized by eddy diffusivity.  This allow us to
make approximate estimates of the magnitudes of eddy diffusivity---in
the context of a 1D model---that produce similar horizontal-mean
behavior as our 3D models.  Such estimates of eddy diffusivity should guide the parameter choices in 1D chemical models like those cited
above.  A comparison between our 3D models and 1D diffusive models also allow us to investigate how the horizontal-mean tracer
depletion relates to the amplitudes of spatial tracer variation.

Therefore, in this section, we present a simple 1D model,
including particle settling, with atmospheric mixing represented
as an eddy diffusivity.

\subsection{System studied}
\label{sec:Syst}
The presence of a superrotating, eastward equatorial jet is a dominant
dynamical feature of many 3D circulation models of hot Jupiters.  This
superrotating jet was first predicted by \citet{Showman2002}, and later
emerges from almost all 3D simulations of hot Jupiters atmospheres
\citep{Cooper2005, Showman2008,Showman2009, Showman2013, Dobbs-Dixon2008, Rauscher2010, Rauscher2012a, Rauscher2012b, Perna2010, Perna2012, Heng2011a, Heng2011, Lewis2010, Kataria2013}
including ours (see Sect. \ref{sec::GlobalCirculation}) and has been
theoretically understood \citep{Showman2011}. A shift of the hottest
point of the planet eastward from the substellar point has been
directly observed in several exoplanets \citep{Knutson2007,Knutson2009,Knutson2012,Crossfield2010} and interpreted as a direct consequence of this jet. Thus we
believe that any study of the day/night cold trap in hot Jupiters
atmospheres must account for this feature.

To include the presence of this jet in our model, we choose as a study
system a vertical column of gas homogeneously advected around the equator by
the superrotating jet. Such a column is transported from day to night
and from night to day with a period $\tau_{\rm adv}=\frac{2\pi
  R_{\rm p}}{u_{\rm jet}}$ where $\tau_{\rm adv}$ is the advective timescale,
$R_{\rm p}$ is the planetary radius and $u_{\rm jet}$ the equatorial jet
velocity. $\tau_{\rm adv}$ is around 48h for HD 209458b.

As in the 3D case, we focus on a hypothetical chemical species which is gaseous on the
dayside and trapped in condensates of size $a$ on the nightside. This species freely diffuses
with a vertical diffusion coefficient $K_{\rm zz}$ on both the dayside
and the nightside. Because we envision the species as condensed on
the nightside, we additionally include downward settling via
Stokes-Cunningham drift on the nightside (Eq.~\eqref{eq::Vf}) but not on the dayside.
The model includes no horizontal dimensions. Rather, we model a single column of gas that is advected from day to night. The horizontal variations therefore translate into a time-dependant settling term. We assume this chemical species to be
a minor constituent of a H$_{2}$-atmosphere and neglect the
latent heat released during the condensation.

\subsection{1D diffusion equation}
As before, $\chi$ is the local mole fraction of the target chemical species~\ie the number of moles of tracer species (whether
in gaseous or condensed form) to the total moles of air in a given
volume. On the dayside, the molecules
of the target chemical species can freely diffuse with a diffusion
coefficient $K_{\rm zz}$, according to the equation:
\begin{equation}
\frac{\partial\chi}{\partial t}-\frac{1}{\rho}\frac{\partial}{\partial z}\left(\rho K_{\rm zz}\frac{\partial \chi}{\partial z}\right)=0
\label{eq::DiffDay}
\end{equation}
with $\rho$ the density of the atmosphere and $z$ the vertical coordinate.

On the nightside, the molecules of the target species are trapped into
particles that both diffuse and settle with the velocity $V_{\rm f}$ described
in~\ref{sec::Velocity}. Thus $\chi$ follows the same equation as Eq.~\eqref{eq::DiffDay}, plus a source term describing the settling:
\begin{equation}
\frac{\partial\chi}{\partial t}-\frac{1}{\rho}\frac{\partial}{\partial z}\left(\rho K_{\rm zz}\frac{\partial \chi}{\partial z}\right)=\frac{1}{\rho}\frac{\partial(\rho\chi V_{\rm f})}{\partial z}
\label{eq::DiffNight}
\end{equation}

We can define the diffusive time scale as $\tau_{\rm d}=\frac{H^2}{K_{\rm zz}}$ and a reference free fall time scale $\tau_{\rm s}=H/V_{\rm s}$ with $H=\frac{k_{B}T}{mg}$ the atmospheric scale height and $V_{\rm  s}$ the Stokes velocity (see \ref{sec::Velocity}). We note that $\tau_{\rm d}$ is a reference time scale and is not equal to the effective free fall time scale for high Knudsen numbers. Assuming hydrostatic balance, we can use the pressure $P$ as the vertical coordinate and using the perfect gas law the system to solve become:

\begin{equation}
\left\{
\begin{aligned} 
&\frac{\partial\chi}{\partial t}-\frac{\partial}{\partial P}\left(\frac{P^{2}}{\tau_{\rm d}}\frac{\partial \chi}{\partial P}\right)=0 & \mbox{on the dayside} \\
\\
&\frac{\partial\chi}{\partial t}-\frac{\partial}{\partial P}\left(\frac{P^{2}}{\tau_{\rm d}}\frac{\partial \chi}{\partial P}\right)=-\frac{1}{\tau_{\rm s}}\frac{\partial \beta P\chi}{\partial P} & \mbox{on the nightside}
  \end{aligned}
  \right.
  \label{eq::System}
\end{equation}

\subsection{Time-dependent solution}
\begin{figure}[h!]
 \center
\includegraphics[height=0.24\textheight]{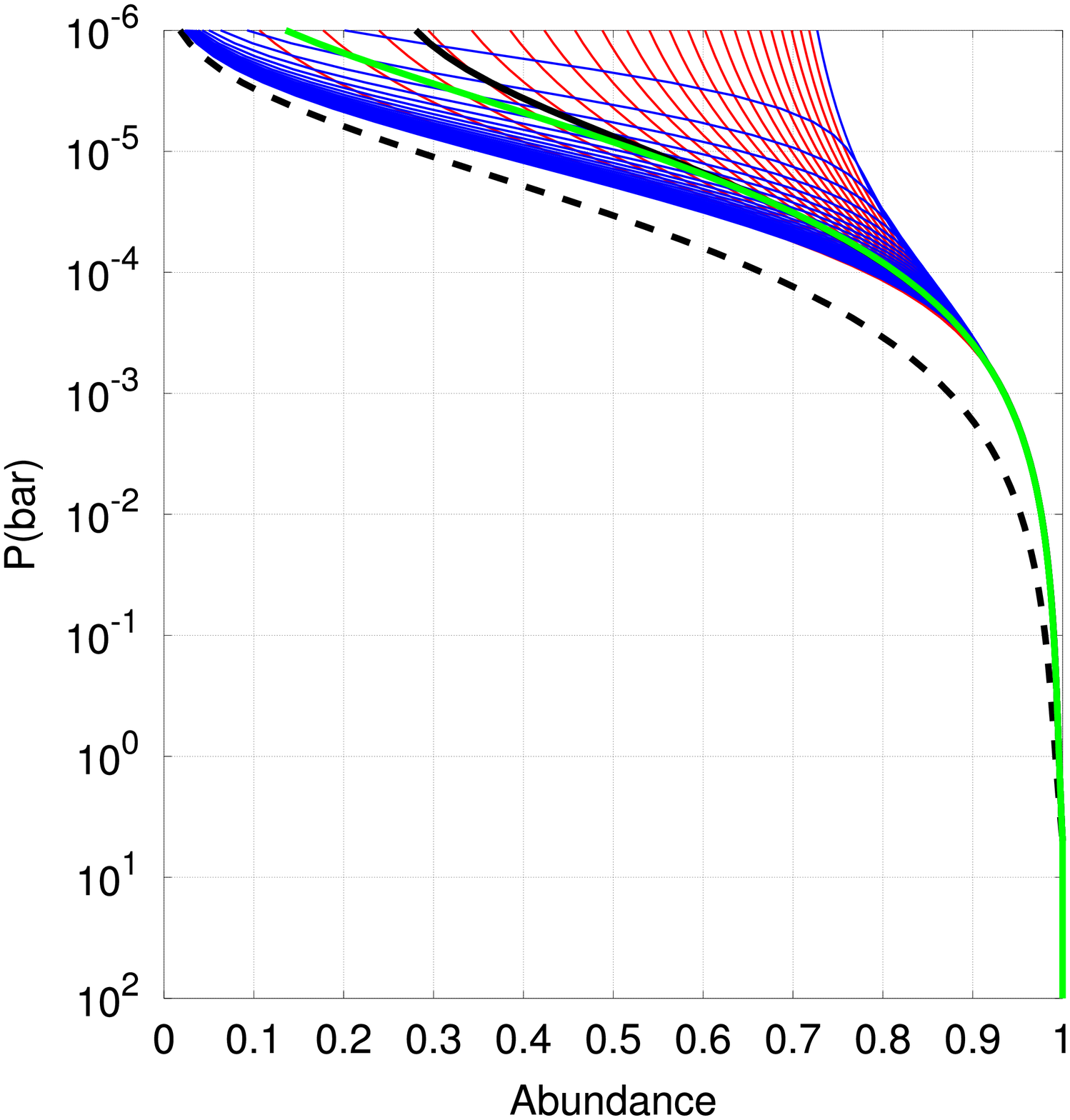}
\includegraphics[height=0.24\textheight]{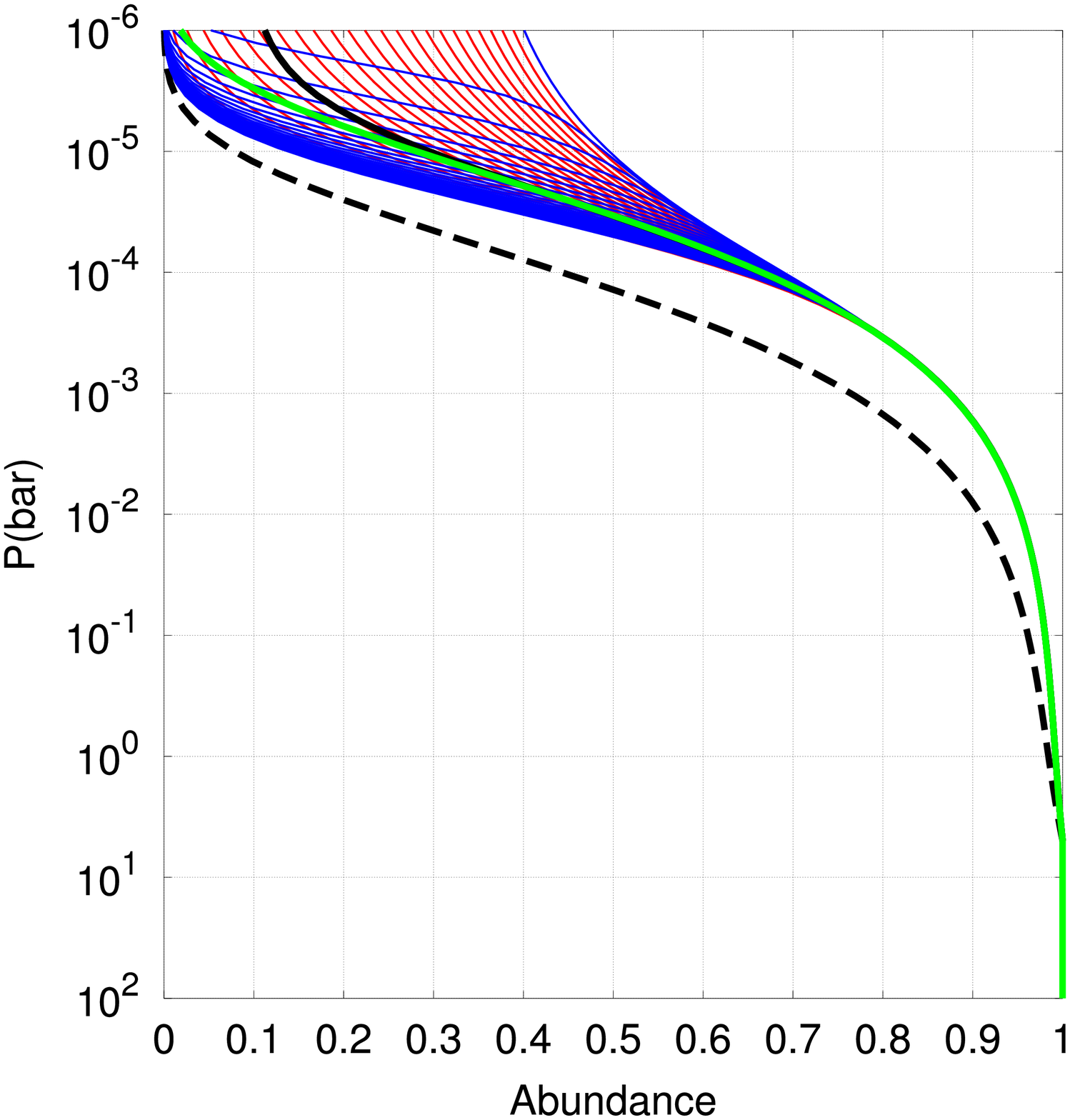}
\includegraphics[height=0.24\textheight]{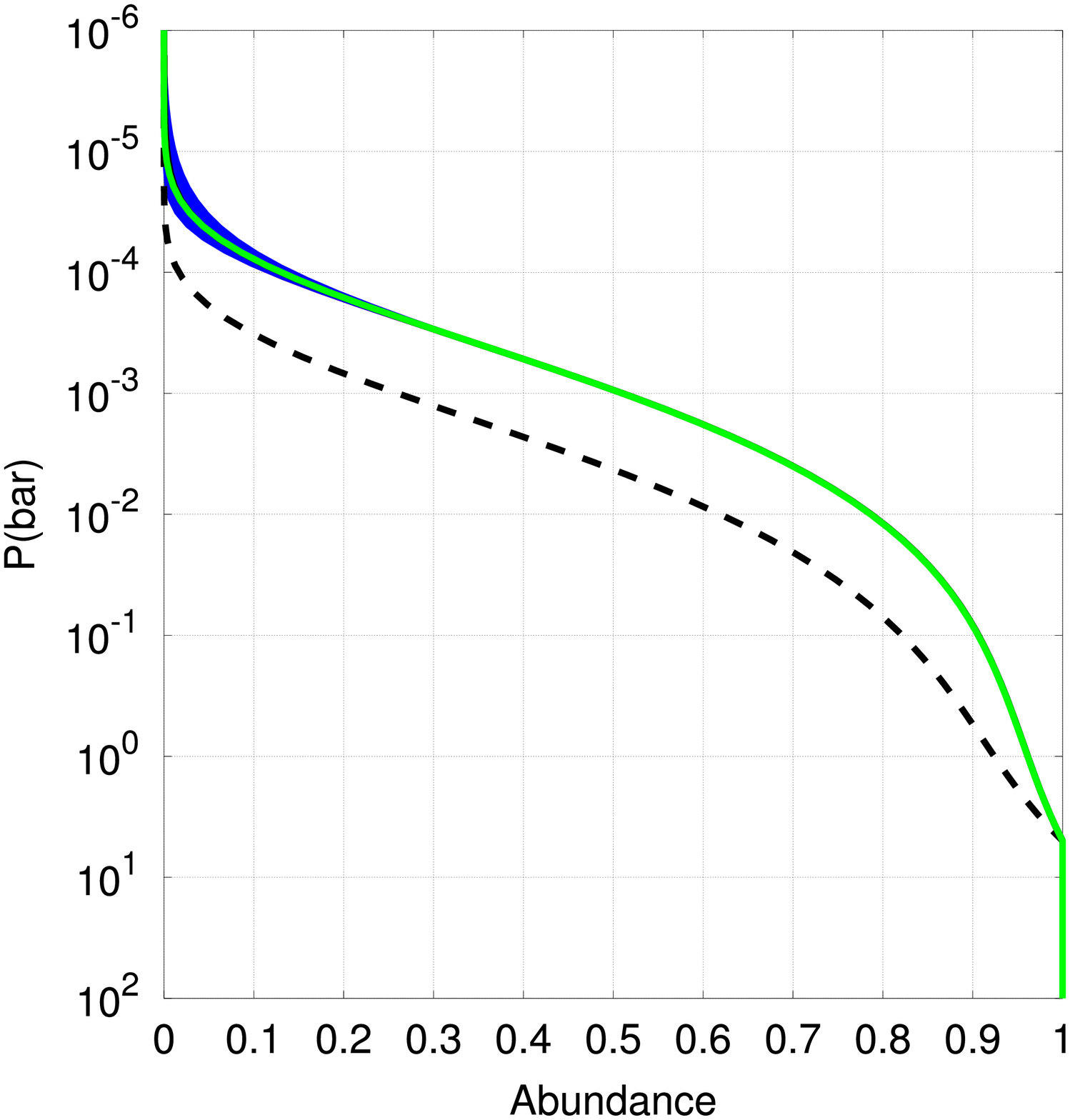}
\caption{Tracer abundance in the advected column of gas from the 1D model as a function of time. From bottom to top we used $K_{zz}=\unit{(10^{4};5\times10^{4};10^{5})\sqrt{\unit{1}\bbar/P}}\meter\squared\reciprocal\second$ and a particle radius on the nightside of $1\micro\meter$. We plot one profile every hour, corresponding to the profiles at equally spaced longitudes. The blue ones are on the nightside, the red ones on the dayside. The curve at the far left (right) of the envelope is the profile at the east (west) terminator. The black curve is the mean over one period. The green line is the analytical model (see Eq. \eqref{eq::ChimAlpha} using $\alpha=1/2$) and the dotted line is the solution for particles that would be constantly falling (both in the dayside and in the nightside).}
\label{fig::Chi-profiles-Kz-square}
\end{figure}

In order to solve the system of equations~\eqref{eq::System} for
$\chi(P,t)$, we need two boundary conditions. We assume the species to
be well mixed below $P_{1}=\unit{1}\bbar$ with a molecular abundance
$\chi_{\rm 0}$. At the top of the atmosphere, we assume that no molecule
crosses the upper boundary (\ie $\frac{\partial\chi}{\partial
  P}\left|_{P=P_{\rm top}}=0\right.$). Then we can solve the system with
an implicit time stepping code using GNU OCTAVE\footnote{http://www.octave.org}, an open-source, free software
equivalent to MATLAB. Assuming that the column of gas spends
$\tau_{\rm adv}/2=\unit{24}\hour$ in each hemisphere, we can reach a
periodic behavior where the initial condition is forgotten. While on
the dayside, there is no settling and, at upper levels, the
tracer diffuses upward.  Thus at a given pressure the molecular
abundance increases with time. This is shown by the red curves in
Fig.~\ref{fig::Chi-profiles-Kz-square}. While on the nightside, the
particles both diffuse vertically and settle downward at their terminal velocity
and thus, at at given pressure level, the abundance decreases with
time. This is shown by the blue curves in
Fig.~\ref{fig::Chi-profiles-Kz-square}.  In the upper atmosphere,
the settling timescale and the diffusion timescale become smaller than
the advective timescale. The particles have time to settle several
scale heights during the time required for the air column to cross
the dayside. Thus, at upper levels, the molecular abundance vary strongly
throughout the diurnal cycle, and the vertical
profiles vary widely around the mean value.

\subsection{Steady-state solutions}
Our system being forced periodically, there is no steady-state
solution \emph{stricto sensu}. However, the mean of the tracer
abundance over one period should remain constant. We can thus
integrate eqs.~\eqref{eq::System} over one period and get an
equation for $\chit$, the mean molecular abundance over one
period. Then, once the periodic state is reached, the mean over one
period of the tracer abundance in our column of gas, $\chit$, is the
same as the mean over longitude at a given time, $\chim$ and we obtain
:
\begin{equation}
P^2\frac{\partial \chim}{\partial P}-\frac{\beta}{2}\frac{\tau_{\rm d}}{\tau_{\rm s}}P\chim=C
\label{eq::diffusion2}
\end{equation}
where the factor $\frac{1}{2}$ appears because the source term is only
integrated during the night. To derive Eq.~\eqref{eq::diffusion2}
we assumed that the mean tracer abundance on the nightside is
close to the global-mean tracer abundance. This is true given than
$\beta\tau_{\rm s}$ and $\tau_{\rm d}$ are much bigger than $\tau_{\rm
  adv}$. This approximation breaks down at low pressure and our
analytical solution diverges from the real one as can be seen by
comparing the black and the green curves in
Fig.~\ref{fig::Chi-profiles-Kz-square}. However, the discrepancy is
small such that the analytic solution is still a good representation
of the time-mean of the full numerical solution of the 1D model everywhere
deeper than the $\sim\unit{10}\micro\bbar$ level.  $C$
is a constant coming from the integration over pressure. When $P$ goes
to $0$, $\chi$ goes to $0$. Moreover we assume that $ \frac{\partial
  \chi}{\partial P}$ does not go to infinity. Then the constant must
be zero and we obtain :
\begin{equation}
\frac{\partial \chim}{\partial P}=\frac{1}{2}\frac{\beta}{P}\frac{\tau_{\rm d}}{\tau_{\rm s}}\chim
\label{eq::dChi}
\end{equation}

To simplify the problem, we neglect the transitional regime for
$\beta$ from Eq.~\eqref{eq::beta} and use the expression :
\begin{equation}
\beta=1+1.656K_{N}
\label{simple-Beta}
\end{equation}
Then choosing a functional form $K_{\rm zz}=K_{\rm zz_{\rm 0}}(P_{\rm 0}/P)^{\alpha}$, where
$\alpha$ is a constant, we can solve equation \eqref{eq::dChi}. For $\alpha\not=0$ and $\alpha\not=1$, we get : 
\begin{equation}
\begin{split}
\chim=&\chi_{r}\exp\left({\frac{1}{2\alpha}\frac{\tau_{\rm d_{\rm 0}}}{\tau_{\rm s}}\frac{P^{\alpha}-P_{r}^{\alpha}}{P_{0}^{\alpha}}}\right)\times\\
&\exp\left(\frac{1}{2(\alpha-1)}\frac{\tau_{\rm d_{\rm 0}}}{\tau_{\rm s}}\frac{1.656k_{\rm B}T}{\sqrt{2}\pi ad^2}\frac{P^{\alpha-1}-P_{\rm r}^{\alpha-1}}{P_{\rm 0}^{\alpha}}\right).
\label{eq::ChimAlpha}
\end{split}
\end{equation}
For the particular case of a constant $K_{\rm zz}$ ($\alpha=0$) the formula becomes :
\begin{equation}
\chim=\chi_{\rm r}\left(\frac{P}{P_{\rm r}}\right)^{\tau_{\rm d_{\rm 0}}/2\tau_{\rm s}}\exp\left(-\frac{\tau_{\rm d_{\rm 0}}}{2\tau_{s}}\frac{1.656k_{\rm B}T}{\sqrt{2}\pi ad^2}\left(\frac{1}{P}-\frac{1}{P_{\rm r}}\right)\right)
\label{eq::Xeq-KzzCste}
\end{equation}
In the case where $K_{\rm zz}$ is inversely proportional to $P$ ($\alpha=1$):
\begin{equation}
\chim=\chi_{\rm r}\exp\left({\frac{1}{2}\frac{\tau_{\rm d_{\rm 0}}}{\tau_{\rm s}}\frac{P-P_{\rm r}}{P_{\rm 0}}}\right)\left(\frac{P}{P_{\rm 0}}\right)^{\frac{1}{P_{\rm 0}}\frac{\tau_{\rm d_{\rm 0}}}{2\tau_{\rm s}}\frac{1.656k_{\rm B}T}{\sqrt{2}\pi ad^2}}
\end{equation}
Where $\chi_{r}$ is the abundance of $\chim$ at $P=P_{\rm r}$ and $\tau_{\rm d_{\rm 0}}\equiv H/K_{\rm zz_{\rm 0}}$. 

\subsection{Diffusivities needed to keep the tracers suspended}

Because particle settling acts to transport condensates downward, the
tracers only exhibit significant abundances in the upper regions
of the atmosphere if the eddy diffusion coefficient exceeds some
critical value, which depends on the particle size of the
condensates.  Here we solve for an approximate analytical expression
for this critical magnitude as a function of particle size and other
parameters.

Assuming a constant vertical diffusion coefficient, we can use
Eq.~\eqref{eq::Xeq-KzzCste} to derive an expression for the
$K_{\rm zz}$ needed to achieve a given molecular abundance $\chi_{\rm lim}$ at a
given pressure $P_{\rm lim}$. Like in~\citet{Spiegel2009}, we use $P_{\rm lim}=\unit{1}\milli\bbar$ and $\chi_{\rm lim}=0.5$. We first note that
Eq.~\eqref{eq::Xeq-KzzCste} is composed of two terms. The first
exponential is given by the Stokes regime ($\beta\approx1$) whereas
the second term is given by the Cuningham regime ($\beta>>1$). As 
can be seen in Fig.~\ref{fig::Vf}, particles smaller than
$\unit{10}\micro\meter$ should be in the Cunhingham regime at
$P=\unit{1}\milli\bbar$. Thus the second term in Eq.~\eqref{eq::Xeq-KzzCste} is dominant and we can
use as a condition :
\begin{equation}
\frac{1}{2}\frac{\tau_{\rm d}}{\tau_{\rm  s}}\frac{1.656k_{B}T}{\sqrt{2}\pi ad^2}\left(\frac{1}{P_{\rm lim}}-\frac{1}{P_{\rm r}}\right)=-\ln(\chi_{\rm lim}/\chi_{\rm r}).
\end{equation}

As in the 3D model, we assume that below $\unit{1}\bbar$ the tracers are no more trapped into condensates and no more subject to settling. Thus we choose $\chi_{\rm r}=1$ at $P_{\rm r}=\unit{1}\bbar$. Then, for $P_{r}\gg P_{\rm lim}$, replacing $\tau_{\rm d}$ and $\tau_{\rm s}$ by
their expressions, using Eq.~\ref{viscosity} for the viscosity and Eq.~\ref{lambdaP} for the mean free path, we obtain a condition on the diffusion coefficient:
\begin{equation}
K_{\rm zz_{\rm lim1}}\sim-\frac{HV_{s}}{2}\frac{1.656k_{\rm B}T}{\sqrt{2}\pi ad^2\ln(\chi_{\rm lim}/\chi_{\rm r})}\frac{1}{P_{\rm lim}}
\end{equation}

For big particles, the Stokes regime become dominant thus we
neglect the second term of Eq.~\eqref{eq::Xeq-KzzCste} and the condition turns to be:

\begin{equation}
K_{\rm zz_{\rm lim2}}\sim\frac{HV_{\rm s}\ln(P_{\rm lim}/P_{1})}{2\ln(\chi_{\rm lim}/\chi_{\rm r})}
\label{Kz2}
\end{equation}

As the relevant range of particle size span several order of magnitudes, a good approximation of $K_{zz_{lim}}$ can be obtained by taking the sum of these two coefficient:
\begin{equation}
K_{\rm zz_{\rm lim}}=K_{\rm zz_{\rm lim1}}+K_{\rm zz_{\rm lim2}}.
\label{eq::Kzlim}
\end{equation}

Again we note that these limits for the diffusion coefficient are
independent of the planet considered. \citet{Spiegel2009} stated that an abundance of
half the solar composition at 1mbar would be necessary to produce an
observable stratosphere. Applying formula~\eqref{eq::Kzlim} with
$\chi_{\rm lim}=0.5$ and $P_{\rm lim}=\unit{1}\milli\bbar$ and assuming a well
mixed layer below $\unit{1}\bbar$ (\ie $P_{\rm r}=\unit{1}\bbar$ and $\chi_{\rm r}=1$) we
obtain $K_{\rm zz_{\rm lim}}=1.4\times 10^{4}\rm\,m^2\,s^{-1}$ for
$a=\unit{0.1}\micro\meter$, $K_{\rm zz_{\rm lim}}=1.4\times 10^{5}\rm
\,m^2\,s^{-1}$ for $a=\unit{1}\micro\meter$ and $K_{\rm zz_{\rm lim}}=
1.6\times 10^{6} \rm\,m^2\,s^{-1}$ for
$a=\unit{10}\micro\meter$. These results are of the same order of
magnitude as the ones found by \citet{Spiegel2009} for the vertical cold trap. This is expected,
since we compare two similar mechanisms: settling and diffusion of
particles. However, this similarity of the results shows that the
day-night cold trap is at least as
important as the vertical one in hot Jupiters atmospheres. Moreover,
the condition on $K_{\rm zz_{\rm lim}}$ we derived is independent of
the planet studied and holds for very hot Jupiters such as WASP-12b or
WASP-33b where the vertical cold trap could be inefficient or
nonexistent.

\section{Effective vertical diffusion coefficient} 
\label{sec::Kzz}

As described previously, 1D models have been extensively used to
investigate chemistry and vertical structure of hot Jupiter
atmospheres, with the 3D dynamics parameterized as vertical
eddy diffusion with a specified diffusivity \citep[e.g ][]{Spiegel2009, Zahnle2009, Zahnle2009a,Youdin2010,Line2010,Line2011, Madhusudhan2011, Moses2011,Venot2012}.

There is no theoretical reason for the mixing by the large scale flow
patterns in hot Jupiters atmospheres to behave like a one-dimensional
diffusion process. However, deriving an {\it a posteriori} effective
diffusion coefficient that describes as closely as possible the
averaged vertical mixing within the atmosphere can be a useful way to
roughly characterize the strength of the vertical fluxes of material
and guide 1D modelers in their choice of vertical mixing parameters.

\begin{figure}[h!]
\includegraphics[width=\linewidth]{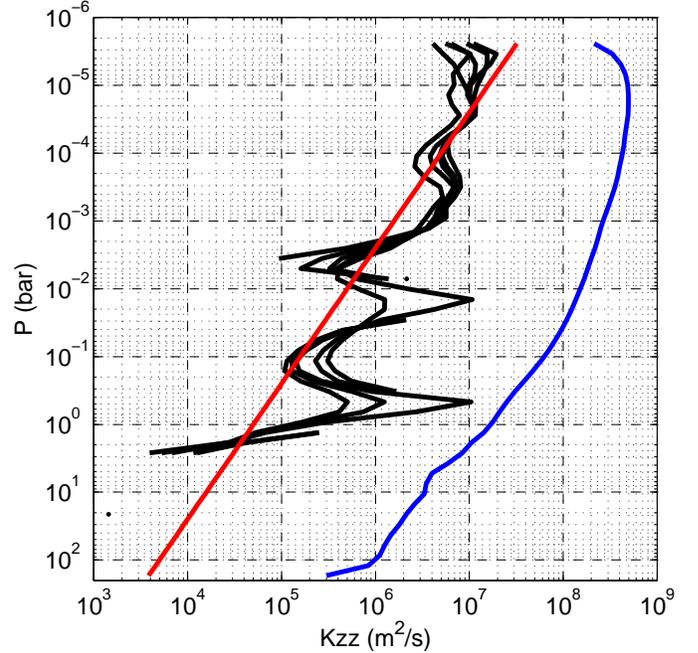}
\caption{Vertical diffusion coefficient from the 3D model estimated from the different tracer fields separately (black lines), from the 1D fit to the 3D model (red line) and the root mean square of the vertical velocity times the vertical scale height, a common estimate of $K_{zz}$ in the literature (blue line) \citep[see][for example]{Lewis2010,Moses2011}.}
\label{fig::Kzz}
\end{figure}

A first way to define a vertical mixing coefficient from our
simulation is to choose the $K_{zz}$ that best reproduces the planet
averaged tracer profiles. The 1D model developed in Sect.
\ref{sec::1DModel} describes the equilibrium between vertical
diffusion of tracers with a specified height-dependent $K_{zz}$ and their
settling on the nightside.  We tune the diffusivities to obtain a
good match between the solutions of our 1D model and the horizontal-mean
tracer abundance versus pressure from the 3D models.  To use Eq.~\eqref{eq::ChimAlpha} we must specify the temperature of the
atmosphere, constant in the analytical model. The temperature appears
in the expression for the Knudsen number and in the expression for the
viscosity of hydrogen. Both quantities are related to the
settling of the particles. Thus the temperature to consider is the
nightside temperature. In our GCM nightside temperatures range from $600$ to
$\approx\unit{1500}\kelvin$ and we decided to use a mean temperature
of 1000 K. However, we note that the derived value of $K_{\rm zz}$ does not
depend strongly on this choice. Then using a $K_{\rm zz}$ value
proportional to the inverse square root of the pressure ($\alpha=0.5$
in Eq.~\eqref{eq::ChimAlpha}), we obtain a remarkably good
agreement between our 1D model (dotted lines in Fig.~\ref{fig::TracerMeanProfile}) and the horizontal average of the 3D model (solid
lines of Fig.~\ref{fig::TracerMeanProfile}). The resulting value for
the vertical mixing coefficient that best fits the different tracer
field used in the simulation is :
\begin{equation}
K_{\rm zz}=\frac{5\times 10^{4}}{\sqrt{P_{\rm bar}}}\rm \,m^2\,s^{-1},
\label{kzz-1d-fit}
\end{equation}
which is valid over a pressure range from $\sim$1 bar to a few
$\mu$bar (see Fig.~\ref{fig::Kzz}). Fundamentally, this represents an ad-hoc fitting of our simulation results, although we find it a good match to the overall mixing properties of our 3D models over a wide range of particle sizes.  Note that the upward-increasing mixing rates captured in Eq.~\eqref{kzz-1d-fit} arise naturally from the fact that the radiative heating rates and vertical velocities tend to increase with decreasing pressure in our models, leading to greater mixing rates at lower pressure. The 3D models adopt the stellar insolation and other properties for HD
209458b, so the results are most germane to that planet; the mixing
rates are likely to be higher for hotter planets and lower for
cooler planets than implied by Eq.~\eqref{kzz-1d-fit}.

Another way to define a one dimensional vertical mixing coefficient
from the three-dimensional simulation is to find the $K_{zz}$ that
leads to an upward diffusive flux of material that matches the
averaged vertical flux produced by the dynamics. This can be written 
\citep[\eg][p.~90]{Chamberlain1987}:
\begin{equation}
K_{\rm zz}=\frac{\left\langle\rho\chi V_{\rm z}\right\rangle}{\left\langle\rho\frac{\partial \chi}{\partial z}\right\rangle}
\label{kzz-gcm}
\end{equation}
where the brackets represent the horizontal average along isobars over
the whole planet. This expression does not necessitate any assumption
on the functional form of $K_{zz}$ nor on the nightside mean
temperature. It also has the advantage that no comparisons or fits to
a 1D diffusion model are necessary; the effective values of $K_{\rm
  zz}$ can be derived directly from the 3D GCM data via
Eq.~\eqref{kzz-gcm}.  As a trade-off, this expression depends on
the vertical tracer gradients that are affected by long term temporal
variability (see Sect. \ref{sec::TimeVariability}), which are not
smoothed out completely given the limited integration time of the
simulation.  This leads to some strong vertical variation of $K_{\rm
  zz}$.  The profile of $K_{\rm zz}$ calculated from
Eq.~\eqref{kzz-gcm} are shown in the black curves of
Fig.~\ref{fig:Kzz}.  Despite the vertical fluctuations, the overall
shape of $K_{\rm zz}$ obtained with this method is close to the
estimate using the planet averaged tracer profile (see the red curve
in Fig.~\ref{fig::Kzz}). Like before, we note that the derived value
does not depend strongly on the particle size, consistent with the
fact that Eq.~\eqref{kzz-1d-fit} leads to a good fit between 1D
models and our 3D model for a wide range of particle sizes.

Several previous studies have attempted to estimate the vertical
diffusion coefficient in hot Jupiters atmospheres \citep{Cooper2006, Moses2011, Heng2011a, Lewis2010}.
\cite{Cooper2006} adopted an estimate for $K_{\rm zz}$ based on the
product of a root-mean-square vertical velocity from their 3D GCMs and
an appropriately chosen vertical length scale following the
formulation of \citet{Smith1998}. \citet{Moses2011} and \citet{Lewis2010} followed a similar procedure but adopted an atmospheric
scale height for the vertical length scale.  These estimates are
crude, although \cite{Cooper2006} showed that this formulation for
$K_{zz}$ allows 1D models to match the full tracer profiles from 3D
GCMs reasonably well. More recently, \citet{Heng2011a} used the
magnitude of the Eulerian mean streamfunction as a proxy for the
strength of the vertical motions and derived a vertical mixing
coefficient of $K_{zz}\sim10^6\rm\,m^2\,s^{-1}$. Note, however, that
the Eulerian-mean velocities are known to be a poor descriptor of
tracer advection rates in planetary atmospheres, since mixing by large scale eddies (potentially resolved by GCMs) often dominates over transport due to the Eulerian-mean circulation~\citep[see, \eg][Chapter 9]{Andrews1987}. 

 Although we confirm that mixing in hot Jupiters
atmospheres is strong, we find a value that is significantly smaller
than the previous ones. In particular, our value is two orders of
magnitude smaller than what is obtained when multiplying the vertical
scale height by the root mean square of the vertical velocity (blue
curve in Fig.~\ref{fig::Kzz}), a common estimate for $K_{\rm zz}$ in the
literature~\citep[\eg][]{Lewis2010,Moses2011}.

As stated in Sect. \ref{sec::Dynamics}, the model does not include
any sub-grid vertical diffusion coefficient. Yet, given the large
values for $K_{\rm zz}$ that we derive from the resolved flow, it seems
unlikely that sub-grid turbulent mixing would contribute significantly
to the total mixing. However, the interaction between small-scale
turbulence and the global flow might not be trivial and a more
detailed study would be required to draw a firm conclusion.

We emphasize that vertical mixing by the global circulation appears to be
planet-wide and differs from region to region.  Although the globally
averaged dynamics seem to be reasonably described by a vertical
mixing coefficient, that is not the case for the local flow in the
simulation. It is therefore difficult to define mixing coefficient
values for particular locations in the planet.

\begin{figure}[h!]
\includegraphics[width=\linewidth]{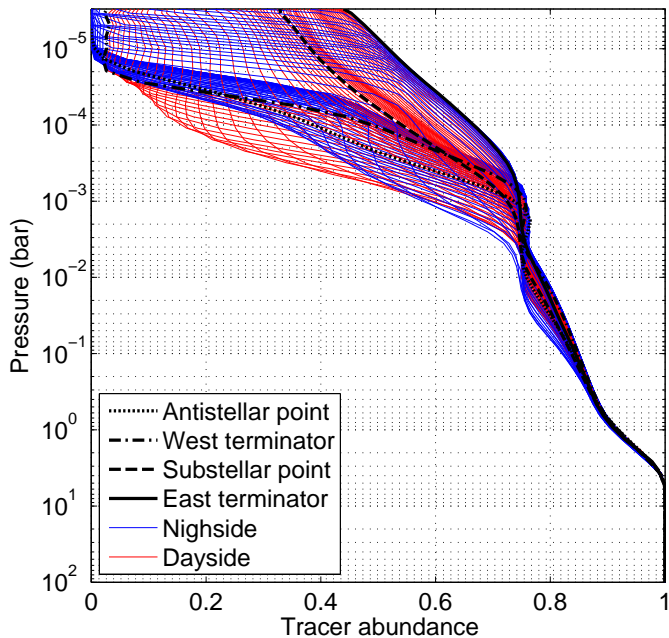}
\caption{Time averaged tracer abundance along the equator for nightside condensate of $\unit{1}\micro\meter$. The profiles are equally spaced in longitude by $2.8\degree$. The dayside profiles are in red whereas the nightside profiles are in blue. We highlight the profiles at the antistellar point (dotted line), the west terminator (dot-dashed line), the substellar point (dashed line) and the east terminator (plain line).}
\label{fig::Abundance-equator}
\end{figure}

Along the equator, where the strong flow efficiently mixes the tracers
longitudinally, we expect a good agreement of our 1D model to the 3D
flow. Indeed, using the value of $K_{\rm zz}$ derived in Sect.
\ref{sec::Kzz} we realize that the spread of the tracer profiles along
the equator (Fig.~\ref{fig::Abundance-equator}) is of the same order
of magnitude as the spread predicted by the 1D model (middle panel
of Fig.~\ref{fig::Chi-profiles-Kz-square}). However, in the 1D
model, profiles from equally sampled longitude are equally spaced in
abundances whereas the profiles obtained from the 3D model are
sometimes packed together and sometimes widely spread, denoting an
unequal strength of the vertical mixing longitudinally.

\section{Applications}
\label{sec::Applications}
\subsection{Presence of a stratosphere on hot Jupiters}
\label{sec::Stratosphere}
TiO is a leading hypothesis for the absorber needed to create temperature inversions in hot Jupiters atmospheres~\citep{Hubeny2003, Fortney2008}.  However, TiO condenses at
temperatures lower than $\approx\unit{2000}\kelvin$. On most hot
Jupiters, while gaseous TiO can be stable on the dayside, it should
condense on the nightside. Thus, our results allow us to address the
question of whether TiO can remain suspended in the atmospheres of hot
Jupiters, and hence whether TiO-induced stratospheres are indeed
viable.  \citet{Spiegel2009} predicted that for a solar abundance of
TiO in the planet, an abundance of 0.5 times the deep abundance at
1~mbar is necessary to maintain a temperature inversion on the
dayside. Our simulations suggest that if TiO condenses into particles
bigger than several micrometers, the day-night cold trap will be
sufficiently efficient to deplete it from the dayside. If, on the
contrary, TiO cannot condense in particles bigger than several
micrometers, it should remain present on the dayside and produce a
stratosphere (see Figs.~\ref{fig::TracerMeanProfile} and \ref{fig::Time-average}).

\begin{figure}[h!]
\center
\includegraphics[width=\linewidth]{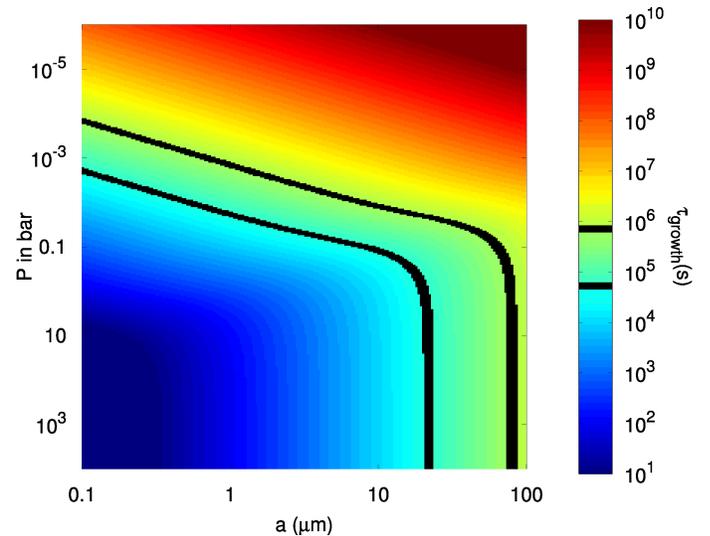}
\caption{Approximate growth time scale of $\mathrm{TiO_{2}}$ grains in HD 209458b. The bottom black line shows the advective timescale and the top one is ten times the advective timescale.}
\label{fig::GrowthTIO2}
\end{figure}

The size of the condensate, a free parameter in our study, results from complex microphysical processes. Once on the nightside,
TiO is over-saturated, and thus Ti-bearing condensates are expected to
appear. We used the formalism of \citet{Woitke2003} to calculate the
characteristic growth time scale of $\mathrm{TiO_{2}}$ particles
assuming that all the titanium is contained in $\mathrm{TiO_{2(g)}}$ and
that this last fully saturates the atmosphere. If the condensate growth time scale exceeds the advective time scale -- the time for the jet to travel across one hemisphere -- the condensate will be back in the dayside before reaching it's full size and will vaporize. Thus, all the particles above the black line in Fig.~\ref{fig::GrowthTIO2} are unlikely to form. The low elemental
abundance of titanium---solar abundance is
$10^{-7}$ compared to H \citep{Lodders2002}---kinetically inhibits
the formation of micrometer size particles above $\unit{10}\milli\bbar$. 

\begin{figure}[h!]
\center
\includegraphics[width=\linewidth]{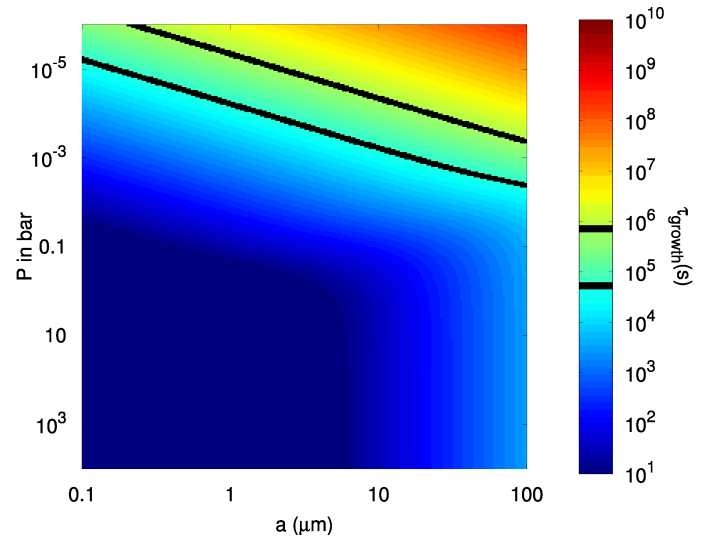}
\caption{Approximate growth time scale of $\mathrm{SiO_{2}}$ grains in HD 209458b. The bottom black line shows the advective timescale and the top one is ten times the advective timescale.}
\label{fig::GrowthSIO2}
\end{figure}

However, titanium is not the only element that can form condensates on
the nightside of hot Jupiters. Silicates are believed to condense and
could incorporate titanium atoms into their grains. Sub-micron size
$\mathrm{TiO_{2}}$ particles could even be used as seeds for the
formation of silicates grains. In that case the relevant time scale is
not tied to the growth of $\mathrm{TiO_{2}}$ grains but rather to the
growth of $\mathrm{SiO_{2}}$ based grains ($\mathrm{MgSiO_{2}}$ for
example). We calculated the growth time scale of
$\mathrm{SiO_{2}}$-based grains using the same formalism as for
$\mathrm{TiO_{2}}$ grains, assuming that $\mathrm{SiO_{2(g)}}$ is
fully saturated and that all the silica atoms are in
$\mathrm{SiO_{2(g)}}$ molecules. As can be seen in Fig.
\ref{fig::GrowthSIO2}, particles as big as $10\,\micro\meter$ can form
at pressures as low as $10\milli\bbar$ in that case.  

The estimates in Fig.~\ref{fig::GrowthSIO2} show that
the growth time of silicate particles is comparable to the time for the jet to cross an hemisphere in the pressure range where the 
stratosphere forms ($\sim$0.01--1$\,$mbar).  This comparison suggests that more detailed calculations, coupling the 3D dynamics to microphysics
that allow self-consistent prediction of particle growth, may
be necessary to obtain a firm conclusion about whether particle growth
timescales are sufficiently long to inhibit loss of TiO from the
atmosphere.

\citet{Spiegel2009} studied the deep cold trap in the deep
layer of the planet, at pressures exceeding tens of bars, where dynamical
mixing rates are probably low. In planets that exhibit such a 
cold trap, mixing TiO upward to altitudes where it could be
affected by the day-night cold trap may be difficult.  On the
other hand, the opacities and therefore temperature structure in
these deep regions are rather uncertain; moreover, hot Jupiters
that are particularly highly irradiated would exhibit
warmer temperatures and would therefore be less likely to
exhibit such a vertical cold trap. On these planets, the
day-night cold trap at low pressure ($P \lesssim 1\,$bar) would
then dominate.

The spatial variations in the gaseous tracer abundance in our
3D models (Fig.~\ref{fig::Time-average}) suggest that the
TiO abundance on the dayside, and hence the stratosphere itself,
could be patchy, with some regions of the dayside exhibiting
a stronger temperature inversion than others.  This would have
interesting consequences for the interpretation of dayside
infrared spectra.

The time variability described in Sect. \ref{sec::TimeVariability}
could affect the presence of the stratosphere, leading to strong temporal
variability in the upper atmospheric temperatures. In particular, the
tracer abundance averaged over much of the dayside exhibits large-amplitude
fluctuations (Fig.~\ref{fig::TracerTimeDependance}),
particularly for larger particle sizes.  This suggests that, at least
for some planets, the TiO abundance could fluctuate between values
large enough to generate a stratosphere and values too small for a
stratosphere to form. The stratosphere itself might then fluctuate
episodically in time, leading to variations of a factor two in the thermal flux emitted by the planet at some wavelength~\citep[see Fig. 12 of][]{Fortney2008}. Although not included in our current models, there is the possibility of feedbacks with the flow itself, since the
presence (or absence) of a stratosphere exerts a significant impact on
the flow structure and vertical mixing rates.  If the feedback is
positive, i.e. if the presence of high temperatures in the upper
atmosphere enhances the mixing, then hot Jupiters could oscillate
between a state with strong vertical mixing and stratospheric heating
by TiO and a state with no stratospheric heating and less vertical
mixing. This is a two-state atmosphere analogous to that described by
\citet{Hubeny2003}. However, this possibility remains speculative and
further models that include the feedback of the tracer field on the flow
are necessary to draw a firm conclusion.

TiO is thought to be the major Ti-bearing gas in hot Jupiters
atmospheres \citep[see][]{Lodders2002}. However, other Ti-bearing
gases, $\mathrm{TiO_{2}}$ being the most abundant, are believed to be
the condensable Ti-bearing species. Thus, a parcel of gas experiencing
a sudden drop in temperature due to its advection to the nightside
might not see all its titanium incorporated into condensates, but rather only
the titanium atoms that already reside in other Ti-bearing
molecules such as $\mathrm{TiO_{2(g)}}$. The reaction between
gaseous species $\mathrm{TiO+H_{2}O\rightleftharpoons TiO_{2}+H_{2}}$
is fast under the conditions relevant to the dayside of hot Jupiters
\citep[see][]{Fortney2008}. However, the reaction might be kinetically
inhibited on the nightside of the planet where the temperature drops
significantly, leading to a smaller depletion of Ti than in the
hypothetical case where $\mathrm{TiO_{(g)}}$ could condense by itself.

\subsection{Clouds in hot Jupiters atmospheres}
\label{sec::Clouds}
The huge day-night temperature difference and cold nightside
temperatures predicted on many hot Jupiters at low pressure
\citep[\eg][]{Showman2009} suggest that, in addition to TiO, a wide
range of other chemical species, including silicates and iron, will
condense on the nightside. Some of them could also stay in a condensed
state in part or all of the dayside hemisphere. The Rayleigh
scattering slope in the transmission spectrum of HD189733b, first
observed by \citet{LecavelierDesEtangs2008} and later confirmed by
numerous observations (see \citet{Pont2013} for a review of the
different observations of this planet) is best fitted by models
including sub-micron sized particles.  Then the strong spatial and
temporal variations observed in our model can also be interpreted as
spatial variation of the cloud coverage in the atmospheres of
hot Jupiters. This could lead to albedo variations along the dayside of
the planet. As the hottest point of the planet is shifted to the east, the western part of the dayside is colder than the eastern one (see Fig.~\ref{fig::TempVelocityField}). Therefore, the western part of the dayside could be colder than the condensation temperature of some species whereas the eastern part could be at higher temperatures, leading to a more cloudy atmosphere west of the substellar point than east of it. Moreover, due to the eastward superrotating jet, material west of the substellar point arrives from the cold nightside -- where condensation is thought to happen --- whereas material east of the substellar point arrives from the hot substellar point --- where the material is thought to have sublimated. Thus, clouds, if present on the dayside of the planet, should form more easily west of the substellar point than east of it, leading to a longitudinal variation of the albedo that contributes to the spatial variation described before. The large amount of data in the visible from the Kepler space telescope is ideal to search for such a spatial and temporal variability in albedo pattern of tidally locked planets.

\subsection{Parameter retrieval}
\label{sec::Parameters}
Atmospheric characteristics of hot Jupiters are usually derived from
disk-integrated fluxes (for secondary eclipses measurements) or
limb-integrated transmission (for transits spectroscopy). Therefore,
interpretation of the data is usually done using one dimensional
atmospheric models, assuming an homogenous atmosphere, both in term of
temperature structure and composition \citep[\eg][among others]{Madhusudhan2009,Lee2012,Benneke2012}. However, given the strong
spatial variability in the tracer distribution both on the dayside
hemisphere and in the limb profiles observed, some future exoplanet spectra --- obtained with more and better data than nowadays --- might be better understood by considering spatial
variation in the atmospheric profiles and chemical composition along
the planet~\citep[see also][on the longitudinal variability in the chemical composition of HD209458b]{Agundez2012}. For example, an inhomogeneous distribution of TiO would lead to strong brightness differences in the emitted flux from the
different locations of the planet, which could affect features such as
the apparent eastward offset of the brightest spot of the
planet. Hazes could also show a similar behavior, leading to planetary
spectra that might be better explained with the combination of two
different (cloudy and cloudless) 1D models---as is the case in recent
models of brown dwarfs \citep[\eg][]{Marley2010, Burrows2011}. The rising technique of
secondary eclipse mapping \citep[see][]{Majeau2012, DeWit2012}, combined with the accuracy of telescopes like EChO~\citep{Tinetti2012} or JWST might soon
allow us to constrain better the spatial inhomogeneities in the disk
of the planet.

\section*{Conclusion}
We presented global, three-dimensional numerical simulations of the
atmospheric circulation of HD 209458b including a passive tracer. This is the first circulation model of
a hot Jupiter to include the dynamical mixing of condensable species. We applied our
model to chemical species that are gaseous on the dayside but
condense on the nightside of the planet and are trapped in particles of a given size. Given the strong day/night
contrast present in hot Jupiters, our model applies to a wealth of
different chemical species such as titanium, vanadium, and
silicate oxides among others.

Prior studies of hot Jupiters
circulation demonstrate the presence of 3D circulation patterns including both strong horizontal and vertical
velocities that are necessary for mixing to happen \citep[\eg][]{Showman2002, Cooper2005, Cooper2006, Showman2008, Showman2013, Rauscher2012a,Rauscher2012b,Rauscher2013, Heng2011a, Heng2011, Heng2012, Lewis2010, Dobbs-Dixon2008, Dobbs-Dixon2010, Dobbs-Dixon2012, Perna2010, Perna2012}. While preliminary attempts
have been made to quantify the mixing rates \citep{Cooper2006, Heng2011}, these methods are approximate and do not lead to a rigorous quantification
of the mixing rate. Here we demonstrate that, although hot Jupiters atmospheres are believed to
be stably stratified (\ie~locally non-convective), they are strongly
mixed. In the presence of a background gradient of chemical species,
large-scale circulation patterns naturally create upward mixing. This
mixing, resolved by the GCM, is strong and likely dominates
over molecular, convective or turbulent mixing.  In HD 209458b, the
mixing is strong enough to keep a condensable species aloft if it
condenses into particles smaller than a few microns on the nightside of
the planet.

The coupling between 3D flow and particle settling leads to strong
spatial and temporal variations in the abundance of a given condensable
species. Around $\unit{0.1}\milli\bbar$, the tracer abundance is homogeneous in
longitude but exhibits a large latitudinal variation, the equator
being more depleted than the poles. Around $\unit{1}\milli\bbar$, at high latitudes,
the day/night contrast becomes important. According to our models,
variability of up to $\sim$50\% in the dayside tracer abundance, and
of up to $\sim$75\% in the tracer abundance along the limb, can occur
for sufficiently large particle sizes ($\sim$$5\rm\,\mu$m). This
variability characteristic periods ranging from
days to $\sim$50--100 days. The observability of such a variation depends on the radiative properties of the considered species and will be quantified in a future work.

These results can be applied to a wide range of molecules in hot
Jupiters atmospheres. Titanium oxide, the best candidate for creating
a temperature inversion in the dayside of hot Jupiters, should
condense on the nightside of most planets. Our results imply that the
day/night cold trap could impede the formation of a stratosphere in
the dayside if TiO condenses into particles bigger than a few microns
on the nightside. Growing particles to such a size seems difficult
when TiO alone is considered due to its small abundance. However, TiO
can be incorporated into condensates from more abundant gases such as
silicate oxides. In that case the day/night cold trap could be strong
enough to impede the formation of a hot stratosphere on the dayside.
Spatial variability of TiO could significantly affect the dayside
temperature structure and exert interesting effects on infrared
spectra and lightcurves. For example, mid-to-high latitudes might keep
enough TiO to create an inversion whereas the equator, more depleted,
might not be able to sustain the inversion. Such a latitudinal
contrast could be observed using the secondary eclipse mapping technique
\citep[\eg][]{Majeau2012, DeWit2012}.  The temporal variability
observed in the model could lead to the appearance and disappearance
of the stratosphere on timescales of $\sim$10--100 earth days. Highly
irradiated planets have significant thermal emission in the Kepler bandpass
\citep[\eg][for Hat-P-7b]{Spiegel2010}. Using the long photometric
series from the Kepler Space Telescope, such a variability might be
observable.  

Our results also apply to silicate hazes. Temporal and
spatial variability in the cloud coverage could strongly affect the
albedo and the thermal emission of the planet. For moderately
irradiated planets, the Kepler spacecraft observes the reflected
light of the star by the planet \citep[\eg][for
  Kepler-7b]{Demory2011}. Thus, the time series from Kepler could be
used to build albedo, and therefore cloud maps of the planet.

Although there is no theoretical reason for the upward mixing driven
by the global circulation to be diffusive, it is interesting to
quantify the averaged vertical mixing with a diffusive model. The
parameterization $K_{\rm zz}$ value of $K_{\rm zz}=5\times10^{4}/\sqrt{P_{\rm bar}}\rm \,m^2\,s^{-1}$ or $K_{\rm zz}=\unit{5\times10^{8}/\sqrt{P_{\rm bar}}}\centi\meter\squared\reciprocal\second$, between $\sim\unit{1}\bbar$ and $\sim\unit{1}\micro\bbar$, can be used in 1D models of HD 209458b.

This study, the first one to include the influence of the dynamics on condensable species in a GCM of a hot Jupiter, confirms that hot Jupiters atmospheres are strongly mixed and that large scale spatial and temporal variability are
expected in any condensable chemical constituents. Today, observers
can already detect longitudinal variations in the emitted thermal flux
of the planet. In the next decade, both longitudinal and latitudinal
variations in thermal emission and albedo of the hot Jupiters will be
observable, expanding the study of weather to extra-solar planets.
 
\section*{Acknowledgement}
The authors wish to thank the ISIMA program (http://isima.ucsc.edu)
where this project was initiated. This project was supported by NASA
Planetary Atmospheres and Origins grants to APS.  We thank Christiane
Helling for discussions on dust formation, Franck Hersant for
discussions on the $K_{zz}$ calculations, Nikole Lewis for sharing her Matlab routines and Franck Selsis for useful
input. We also acknowledge the anonymous referee whose numerous comments improved the clarity of this paper.

\bibliography{./Parmentier2013a.bib}

\begin{appendix} 
\section{Departure from the Cunningham velocity}
\label{sec::Appendix}
The Stokes-Cunningham velocity defined in Eq.~\eqref{eq::Vf} is derived under the assumption of low Reynolds number. Therefore it is not valid for turbulent flow and other expressions may be used when the Reynolds number increases.
 Here we derive better laws for intermediate and large Reynolds number. 

\subsection{Low Knudsen number}
For small Reynolds number and small Knudsen number, the drag force exerted by a fluid on a sphere at rest is considered proportional to the kinetic energy of the fluid and the projected area of the sphere. The coefficient of proportionality, or drag coefficient, $C_{\rm D}$ is given by :
\begin{equation}
C_{\rm D}=\frac{F_{\rm drag}}{\rho V^{2}\pi a^{2}/2}
\end{equation}
Then, equating gravity and drag forces leads to the settling velocity of a particle in an atmosphere :
\begin{equation}
V_{f}^2C_D=\frac{8a}{3}\frac{\rho_{\rm p}-\rho}{\rho}
\label{eq::equilibrium}
\end{equation}
Where $\rho_{\rm p}$ is the density of the particle. For small Reynolds numbers and high Knudsen number, $C_{D}=24$ is constant and the settling velocity is the Stokes velocity.
When increasing the Reynolds number, the non linear terms of the Navier-Stokes equation become important and $C_{D}$ is no longer constant. We used tabulated values of the drag coefficient as a function of the Reynolds number given by \cite{Pruppacher}. We assume that $C_{\rm D}=24$ when the Reynolds number reaches 1 to stay consistent with Stokes flow and that $C_{\rm D}$ reaches its asymptotic value, $C_{\rm D}=0.45$, when $\NRe\equiv 2R_{\rm e}=1000$ and fit the relationship :

\begin{alignat}{3}
&\log_{10}(\NRe)=&-&1.215047+ 0.923242\log_{10}(C_D\NRe^2)\nonumber\\
&&-&0.031293\log_{10}(C_{D}\NRe^2)^2
\label{fit}
\end{alignat}

Then we follow the same method as ~\cite{Ackerman2001}. Noting that :
\begin{equation}
C_{\rm D}\NRe^2=\frac{32\rho ga^3(\rho_p-\rho)}{3\eta^2}
\end{equation}
is independent of the velocity, we use the fit of Eq.~\ref{fit} and extract the velocity :

\begin{equation}
V_{\rm f}=\frac{\eta}{2\rho a}10^{-1.21}\left(\frac{32\rho g a^3\Delta\rho}{3\eta^2}\right)^{0.92}\left(\frac{32\rho g a^3\Delta\rho}{3\eta^2}\right)^{-0.062}
\label{V2}
\end{equation}

\subsection{High Knudsen number}
In the free-molecular regime, calculations have been made by \cite{Probstein} leading to an expression for the drag coefficient:
\begin{equation} 
C_{\rm D}=\frac{2}{3\sa}\sqrt{\pi}+\frac{2\sa^2+1}{\sa^3\sqrt{\pi}}\exp(-\sa^2)+\frac{4\sa^4+4\sa^2-1}{2\sa^4}erf(\sa)
\label{eq::drag}
\end{equation}
where $s_a$ is the ratio of the object velocity over the thermal speed of the gas ($V_{\rm T}=\sqrt{\frac{2k_BT}{m}}=\frac{\sqrt{\pi}}{2}\bar c$ where $\bar c$ is the sound speed) and $erf$ is the error function.

For velocities much smaller than the sound speed, $s_{a}\to0$ and we can use an equivalent of the error function in 0:

\begin{equation}
erf(\sa) = \frac{2}{\sqrt(\pi)}e^{-\sa^2}(s_a+\frac{2}{3}\sa^3)+o(\sa^4)
\end{equation}

Using this equation inside Eq.~\eqref{eq::drag} and taking the limit $s_{a}\to0$, the term $e^{-\sa^{2}}$ goes to 1 and the terms proportional to $\frac{1}{s_a^3}$ cancels out leading to:
\begin{equation}
C_{\rm D}= \left(\frac{2\sqrt{\pi}}{3}+\frac{16}{3\sqrt{\pi}}\right)\frac{V_{\rm T}}{V}
\end{equation}

For velocities much greater than the sound speed, the limit of Eq.~\eqref{eq::drag} when $s_{a}\to\infty$ is:
\begin{equation}
C_D\sim 2
\end{equation}


In order to simplify Eq.~\eqref{eq::drag} we use the following expression for the drag coefficient at high Knudsen number :

\begin{equation}
C_{\rm D}= \left(\frac{2\sqrt{\pi}}{3}+\frac{16}{3\sqrt{\pi}}\right)\frac{V_{\rm T}}{V}+2
\end{equation}


Our approximation fits correctly the exact expression in the limit of low and high velocities. In between the difference to the exact expression is at most $30\%$. Replacing $C_D$ by its value in Eq.~\eqref{eq::equilibrium} we obtain a second order equation for the velocity :
\begin{equation}
2\left(\frac{V}{V_{\rm T}}\right)^2+\left(\frac{2\sqrt{\pi}}{3}+\frac{16}{3\sqrt{\pi}}\right)\frac{V}{V_{\rm T}}-\frac{8}{3}ag\frac{\Delta\rho}{\rho}=0
\end{equation}
which leads to :


\begin{equation}
V_{\rm f}=\frac{A}{4}V_{\rm T}\left(\sqrt{1+\frac{96}{A^2\sqrt{\pi}}\frac{V_{\rm stokes}}{V_{\rm T}}K_N}-1\right)
\end{equation}
with $A=\left(\frac{2\sqrt\pi}{3}+\frac{16}{3\sqrt{\pi}}\right)$. When the speed becomes small compared to the sound speed ($V_{\rm stokes}\ll V_{\rm T}$) we obtain :
\begin{equation}
V_{\rm f}=\frac{96}{8\sqrt{\pi} A}K_{\rm N}V_{\rm stokes}\approx1.61K_{\rm N}V_{\rm stokes}
\end{equation}
which is in good agreement with Eq.~\eqref{eq::Vf}, derived for high Knudsen number and small Reynolds numbers.

\subsection{Comparison with the Cunningham velocity}
\begin{figure}[h!]
\centering
\includegraphics[width=\linewidth]{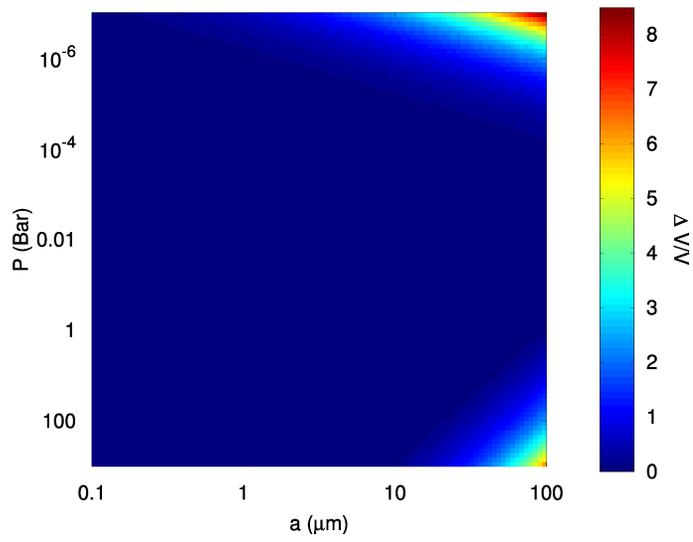}
\caption{Ratio of the Cunningham velocity to the more sophisticated model of the particle settling velocity considered in the Appendix.}
\label{fig::VelocityRatio}
\end{figure}

Figure \ref{fig::VelocityRatio} shows the ratio of the Cunningham
velocity (see Eq.~\eqref{eq::Vf}) to the ones we just derived. The
difference is noticeable only for particles of the order of
$\unit{100}\micro\meter$ at pressures less than the
$\unit{10^{-4}}\bbar$ level and exceeding the $\unit{10}\bbar$
level. This difference is always less than one order of magnitude and
concern only a tiny portion of the parameter space which has little
relevance to our study (the largest particle sizes considered in our 3D
models is $10\,\mu$m). Thus we decided to neglect this discrepancy in the main
study.

\end{appendix}
\end{document}